\documentclass[twocolumn]{aastex62}

\usepackage{graphicx}
\usepackage{color}
\usepackage{amsmath}  
\usepackage{verbatim}     
\usepackage{hyperref}
\usepackage[noabbrev,capitalise]{cleveref}  
\usepackage{natbib}
\usepackage[mathscr]{euscript} 
\usepackage{xspace}

\makeatletter
\usepackage{etoolbox}
\patchcmd\H@refstepcounter{\protected@edef}{\protected@xdef}{}{}
\makeatother

\newcommand{\sn}{SNIa}

\newcommand{\h}{\ensuremath{\text{H}_0}}

\newcommand{\ha}{\ensuremath{\text{H}_\alpha}}
\newcommand{\un}[1]{~\text{#1}}  
\newcommand{\sfh}{4p\ensuremath{\tau}-model} 
\newcommand{\hr}[1]{Hubble residual{#1}}   
\newcommand{\nuc}[2]{\ensuremath{^{#2}\text{#1}}}

\newcommand{\globalCorr}{-0.23}
\newcommand{\globalCorrSig}{2.1\ensuremath{\sigma}}
\newcommand{\localCorr}{-0.21}
\newcommand{\localCorrSig}{1.8\ensuremath{\sigma}}

\newcommand{\ageStepLocation}{\ensuremath{8\un{Gyr}}}
\newcommand{\meanYoung}{\ensuremath{0.005 \pm 0.018 \un{mag}}}
\newcommand{\meanOld}{\ensuremath{-0.109 \pm 0.035 \un{mag}}}
\newcommand{\ageStep}{\ensuremath{0.114 \pm 0.039 \un{mag}}}
\newcommand{\ageStepSig}{\ensuremath{2.9\sigma}}
\newcommand{\youngCorr}{\ensuremath{-0.07}}
\newcommand{\oldCorr}{\ensuremath{-0.09}}

\newcommand{\pcTwoCorr}{\ensuremath{-0.06}}
\newcommand{\pcThreeCorr}{0.21}

\newcommand{\pcFourCorr}{\ensuremath{-0.15}}
\newcommand{\pcOneCorr}{0.44}
\newcommand{\pcOneGlobalCorr}{0.40}

\newcommand{\pcOneP}{\ensuremath{3.86 \times 10^{-6}}}
\newcommand{\pcOneSig}{\ensuremath{4.7\sigma}}

\newcommand{\pcOneGlobalSig}{\ensuremath{4.0\sigma}}

\newcommand{\pcOneSlope}{\ensuremath{0.051}}
\newcommand{\pcOneSlopeUncert}{\ensuremath{\pm 0.011 \un{mag}}}
\newcommand{\pcOneSlopeFull}{\ensuremath{\pcOneSlope \pcOneSlopeUncert}}
\newcommand{\pcOneIntercept}{\ensuremath{-0.012}}
\newcommand{\pcOneInterceptUncert}{\ensuremath{\pm 0.016 \un{mag}}}
\newcommand{\pcOneInterceptFull}{\ensuremath{\pcOneIntercept \pcOneInterceptUncert }}
\newcommand{\pcOneGlobalSlope}{\ensuremath{0.045}}
\newcommand{\pcOneGlobalSlopeUncert}{\ensuremath{\pm 0.011 \un{mag}}}
\newcommand{\pcOneGlobalSlopeFull}{\ensuremath{\pcOneGlobalSlope \pcOneGlobalSlopeUncert}}
\newcommand{\pcOneGlobalIntercept}{\ensuremath{-0.012}}
\newcommand{\pcOneGlobalInterceptUncert}{\ensuremath{\pm 0.016 \un{mag}}}
\newcommand{\pcOneGlobalInterceptFull}{\ensuremath{\pcOneGlobalIntercept \pcOneGlobalInterceptUncert}}

\newcommand{\HrSigma}{\ensuremath{0.17 \un{mag}}}
\newcommand{\HrSigmaLocalTrend}{\ensuremath{0.15 \un{mag}}}
\newcommand{\HrSigmaGlobalTrend}{\ensuremath{0.16 \un{mag}}}

\newcommand{\bootstrapN}{100,000\xspace}

\newcommand{\maxCorr}{0.43} 

\newcommand{\deltaMagLocal}{\ensuremath{0.021 \un{mag}}}
\newcommand{\deltaMagGlobal}{\ensuremath{0.028 \un{mag}}}
\newcommand{\hShiftPercentLocal}{1.0\%}
\newcommand{\hShiftPercentGlobal}{1.3\%}
\newcommand{\deltaPCNeeded}{\ensuremath{\sim 3.5}} 
\newcommand{\fractionLargerPCNeeded}{6} 

\begin{document}

\title{Think Global, Act Local: The Influence of Environment Age and Host Mass on Type Ia Supernova Light Curves}
\shorttitle{The Influence of Age and Mass on SNIa Light Curves}

\author[0000-0002-1873-8973]{B. M. Rose}
\affiliation{University of Notre Dame, Center for Astrophysics, Notre Dame, IN 46556}
\author[0000-0003-4069-2817]{P. M. Garnavich}
\affiliation{University of Notre Dame, Center for
Astrophysics, Notre Dame, IN 46556}
\author[0000-0002-8518-6638]{M. A. Berg}
\affiliation{University of Notre Dame, Center for
Astrophysics, Notre Dame, IN 46556}
\correspondingauthor{B. M. Rose}
\email{brose3@nd.edu}
\shortauthors{Rose, Garnavich, \& Berg}


\date{\today}
\received{November 12, 2018}
\revised{January 29, 2019}
\accepted{February 4, 2019}
\submitjournal{ApJ}

\begin{abstract}
The reliability of Type Ia supernovae (SNIa) may be limited by the imprint of their galactic origins. To investigate the connection between supernovae and their host characteristics, we developed an improved method to estimate the stellar population age of the host as well as the local environment around the site of the supernova. We use a Bayesian method to estimate the star formation history and mass weighted age of a supernova's environment by matching observed spectral energy distributions to a synthesized stellar population. Applying this age estimator to both the photometrically and spectroscopically classified Sloan Digital Sky Survey II supernovae (N=103) we find a $0.114 \pm 0.039~{\rm mag}$ ``step'' in the average Hubble residual at a stellar age of $\sim 8~\text{Gyr}$; it is nearly twice the size of the currently popular mass step. We then apply a principal component analysis on the SALT2 parameters, host stellar mass, and local environment age. We find that a new parameter, PC$_1$, consisting of a linear combination of stretch, host stellar mass, and local age, shows a very significant ($4.7\sigma$) correlation with Hubble residuals. There is a much broader range of PC$_1$ values found in the Hubble flow sample when compared with the Cepheid calibration galaxies. These samples have mildly statistically different average PC$_1$ values, at $\sim 2.5\sigma$, resulting in at most a 1.3\% reduction in the evaluation of H$_0$. Despite accounting for the highly significant trend in SNIa Hubble residuals, there remains a 9\% discrepancy between the most recent precision estimates of H$_0$ using SNIa and the CMB.
\end{abstract}


\defcitealias{Gupta2011}{G11}
\defcitealias{Campbell2013}{C13}

\section{Introduction}\label{intro}

For decades, astronomers have been developing methods to better understand the variation in peak luminosities of Type Ia supernovae (\sn{}) and improve their use as precision distance indicators. In \citeyear{Phillips1993}, \citeauthor{Phillips1993} identified a relationship between peak magnitude and the rate of fading in the light curves of \sn{}. A connection between supernova color and peak luminosity was also shown to improve distance estimates of \sn{} \citep{Riess1996, Tripp1999, Phillips1999}. \sn{} quickly became useful cosmological probes, measuring the expansion rate of the universe \citep{Hamuy1995, Riess1995}, the density of matter \citep{Garnavich1998a,Perlmutter1998}, and the accelerated expansion of the universe \citep{Riess1998,Perlmutter1999}.

 The principle energy source that powers a \sn{} light curve derives from the decay chain of \nuc{Ni}{56} that is synthesized in the runaway nuclear fusion at the start of the explosion \citep{Arnett1982}. The radioactive nickel yield appears to vary by a factor of eight over the extreme range of typical \sn{} luminosities. The mass of radioactive elements helps regulate the rate of recombination in iron group elements, and this results in the famous ``Phillips relation'' between the light curve decline rate and luminosity \citep{Kasen2007}.
 
 The reason some \sn{} synthesize nearly a solar mass of radioactive nickel while others are powered by a tenth of a solar mass remains uncertain. Models suggest that the speed of the fusion front moving through the white dwarf has a major influence on the radioactive yield.  The transition between deflagration (subsonic fusion) and detonation (supersonic fusion) may vary from supernova to supernova and this could explain the diversity in their luminosities. The variation in nickel yield appears to be influenced by host properties as \citet{Hamuy1996c} and \citet{Hamuy2000} noted that low-luminosity \sn{} tend to occur in passive galaxies like large ellipticals. This observation has been confirmed and expanded in several subsequent studies \citep{Gallagher2005,Sullivan2006} which clearly demonstrate that host galaxies provide an important clue to the progenitors and explosion mechanisms of \sn{}.
 
Metal abundance and population age are global properties of galaxies that could conceivably have an impact on a supernova's \nuc{Ni}{56} yield. Host properties that correlate with age or metallicity, such as galaxy mass, could also influence the character of a supernova explosion. For example, \citet{Timmes2003} proposed that the metal abundance during the main sequence stage could affect the neutron fraction in the resulting white dwarf stars. They predicted that high metal abundance populations will generate low radioactive yields and, thus, faint supernovae. Applying the galactic mass-metallicity relationship to this finding, \sn{} from large galaxies would then be systematically fainter just as was seen in \citet{Hamuy2000}. \citet{Bravo2010} derived a similar luminosity-metallicity relationship, but 3D calculations by \citet{Ropke2004} suggest the effect is much smaller than originally envisioned.  An observational test of this hypothesis by \citet{Gallagher2008} looked at \sn{} in elliptical galaxies and found trends with both age and metallicity, although this was disputed by \cite{Howell2009}.

A sensitive test of the environmental effects on \sn{} is to study the scatter in a \sn{} Hubble diagram after light curve shape and color corrections have been applied. This type of analysis is also important for constraining systematic errors in cosmological measurements.  Hubble residuals 
are the difference between the luminosity distance and the expected distance from the best fit cosmology and are most powerful in the ``Hubble flow'' where peculiar velocities are small compared with the expansion velocity. \citet{Woosley2007} and \citet{Kasen2009} showed that explosion asymmetries, metallicity variations, kinetic energy variations, and other explosion parameters can produce a dispersion in the Phillips relation for a fixed radioactive yield. These relationships present the possibility of using \hr{s} to probe supernova physics. Research over the past several years indicates that some host galaxy properties correlate with Hubble residuals \citep[e.g.][]{Sullivan2010,Lampeitl2010,Childress2013b}.

Surprisingly, the strongest correlation between \hr{s} and galaxy properties appears to be with the host stellar mass. The effect is called the ``mass step'' because at $\sim 10^{10} \un{M}_{\sun}$ there appears to be a jump in the average Hubble residual. This led to extensive work on understanding the physical properties such as population age and metallicity that could underlie the mass correlation \citep[e.g.][]{Gupta2011, Hayden2013, Moreno-Raya2016a, Moreno-Raya2016b}.  \citet{Childress2014} has proposed that the mass correlation is really an age effect amplified by galaxy ``downsizing.'' 
This research has been fruitful, but not definitive. Ongoing  analyses of \sn{} data sets continue to debate the significance of these effects \citep[e.g.][for LOSS and Foundation respectively]{Graur2016a,Jones2018}.  

\citet{Rigault2013} and \citet{Rigault2018} looked at star formation and specific star formation rates respectively at the sites of \sn{} explosions by measuring \ha\ emission strength using spatially resolved spectroscopy. They found a very significant step in corrected \sn{} peak luminosity as a function of the specific star formation rate at the location of the explosion. This research identified a set of \sn{} with a 0.2 mag offset in \hr{} that are generally found in regions of lower star formation rate. 
There is still not a consensus on the impact of local star formation rates on \hr{} especially when these trends are measured using UV observations \citep{Rigault2015, Jones2015}. 

A very recent study \citep{Jones2018} on a large low-redshift data set \citep[the Foundation sample,][]{Foley2017} compared \hr{} with host galaxy stellar mass, local environment mass (the region within a radius of $1.5\un{kpc}$ of the \sn{}), host galaxy rest frame $u-g$ color, and local environment rest frame $u-g$ color. The rest frame $u-g$ color is a simple way to estimate recent star formation, and therefore, a crude age estimator. They found that the local environment contained no new information compared to the global parameters. 



In addition, there appears to be a tension between the \h{} estimated from the cosmic microwave background observations \citep{Planck2016, Planck2018} and estimates based on \sn{}. These precision measurements now disagree by $3.8\sigma$ \citep{Riess2016, Riess2018b} due to either new physics or a systematic bias in one of the measurements.
The \sn{} host environment is of particular importance to the precision measurements of the Hubble constant (\h{}). The Cepheid variables used to calibrate the \sn{} peak absolute magnitude are massive stars, so are observed only in star-forming galaxies. In contrast, supernovae discovered in unbiased searches are found in all types of galaxies.  A mass step correction between the Cepheid calibrated hosts and the Hubble flow galaxies is currently applied to the data (at $\pm 0.03\un{mag}$) and provides a relatively minor tweak to the value of \h{} derived from supernovae. So, at present, host environment is not a major contributor to the uncertainty budgets of cosmological measurements.


Here, we scrutinize the relation between \hr{} and the age of the stellar population derived from host galaxy colors. We develop a technique to translate multi-band galaxy photometry into an estimate of the star formation history and finally an average age for the stellar population. In principle, colors should be a better indicator of \sn{} progenitor age than \ha{}. This is because after a single burst of star formation, \ha{} emitting HII regions are dissipating just as \sn{} are beginning to explode.
We apply our technique to both the global photometry of \sn{} host galaxies and to the specific populations at the sites of the explosions. We then compare the local and global ages with \sn{} characteristics, and other host properties to determine the parameters that have the largest impact on the measured \hr{}. Finally, we investigate how our estimated ages may influence the current \sn{} measurements of \h{} and its error budget.

\section{Data}\label{sec:data}

For our primary analysis, we use \sn{} that were discovered with the SDSS-II Supernova survey \citep{Sako2008}.
We selected both spectroscopic and photometric classified \sn{} that passed cosmology cuts as described and released by \cite{Campbell2013}. This cosmological data set, including their SALT2 \replaced{\citep{Guy2007}}{\citep{Guy2007,Guy2010}} calibration parameters, are\dataset[available online]{http://www.icg.port.ac.uk/stable/campbelh/SDSS_Photometric_SNe_Ia.fits}.
Hereafter we will refer to \cite{Campbell2013} as \citetalias{Campbell2013}.
For our analysis, a few additional cuts were applied. For quality, we limited our analysis to objects whose photometric uncertainty is less than $1.5~\text{mag}$. These cuts are dominated by low quality $u$-band magnitudes. In the end, the resulting $g$- and $i$-band maximum uncertainties are less than $0.3\un{mag}$, and less than $0.16\un{mag}$ for $r$-band. In addition we removed objects that had a \hr{} greater than $0.7\un{mag}$. Looking at Figure 16 of \citetalias{Campbell2013}, these objects are likely core collapse supernovae that passed the color-magnitude cut.

\added{This data set does not use the most recent standardization tools, such as BEAMS with Bias Corrections applied to the Pantheon data set \citep{Kessler2017, Scolnic2018}. Restricting ourselves to SDSS photometry and low-redshift events avoids several of the biases mitigated in the Pantheon analysis while still providing a large, uniform sample.}

The software developed for the data aggregation and analysis described in this paper is available at \url{https://github.com/benjaminrose/mc-age}.

\subsection{Local Environment Photometry}

The photometry of the environment around the site of the supernovae are taken from the ``Scene Modeling'' method described in \cite{Holtzman2008}. The method incorporates all the images taken during the survey to build a photometric model at the location of the transient. The resulting photometry at the site is a convolution of the galaxy on the scale of the typical seeing. By applying this fixed angular aperture we always get the most compact local environment possible. Finally, in order to keep the environment truly ``local,'' a redshift cutoff was imposed, $z < 0.2$. With the average seeing for SDSS being $1.4''$, the maximum extent of a galaxy's local environment was $3\un{kpc}$ in radius. \added{At higher redshifts the angular resolution encompasses a majority of typical galaxy and there is little difference between local and global properties.} The typical size of the projected aperture defining the environment at the supernova location was $1.5\un{kpc}$. 

\added{Since core collapse supernovae are less luminous than \sn{}, their contamination percentage increases in the low redshift portion of any data set. Anticipating this higher percentage of core collapse supernova (CC) interlopers, we added further \hr{} cut to minimize the CC contamination. In addition, the statistics used in this work are robust against the $\gtrsim 3.9\%$ CC contamination estimated in \citetalias{Campbell2013}. A detailed study of CC contamination affecting \sn{} standardization with host galaxy properties should be done since the ratio of \sn{} and CC is highly dependent on host properties.}

This results in a final data set of 103 \sn{}.
A partial list of the final data set used in the local environment analysis is visible in \cref{tab:local}.

\begin{deluxetable*}{ccc|ccccc|ccccc|cc}
\tablecolumns{7}
\tablewidth{0pt}
\tablecaption{Local environment data for \cite{Campbell2013} \sn{} \label{tab:local}} 
\tablehead{
    \colhead{SDSS ID} & \colhead{redshift} & \colhead{uncert.}
    & 
    \colhead{u} & \colhead{g} & \colhead{r}  & \colhead{i} & \colhead{z}
    &
    \colhead{$\sigma_u$} & \colhead{$\sigma_g$} & \colhead{$\sigma_r$}  & \colhead{$\sigma_i$} & \colhead{$\sigma_z$}
    &
    \colhead{HR} & \colhead{uncert.} \vspace{-0.5em}
    \\
    \colhead{} & \colhead{} & \colhead{$\times 10^{-5}$}
    & 
    \colhead{[mag]} & \colhead{[mag]} & \colhead{[mag]}  & \colhead{[mag]} & \colhead{[mag]}
    &
    \colhead{[mag]} & \colhead{[mag]} & \colhead{[mag]}  & \colhead{[mag]} & \colhead{[mag]}
    &
    \colhead{[mag]} & \colhead{[mag]}
    }
    
\startdata
762  & 0.19138 & 2.4  & 24.65 & 23.82 & 22.95 & 22.61 & 22.18 & 0.79 & 0.09 & 0.04 & 0.04  & 0.13 &  0.15 & 0.08 \\
1032 & 0.12975 & 3.3  & 24.92 & 24.74 & 23.73 & 23.32 & 22.87 & 0.61 & 0.16 & 0.09 & 0.11  & 0.26 & -0.15 & 0.12 \\
1371 & 0.11934 & 2.6  & 23.22 & 21.52 & 20.42 & 20.00 & 19.62 & 0.16 & 0.01 & 0.00 & 0.006 & 0.01 & -0.13 & 0.06 \\
1794 & 0.14191 & 6.3  & 23.76 & 23.77 & 23.09 & 22.89 & 22.81 & 0.43 & 0.11 & 0.08 & 0.06  & 0.24 &  0.27 & 0.08 \\
2372 & 0.18046 & 2.2  & 24.81 & 22.85 & 21.84 & 21.40 & 21.02 & 0.87 & 0.03 & 0.01 & 0.01  & 0.03 & -0.12 & 0.06 
\enddata
\tablecomments{Full machine readable data set is\dataset[available in the online journal]{https://raw.githubusercontent.com/benjaminrose/MC-Age/master/data/campbell_local.tsv}.}


\end{deluxetable*}





\subsection{Global Photometry}

In addition to the photometry of the local environment, we also study the correlation between the supernova characteristics and the host properties as a whole (hereafter the global properties).  From the global host properties we can compare our results directly to the work presented in \citet{Gupta2011}, hereafter \citetalias{Gupta2011}, to check if our age estimator is more sensitive to younger stellar populations.  Secondly, we can contrast the local and global properties of hosts to check if there is more information contained in the local environment rather than the more easily studied global properties.

\begin{deluxetable*}{cc|ccccc|ccccc}
\tablecolumns{12}
\tablewidth{0pt}
\tablecaption{Data used for validation with \cite{Gupta2011} ages\label{tab:gupta}} 
\tablehead{
    \colhead{SDSS ID} & \colhead{redshift}
    & 
    \colhead{u} & \colhead{g} & \colhead{r}  & \colhead{i} & \colhead{z}
    &
    \colhead{$\sigma_u$} & \colhead{$\sigma_g$} & \colhead{$\sigma_r$}  & \colhead{$\sigma_i$} & \colhead{$\sigma_z$} \vspace{-0.5em}
    \\
    \colhead{} & \colhead{}
    & 
    \colhead{[mag]} & \colhead{[mag]} & \colhead{[mag]}  & \colhead{[mag]} & \colhead{[mag]}
    &
    \colhead{[mag]} & \colhead{[mag]} & \colhead{[mag]}  & \colhead{[mag]} & \colhead{[mag]}
    }
    
\startdata
1166 & 0.3820 & 22.54 & 21.83 & 20.04 & 19.38 & 19.03 & 0.48 & 0.11 & 0.04 & 0.03 & 0.08 \\
1580 & 0.1830 & 24.99 & 20.41 & 19.21 & 18.72 & 18.28 & 1.35 & 0.03 & 0.02 & 0.02 & 0.04 \\
2165 & 0.2880 & 22.82 & 23.35 & 22.04 & 22.22 & 21.28 & 0.38 & 0.24 & 0.12 & 0.20 & 0.36 \\
2422 & 0.2650 & 23.64 & 22.86 & 22.00 & 21.95 & 22.01 & 0.90 & 0.23 & 0.17 & 0.24 & 0.76 \\
2789 & 0.2905 & 22.01 & 20.92 & 19.42 & 18.84 & 18.39 & 0.35 & 0.05 & 0.02 & 0.02 & 0.06 \\
\enddata
\tablecomments{Full machine readable data set is\dataset[available in the online journal]{https://raw.githubusercontent.com/benjaminrose/MC-Age/master/data/gupta-global.tsv}.}

\end{deluxetable*}

The final analysis of \citetalias{Gupta2011} included 206 \sn{} and hosts. 
We looked at the 76 objects that passed our redshift and other quality cuts to use in our method validation.
A sample of the data set used for this validation is visible in \cref{tab:gupta}.

\begin{deluxetable*}{c|ccccc|ccccc}
\tablecolumns{11}
\tablewidth{0pt}
\tablecaption{Global host data for \cite{Campbell2013} \sn{} \label{tab:global}} 
\tablehead{
    \colhead{SDSS ID}
    & 
    \colhead{u} & \colhead{g} & \colhead{r}  & \colhead{i} & \colhead{z}
    &
    \colhead{$\sigma_u$} & \colhead{$\sigma_g$} & \colhead{$\sigma_r$}  & \colhead{$\sigma_i$} & \colhead{$\sigma_z$} \vspace{-0.5em}
    \\
    \colhead{}
    & 
    \colhead{[mag]} & \colhead{[mag]} & \colhead{[mag]}  & \colhead{[mag]} & \colhead{[mag]}
    &
    \colhead{[mag]} & \colhead{[mag]} & \colhead{[mag]}  & \colhead{[mag]} & \colhead{[mag]}
    }
    
\startdata
762     & 20.34  & 18.50  & 17.46 &  17.01 &  16.70 &  0.13 &   0.01 &   0.01 &   0.01 &   0.02 \\
1032   & 21.49  & 19.40  & 18.30 &  17.83 &  17.47 &  0.19 &   0.01 &   0.01 &   0.01 &   0.02 \\
1371   & 20.60  & 18.62  & 17.55 &  17.10 &  16.68 &  0.08 &   0.01 &   0.01 &   0.01 &   0.01 \\
1794   & 22.37  & 20.76  & 20.37 &  20.10 &  20.12 &  0.45 &   0.05 &   0.06 &   0.08 &   0.29 \\
2372   & 21.79  & 20.50  & 19.53 &  19.02 &  18.59 &  0.24 &   0.03 &   0.02 &   0.02 &   0.05 \\
\enddata
\tablecomments{Full machine readable data set is\dataset[available in the online journal]{https://raw.githubusercontent.com/benjaminrose/MC-Age/master/data/campbell_global.tsv}.}

\end{deluxetable*}

For each of the 103 hosts where we had local environment photometry, we gathered the SDSS-DR12 model magnitudes for the estimate of the global properties. 
A sample of this data set can be seen in \cref{tab:global}.


\subsection{Photometry of Nearby Galaxies}

\begin{deluxetable*}{cc|c|ccccc|ccccc}
\tablecolumns{13}
\tablewidth{0pt}
\tablecaption{SEDs and redshifts for nearby galaxies\label{tab:nearby}} 
\tablehead{
    \multicolumn{3}{c}{} & 
    \multicolumn{5}{c}{Local SED} &
    \multicolumn{5}{c}{Global SED}
    \\ 
    \cline{5-7} \cline{10-12}
    \colhead{Host Galaxy} & \colhead{\sn{}}
    & 
    \colhead{redshift}
    & 
    \colhead{u} & \colhead{g} & \colhead{r}  & \colhead{i} & \colhead{z}
    &
    \colhead{u} & \colhead{g} & \colhead{r}  & \colhead{i} & \colhead{z}
    }
\startdata
M101 & 2011fe & 0.000804 & 14.73 & 13.28 & 12.87 & 12.65 & 12.71 & 11.63 & 10.14 & 9.56 & 9.24 & 9.02\\
NGC 1015 & 2009ig & 0.008797 & 14.54 & 12.93 & 12.19 & 11.78 & 11.53 & 18.49 & 16.91 & 16.19 & 15.74 & 15.54\\
NGC 1309 & 2002fk & 0.007138 & 15.10 & 13.96 & 13.46 & 13.18 & 13.01 & 12.93 & 11.97 & 11.54 & 11.33 & 11.19\\
NGC 3021 & 1995al & 0.00535 & 15.24 & 14.05 & 13.44 & 13.12 & 12.86 & 13.76 & 12.61 & 12.02 & 11.71 & 11.44\\
NGC 3370 & 1994ae & 0.004276 & 17.08 & 15.97 & 15.50 & 15.23 & 15.06 & 13.63 & 12.58 & 12.06 & 11.79 & 11.56\\
NGC 3447 & 2012ht & 0.003556 & 17.53 & 16.33 & 15.88 & 15.65 & 15.55 & 14.78 & 13.82 & 13.41 & 13.23 & 13.07\\
NGC 3972 & 2011by & 0.002799 & 16.85 & 15.69 & 15.09 & 14.71 & 14.60 & 14.07 & 12.86 & 12.24 & 11.88 & 11.66\\
NGC 3982 & 1998aq & 0.00371 & 15.28 & 14.16 & 13.72 & 13.43 & 13.27 & 12.95 & 11.92 & 11.45 & 11.17 & 10.98\\
NGC 4424 & 2012cg & 0.00162 & 13.93 & 12.90 & 12.35 & 11.88 & 11.85 & 12.97 & 11.87 & 11.28 & 10.78 & 10.77\\
NGC 4536 & 1981B & 0.00603 & 16.51 & 15.30 & 14.77 & 14.50 & 14.43 & 12.60 & 11.32 & 10.64 & 10.28 & 10.03\\
NGC 4639 & 1990N & 0.00364 & 18.04 & 16.91 & 16.46 & 16.14 & 16.12 & 13.40 & 12.11 & 11.46 & 11.12 & 10.89\\
NGC 5584 & 2007af & 0.005525 & 17.01 & 15.94 & 15.47 & 15.16 & 15.26 & 13.39 & 12.41 & 11.92 & 11.59 & 11.64\\
NGC 7250 & 2013dy & 0.0039 & 15.35 & 14.63 & 14.17 & 14.08 & 13.88 & 14.27 & 13.48 & 13.04 & 12.93 & 12.72\\
UGC 9391 & 2003du & 0.00649 & 17.80 & 16.79 & 16.39 & 16.21 & 16.06 & 16.05 & 15.15 & 14.78 & 14.62 & 14.42
\enddata
\tablecomments{The uncertainties in the photoemtry are $\pm$0.03.}
\end{deluxetable*}

The SDSS model magnitudes are unreliable for galaxies with a large angular extent.  For the nearby galaxies 
with distances calibrated with Cepheid variable stars, we used aperture photometry to obtain both the local and global magnitudes. Images of the large galaxies were downloaded from the SDSS~DR12 and individual apertures were designed to capture 90\%\ of the combined light in all the filters. After masking out stars projected on the galaxy, the aperture was then applied to the image of each filter. The magnitude was then calibrated using nearby stars. The photometry for these galaxies can be seen in \cref{tab:nearby}.

\subsection{Supernova Properties} \label{sec:cosmo}

 We use the \citetalias{Campbell2013} supernova sample to provide light curve properties and \hr{} information. We use the Malmquist bias corrected distances derived from the best fit cosmology (\h$~=73.8~\text{km s}^{-1}~\text{Mpc}^{-1}$, $\Omega_{M}=0.24$, and $\Omega_{\Lambda} = 0.76$). When needed, we use these values for our assumed cosmology. For the maximum redshift in our sample, the Malmquist bias correction is $\sim 0.01\un{mag}$. However, \citetalias{Campbell2013} noted that the stretch correction coefficient, $\alpha$ they found for their full sample was significantly larger than typical and larger than the $\alpha$ derived from their spectroscopically classified sub-sample. After our cuts, we found a significant correlation between the supernova stretch parameters, $x_1$, and the \citetalias{Campbell2013} \hr{s} which would likely result in spurious correlations with our host galaxy analysis.  We corrected the \citetalias{Campbell2013} \hr{s} using their spectroscopically derived $\alpha$ value of 0.16 and no longer detected a significant correlation between $x_1$ and our sample's \hr{s}. The resulting Hubble residuals can also be found in \cref{tab:local}. 

For the nearby \sn{} used in the Cepheid calibration of \sn{} peak luminosity, we obtained light curves from the SNANA database\footnote{\url{http://snana.uchicago.edu}} and fit them using SALT2.4 implemented from the {\tt sncosmo}\footnote{\url{https://sncosmo.readthedocs.io/}}. The model also corrected for Milky Way dust extinction from the dust maps of \citet{Schlegel1998} and \citet{Schlafly2011} via {\tt sfdmap}\footnote{\url{https://github.com/kbarbary/sfdmap}}. 


\section{Stellar Population Model} \label{sec:model}


A direct estimate of the age of the stellar population requires a robust model for the observed population.
Flexible Stellar Population Synthesis (FSPS) \citep{Conroy2009, Conroy2010} takes a star formation history and outputs either a spectrum or a redshifted spectral energy distribution (SED) of the resulting stellar population.
The version of FSPS we used (commit \href{https://github.com/cconroy20/fsps/commit/ae31b2f63d865354ce944e5c22eba6e93e01e67d}{ae31b2f} from November 2016) uses the MIST isochrones \citep{Dotter2016, Choi2016} and the MILES spectral libraries \citep{Falcon-Barroso2011}.

\subsection{FSPS settings}

Many of the FSPS parameters were set at their default values, but a 
number of key settings were adjusted to produce the desired model space. 

To control the metallicity of our stellar population, we set \texttt{zcontinuous = 2}. This setting convolves the SSPs (simple stellar populations) with a metallicity distribution function. The metallicity distribution is defined as
\begin{equation}
    (Z e^{-Z})^{\texttt{pmetals}}
\end{equation}
with 
\begin{equation}
    Z \equiv \frac{z}{z_{\sun}10^{\log(z/z_{\sun})}}
\end{equation}
where $z$ is the metallicity in linear units and $z_{\sun} = 0.019$.
This metallicity distribution is governed by two more FSPS parameters: \texttt{pmetals} and \texttt{logzsol} (i.e. $\log(z/z_{\sun})$). We left \texttt{pmetals} at its default value of $2$ and during the fitting process \texttt{logzsol} was allowed to vary but was marginalized over when the age probability distribution was determined.

The next set of parameters govern the treatment of dust. We used the default power law dust model as explained in \cite{Conroy2009}, based off of \cite{Charlot2000}.
The attenuation curve of a star, as a function of stellar age, is defined as
\begin{equation}
    \tau_{\lambda}(t) =
    \begin{cases} 
    \tau_1 (\lambda/5500\un{\AA})^{-0.7} & t \leq 10^7\un{yr}\\
    \tau_2 (\lambda/5500\un{\AA})^{-0.7} & t > 10^7\un{yr}\\
    \end{cases}
\end{equation}
where $\tau_1$ and $\tau_2$ are the attenuation around a young stellar population and in the ISM respectively. See \cite{Charlot2000} Figure 1 for a visual representation. In FSPS these two parameters are controlled via the \texttt{dust1} and \texttt{dust2} variables. For this analysis, \texttt{dust1} was set to two times \texttt{dust2}, and \texttt{dust2} was allowed to vary to match the observations. \cite{Conroy2009} claims good values of \texttt{dust1} and \texttt{dust2} are 1.0 and 0.3 respectively, while \cite{Charlot2000} prefers values of \texttt{dust1} and \texttt{dust2} of 1.0 and 0.5 respectively. The allowed range for {\tt dust2} in this analysis is explained in \cref{sec:priors} and is consistent with these recommendations.

A few  of the host SEDs show an unusual, e.g. SN4019. 
Adding nebular emission (setting {\tt add\_neb\_emission = True} and \texttt{cloudy\_dust = True}) adjusts the $r$-band magnitude for a young stellar population and allows the model to match this observed feature. This characteristic is shown to be achievable in the self-consistency validation test number 3, as explained in \cref{sec:circle}. 




Finally, FSPS outputs the luminosity of $1\un{M}_{\sun}$, so an extra constant, $\delta$, is used to scale the output SED of FSPS to match the observed SEDs.

\subsection{Star Formation History}

FSPS has many inputs for describing the star formation history of a galaxy. The \texttt{sfh} parameter allows the user to select the functional form of star formation history. \citetalias{Gupta2011} used the simple $\tau$-model: the star formation rate is proportional to $e^{-t/\tau}$, with $t$ being the time since the start of star formation and $\tau$ is a free parameter. \citetalias{Gupta2011} fit both $\tau$ and the length of star formation history. This is the simplest model, which is important when fitting a small number of free parameters. However, such a simple prescription makes it difficult to create a model with both old stars and recent star formation.

\citet{Simha2014} investigated the ability of several star formation history models to match simulated galaxies. 
This research looked at the simple $\tau$-model, a linear-exponent model\footnote{This is the same as FSPS's delayed $\tau$-model, \texttt{sfh = 4}.}, and a four parameter $\tau$-model. A visual comparison is presented in Figures 3--5 of \citeauthor{Simha2014}. According to the calculations in \citeauthor{Simha2014}, the simple $\tau$-model can overestimate the age by $\sim 1\text{--}2 \un{Gyr}$, particularly for younger populations. Since we expect some supernovae to explode in young ($\lesssim 2\un{Gyr}$) stellar populations, we decided to use a four parameter $\tau$-model.

The main feature of the four-parameter $\tau$-model (\sfh{}) is that it separates the properties of early and late time star formation. 
This model can describe a wide range of star formation histories: an early burst, a history dominated by recent star formation, or both an early burst and recent star formation. 
The \sfh{} is a piecewise combination of a linear-exponent star formation history then a linearly rising or falling star formation. This model is used by FSPS when \texttt{sfh = 5}. Mathematically the \sfh{} can be written as
\begin{equation}
\label{eqn:sfh}
    \Psi_{\star}(t) \propto 
    \begin{cases} 
     (t - t_0)~ e^{-(t-t_0)/\tau} & t_{0} \leq t \leq t_{i} \\
     \Psi_{\star}(t_{i}) + m_{\text{sf}}~\mathscr{R}(t-t_{i}) ~~ & t_{\text{i}} < t \leq \mathscr{A}(z)\\
     0 & \text{else}
    \end{cases}
\end{equation}
where $t_0$ is when the star formation started, $\tau$ is the e-folding parameter, $t_i$ is the star formation transition time, $m_{\text{sf}}$ is the slope of the late time star formation, $\mathscr{R}$ is the ramp function, and $\mathscr{A}(z)$ is the redshift dependant time when the observed light was emitted.
Note that the equation above allows negative values of $\Psi$, which is nonphysical. So we add an extra constraint that forces $\Psi(t)$ to be $0$ if calculated to be negative.
The four variables in the equation ($t_0$, $\tau$, $t_i$, $m_{\text{sf}}$) are the free parameters that give this model its name. These correspond to the FSPS parameters \texttt{sf\_start}, \texttt{tau}, \texttt{sf\_trunc}, and \texttt{sf\_slope} respectively. Also, this function takes $t$ as the time from the start of the universe.
A sample of various star formation histories calculated from the tau model can be seen in \cref{fig:sfh}.

\subsection{Calculating Ages}

\begin{figure}
    \centering
    \includegraphics[width=3.2in]{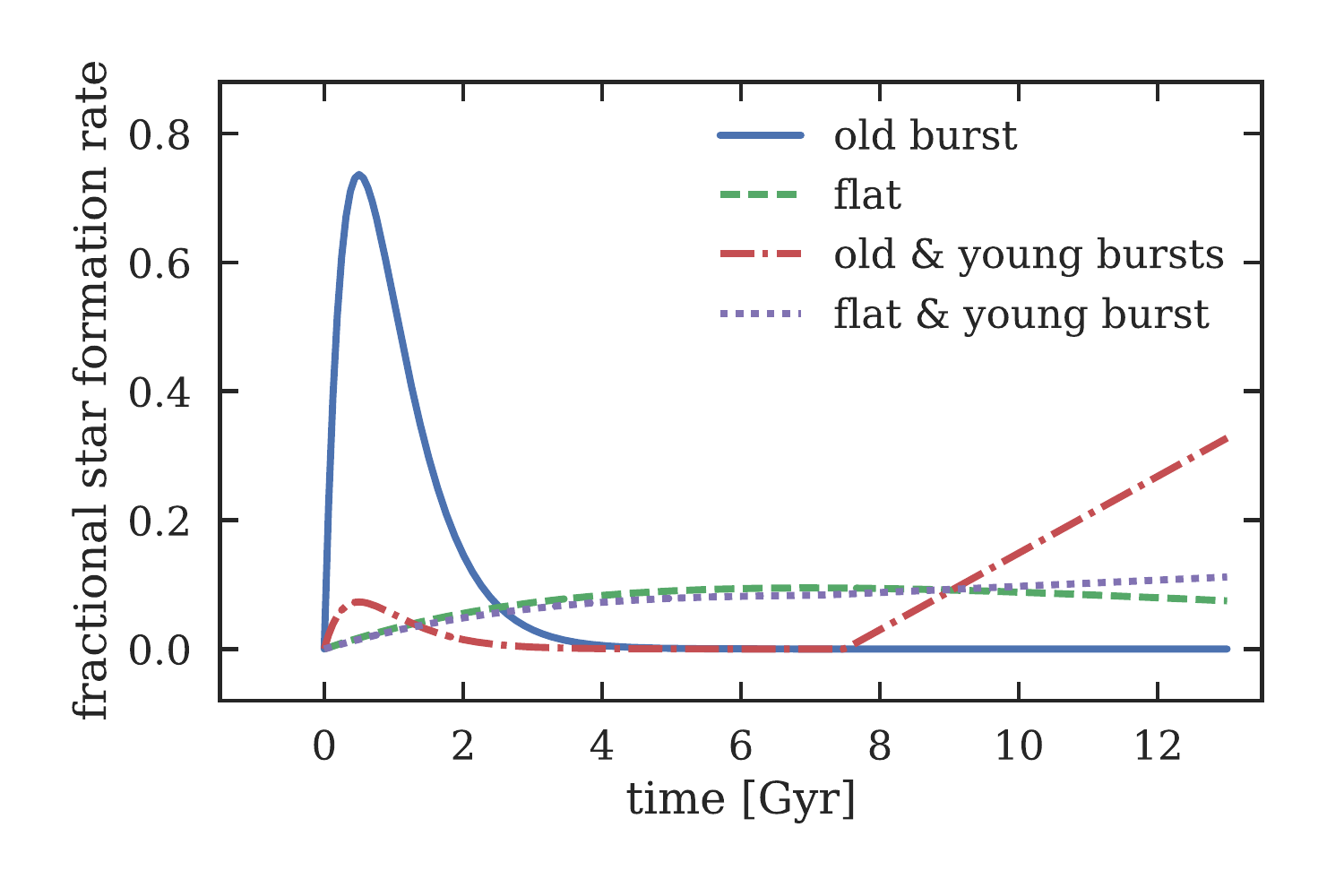}
    \caption{An example of several four-parameter $\tau$-model star formation histories. These star formation histories are normalized to produce the same total stellar mass. The old bursts have a $\tau = 0.5$ and ``flat" histories have a $\tau = 7.0$. The two examples of young bursts show how the number of young stars produced after $t = t_i$ depends heavily on the amount of previously formed stars even for the same $m_{\text{sf}}$.}
    \label{fig:sfh}
\end{figure}

For any set of star formation parameters we determine the average population age. The mass weighted average age is:
\begin{equation}
\label{eqn:age}
    \langle A \rangle_{\text{mass}} = 
    \mathscr{A}(z) - t_0 -
    \dfrac{
        \int_{t_0}^{\mathscr{A}(z)} (t-t_{0}) \Psi_{\star}(t)dt
            }{
        \int_{t_0}^{\mathscr{A}(z)} \Psi_{\star}(t)dt
    }
\end{equation}
where all of the variables are the same as described in \cref{eqn:sfh}
so $t - t_{\text{0}}$ is simply the length of star formation. 
In the integral, a variable substitution of $t - t_0 \rightarrow t$ transforms the time zero point from the beginning of the universe to the start of star formation. We would also need to transform $\Psi_{\star}(t) \rightarrow \Psi'_{\star}(t)$ so that $\Psi'_{\star}$ assumes $t=0$ is the start of star formation. This makes the  equation for the mass weighted average age to be 
\begin{equation*}
\langle A \rangle_{\text{mass}} = \mathscr{A}(z) - t_0 - \dfrac{\int_{0}^{\mathscr{A}(z) - t_0} t \Psi'_{\star}(t)dt}{\int_{0}^{\mathscr{A}(z)- t_0} \Psi'_{\star}(t)dt}
\end{equation*}
or the equation used in \citetalias{Gupta2011}. In this paper, the age of a stellar population will refer to the mass weighted average described here.

The model using the star formation parameters $t_0=8.0$, $\tau=0.1$, $t_i=12$, and $m_{\text{sf}}=20$, produces a population with an average age of $437\un{Myr}$. This demonstrates that our SFH prescription can generate a dominant young population when \sn{} are expected to start exploding. A small $\tau$ is needed to keep the number of old stars from building up over the cosmic time and dominating over the recent linear star formation period. An example of this can be seen by the ``old \& young burst'' star formation history in \cref{fig:sfh}.

\section{Determining the most Probable SFH} \label{sec:bayes}

Using Bayesian statistics and a Markov chain Monte Carlo (MCMC) sampling method we determine the probability of each free parameter described in \cref{sec:model}: $\log(z/z_{\sun})$, $\tau_2$, $\tau$, $t_0$, $t_i$, $m_{\text{sf}}$, and $\delta$. Data modeling and parameter estimation is often done with Bayesian statistics because it calculates the probability of the model parameters given the observed data by using Bayes' Theorem:
\begin{equation}\label{eqn:bayes-thm}
P(\theta|D) = \frac{P(D|\theta)P(\theta)}{P(D)}
\end{equation}
where $\theta$ is a given model's parameters and $D$ represents the observed data. Each probability is:
\begin{description}
\item[$P(\theta|D)$] The \textit{posterior}, which is the probability of the model parameters given the data.
\item[$P(D|\theta)$] The standard \textit{likelihood} function, $\mathscr{L}(D|\theta)$.
\item[$P(\theta)$] The model \textit{prior}, which describes what we know about the model before considering the data $D$, such as model parameter limits.
\item[$P(D)$] The model \textit{evidence}, which in practice amounts to a normalization term.
\end{description}

For MCMC sampling, only relative probabilities are needed, so $P(D)$ is generally ignored and \cref{eqn:bayes-thm} becomes a proportionality, not an equality. It is common to use flat priors, $P(\theta) \propto 1$. In this case Bayes' Theorem simplifies to a standard frequentist likelihood estimation, $P(\theta|D) \propto \mathscr{L}(D|F)$. More generally, some prior information is used and Bayes' Theorem becomes 
\begin{equation}
P(\theta|D) \propto P(\theta) \times \mathscr{L}(D|\theta).
\end{equation}
To find the maximum posterior probability of the model parameters, we need only to know the priors, $P(\theta)$, and the likelihood of the data, $\mathscr{L}(D|\theta)$. A more complete description of Bayesian statistics and MCMC sampling  is available in \cite{VanderPlas2014} and \cite{Hogg2017}.

\subsection{Likelihood}



This method uses a standard log-likelihood function for data with Gaussian uncertainties. Summing over each filter, we compare the observed apparent magnitude ($m_i$) with the resulting apparent magnitude from FSPS ($m_{\text{FSPS},i}$) plus a scaling factor ($\delta$) to account for FSPS's $1\un{M}_{\sun}$ output. Mathematically this is written as:
\begin{equation}
    \ln(\mathscr{L}) \propto \sum_i\frac{(m_i - (m_{\text{FSPS},i} + \delta))^2}{\sigma_i^2} + \ln\left(2 \pi\sigma_i^2\right)
\end{equation}
with $\sigma_i$ as the uncertainty in each $m_i$ measurement. 

\subsection{Priors}
\label{sec:priors}

For five of the variables we use bounded flat tops as described below:
\begin{align}\label{eqn:priors}
2.5 &< t_i < \mathscr{A}(z) \nonumber \\
0.1 &< \tau < 10 \\
-1.520838 &< \phi < 1.520838 \nonumber \\
0.5 &< t_{0} < t_i - 2.0 \nonumber \\
-45.0 &< \delta  < -5.0 \nonumber
\end{align}


Since a flat distribution in a slope parameter preferentially searches the high values space,\footnote{A mathematical description is available in \citet[Appendix A]{VanderPlas2014}; he also has a nice graphical example on his \href{http://jakevdp.github.io/blog/2014/06/14/frequentism-and-bayesianism-4-bayesian-in-python/\#The-Prior}{website.}} the MCMC was performed over $\phi$, the angle the ramp function makes with respect to the $x$-axis; therefore, $m_{\text{sf}} = \arctan(\phi)$. The prior bound above, $-1.520838 < \phi < 1.520838$, corresponds to $-20 < m_{\text{sf}} < 20$.

In addition, $\log(z/z_{\sun})$ uses a Gaussian distribution with $\mu = -0.5\un{dex}$ and $\sigma = 0.5\un{dex}$ limited to the range of $-2.5  < \log(z/z_{\sun}) < 0.5$. This is a common assumption as seen in \cite{Belczynski2016}. 
Our model reaches a lower metallicity than the grid search used in \citetalias{Gupta2011}.

For the ISM dust parameter, $\tau_2$, we assume a Gaussian prior on the top bounds. The Gaussian distribution is defined by $\mu = 0.3$ and $\sigma = 0.2$, but only values between $0$ and $0.9$ are accepted. This allows for some variability but keeps the values near the $0.3$ and $0.5$ as recommended by \cite{Conroy2009} and \cite{Charlot2000} respectively.

\section{Validation}

Following the statistical method described in \cref{sec:bayes}, we derive probability distributions for the model parameters defined in \cref{sec:model}. Using \cref{eqn:age} at each step in the MCMC chain, we build a probability distribution function for the age marginalized over metallicity and host galaxy dust.



\subsection{Self-consistency}\label{sec:circle}

\begin{deluxetable}{c|cccccc|c}
\tablecolumns{8}
\tablewidth{0pt}
\tablecaption{SFH parameters used in the self consistency test\label{tab:circle-sfh}} 
\tablehead{
    \colhead{ID}
    & 
    \colhead{$\log(z/z_{\sun})$} & \colhead{$\tau_2$} & \colhead{$\tau$} & \colhead{$t_0$}  & \colhead{$t_i$} & \colhead{$\phi$}
    & 
    \colhead{age}\vspace{-0.5em}
    \\
    \colhead{} & \colhead{} & \colhead{} & \colhead{[Gyr$^{-1}$]} & \colhead{[Gyr]}  & \colhead{[Gyr]} & \colhead{[rad]} & \colhead{[Gyr]}
    }
    
\startdata
1 & -0.5 & 0.1 & 0.5 & 1.5 & 9.0 & -0.785 & 10.68\\
2 & -0.5 & 0.1 & 0.5 & 1.5 & 9.0 & 1.504 & 1.41\\
3 & -0.5 & 0.1 & 7.0 & 3.0 & 10.0& 1.504 & 1.75\\
4 & -0.5 & 0.1 & 7.0 & 3.0 & 13.0 & 0.0 & 4.28\\
5 & -1.5 & 0.1 & 0.5 & 1.5 & 9.0 & -0.785 & 10.68\\
6 & -0.5 & 0.8 & 7.0 & 3.0 & 10.0 & 1.504 & 1.75\\
7 & -0.5 & 0.1 & 0.5 & 1.5 &  6.0 & 1.504 & 2.40\\
8 & -0.5 & 0.1 & 0.1 & 8.0 & 12.0 & 1.52 & 0.437\\
\enddata
\tablecomments{All models are at a $z=0.05$.}
\end{deluxetable}

\begin{deluxetable}{c|ccccc|c}
\tablecolumns{7}
\tablewidth{0pt}
\tablecaption{SEDs for the self consistency test\label{tab:circle-sed}} 
\tablehead{
    \colhead{ID}
    & 
    \colhead{u} & \colhead{g} & \colhead{r}  & \colhead{i} & \colhead{z}
    & 
    \colhead{$\langle A \rangle$}\vspace{-0.5em}
    \\
    \colhead{} 
    & 
    \colhead{[mag]} & \colhead{[mag]} & \colhead{[mag]}  & \colhead{[mag]} & \colhead{[mag]}
    & 
    \colhead{[Gyr]}
    }
    
\startdata
1 & 20.36 & 18.76 & 17.99 & 17.67 & 17.39 & $8.5\pm1.5$\\
2 & 20.31 & 18.74 & 17.98 & 17.66 & 17.39 & $7.7\pm1.5$\\
3 & 16.15 & 15.43 & 15.40 & 15.19 & 15.21 & $1.4\pm0.5$\\
4 & 17.65 & 16.74 & 16.49 & 16.26 & 16.16 & $4.2\pm1.0$\\
5 & 19.69 & 18.29 & 17.70 & 17.45 & 17.29 & $7.2\pm1.6$\\
6 & 17.66 & 16.58 & 16.25 & 16.01 & 15.86 & $2.7\pm0.8$\\
7 & 17.62 & 16.80 & 16.57 & 16.34 & 16.26 & $4.5\pm1.4$\\
8 & 19.72 & 18.37 & 17.88 & 17.68 & 17.56 & $4.4\pm1.1$\\
\enddata
\tablecomments{SEDs were scaled with a $\delta = -25\un{mag}$.}
\end{deluxetable}

The first validation of this newly developed age estimator was to verify that it was self-consistent, i.e. it could correctly estimate the star formation parameters from an SED generated by FSPS.

\begin{figure}
    \centering
    \includegraphics[width=3in]{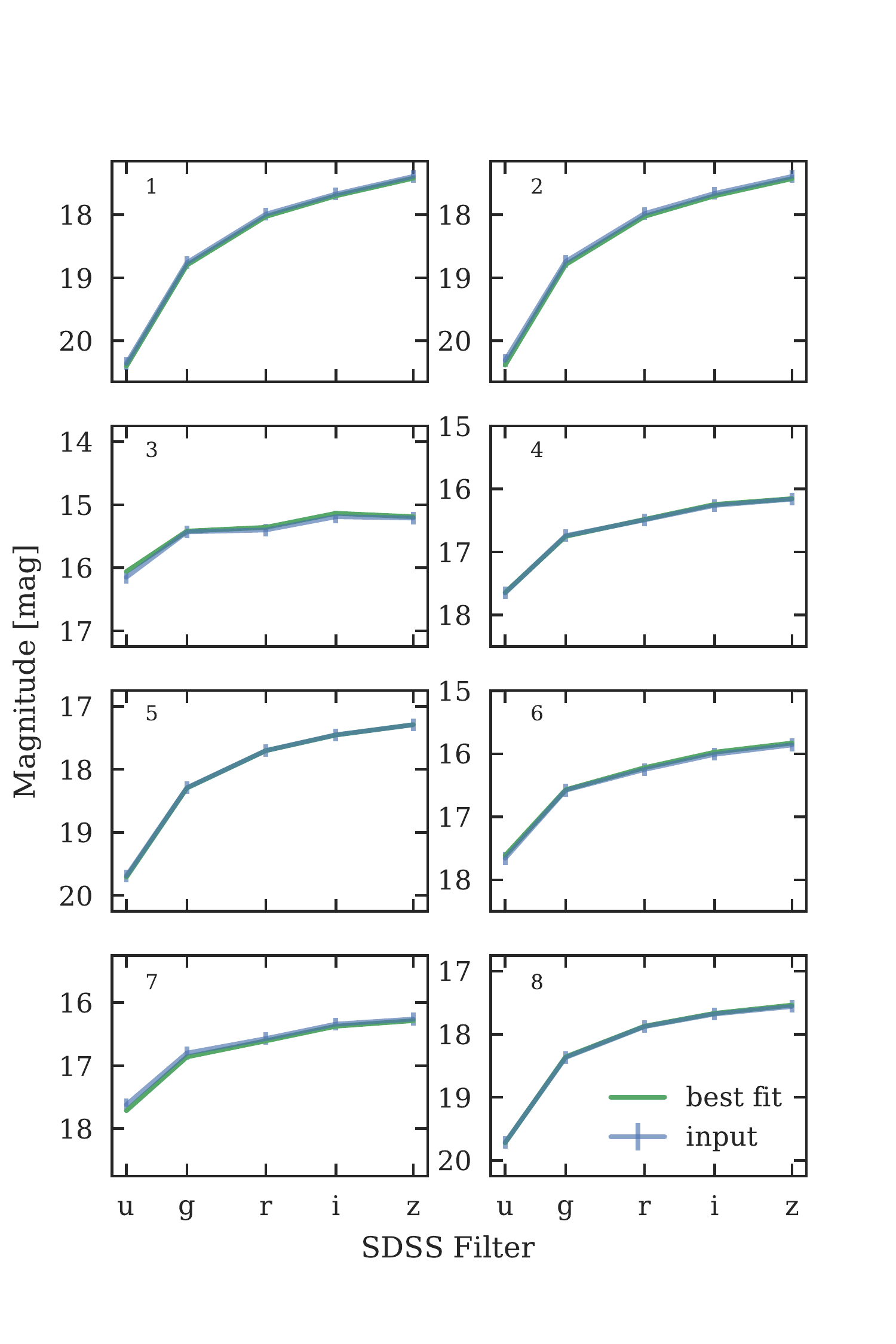}
    \caption{The FSPS produced SEDs (blue lines labeled inputs) from the initial parameters (available in \cref{tab:circle-sfh}) as well as the best fit values derived by our new Bayesian age estimator. For these eight stellar populations, covering most of our model space, the fits are excellent matches to the input SEDs.}
    \label{fig:sed}
\end{figure}

Eight different models were used in this test. The initial metallicity, dust, and star formation parameters can be seen in \cref{tab:circle-sfh}. These values were put into FSPS, at a redshift of $z=0.05$, to generate observed SEDs. The resulting SEDs, \cref{tab:circle-sed}, were then analyzed with our Bayesian estimator. This set of models produced old populations ($\sim 10.5 \un{Gyr}$) and very young populations ($\sim 0.5 \un{Gyr}$). They explored the effect of metallicity (Model 5) and dust (Model 6). Model 2 also looked at a ``mixed'' population with an old burst of star formation and a strong increase of star formation to the present epoch, a stellar population that is not possible with a simpler star formation history. 

This method can model a young stellar population, (like Model 3), but is unable to recover a star-burst or mixed populations (Models 8 and 2 respectively) based on SED fitting. The ``old \&\ young burst'' of Model 2 is not particularly blue and as such, we identify an age of $4.4 \un{Gyr}$. We do better with Model 3 where we estimate the correct age of 1.4~Gyr with a small uncertainty.

\begin{figure*}
    \centering
    \includegraphics[width=6in]{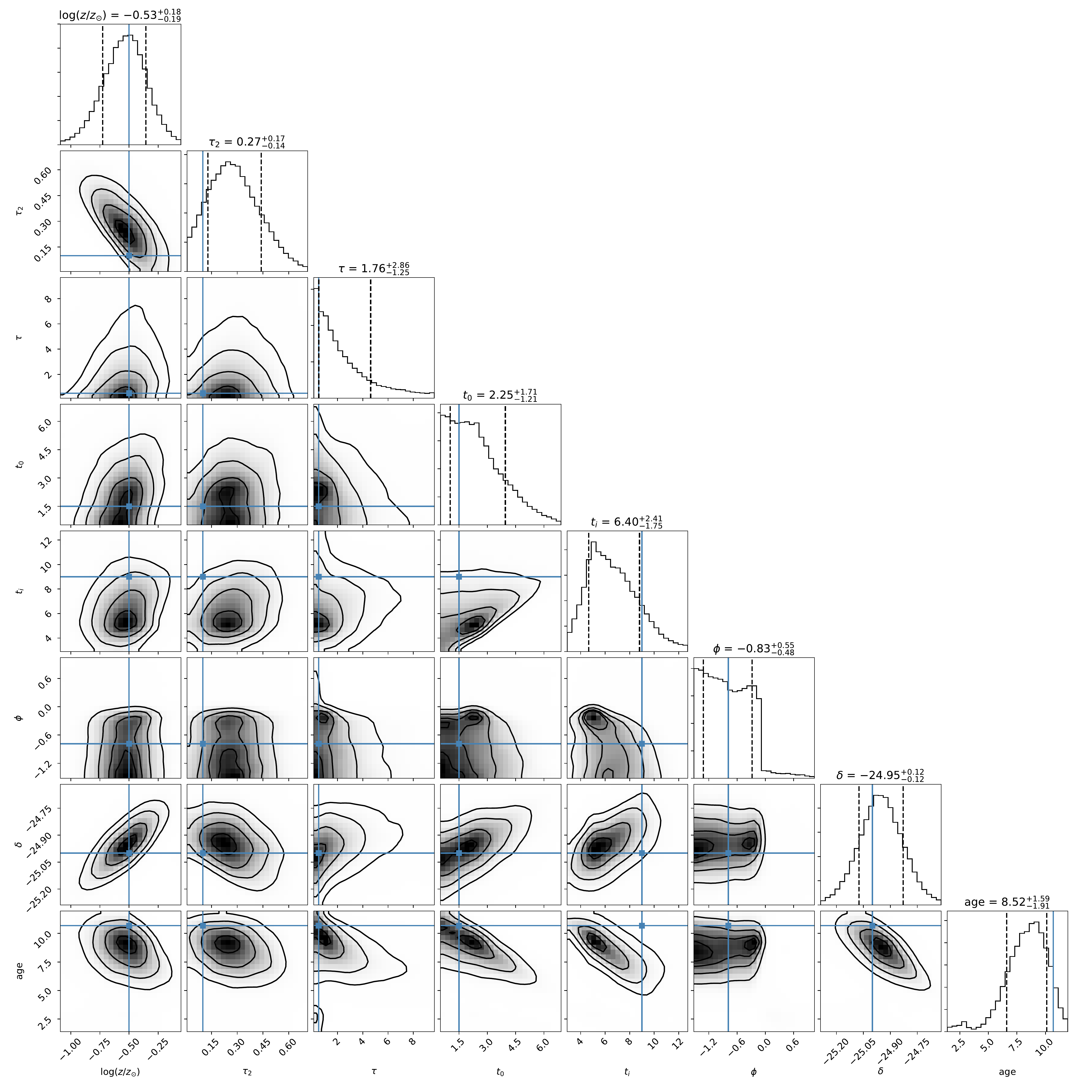}
    \caption{A corner plot of the posterior distribution for the first model in the self-consistency test. The input values are plotted as blue lines. The vertical dashes lines represent the 68\% credible region for each parameter. \added{The solid blue lines represent the input values as stated in \cref{tab:circle-sfh}.} The posterior provides a good estimate of the true parameters, including the average age of the stellar population.
    The corresponding figure for each model (8 images) are available as a Figure Set in the online journal.}
    \label{fig:circlePosterior}
\end{figure*}

\Cref{fig:sed} shows the FSPS output SEDs of the initial input star formation parameters and the best-fit parameters from our analysis. The SEDs are fit very well across all the models used in this self-consistency test.
In addition to fitting the SEDs, our analysis was able to approximate the underlying parameters. An example corner plot of the posterior probabilities is displayed in \cref{fig:circlePosterior}, and the the full set of figures is available online.

\figsetstart
\figsetnum{\thefigure} 
\figsettitle{Self-consistency posterior results.}

    \figsetgrpstart
        \figsetgrpnum{\thefigure.1}
        \figsetgrptitle{Self-consistency Test 1}
        \figsetplot{C1-truths-0717.pdf}
        \figsetgrpnote{Corner plot of posterior distributions from the self-consistency validation, model 1.}
    \figsetgrpend


    \figsetgrpstart
        \figsetgrpnum{\thefigure.2}
        \figsetgrptitle{Self-consistency Test 2}
        \figsetplot{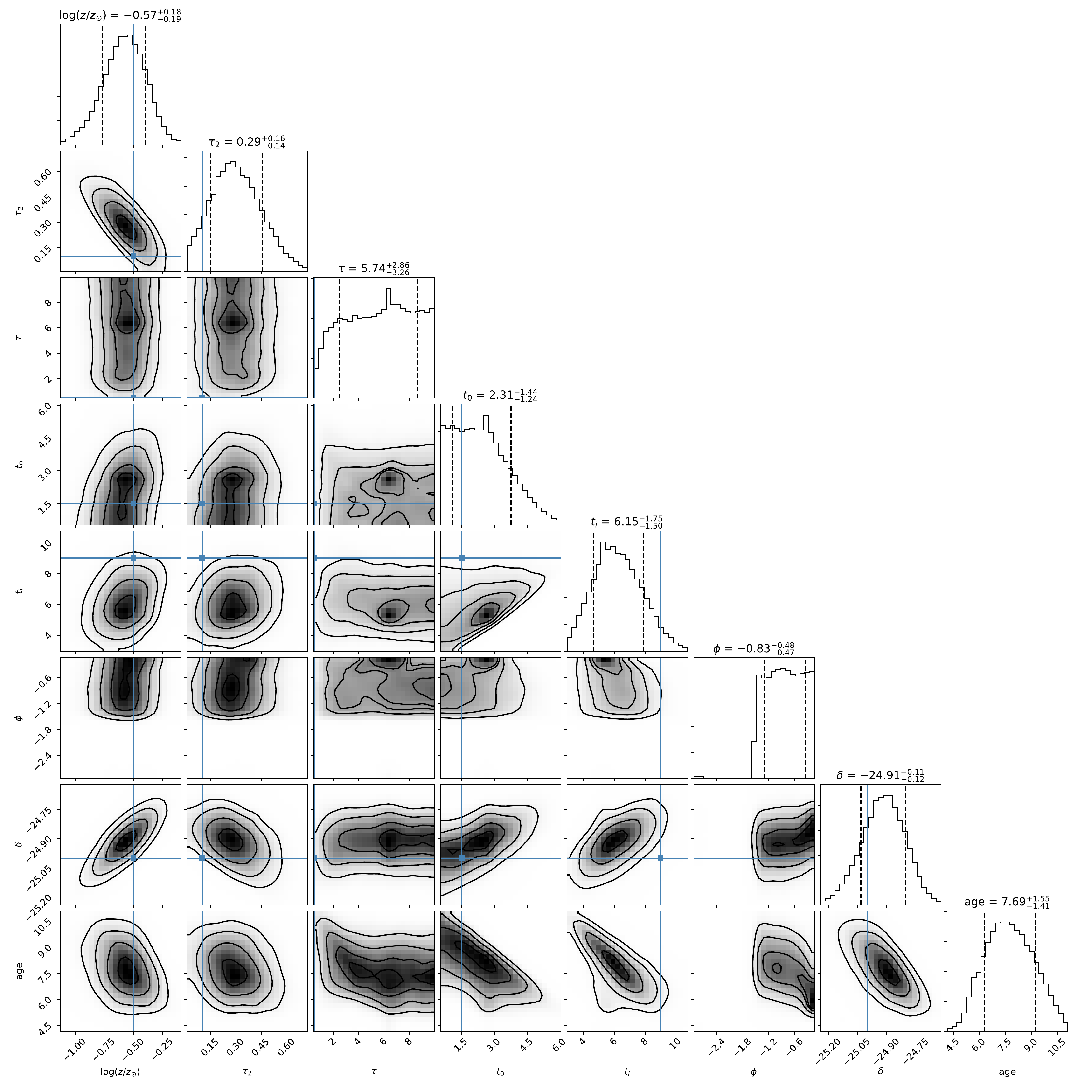}
        \figsetgrpnote{Corner plot of posterior distributions from the self-consistency validation, model 2.}
    \figsetgrpend
    
    \figsetgrpstart
        \figsetgrpnum{\thefigure.3}
        \figsetgrptitle{Self-consistency Test 3}
        \figsetplot{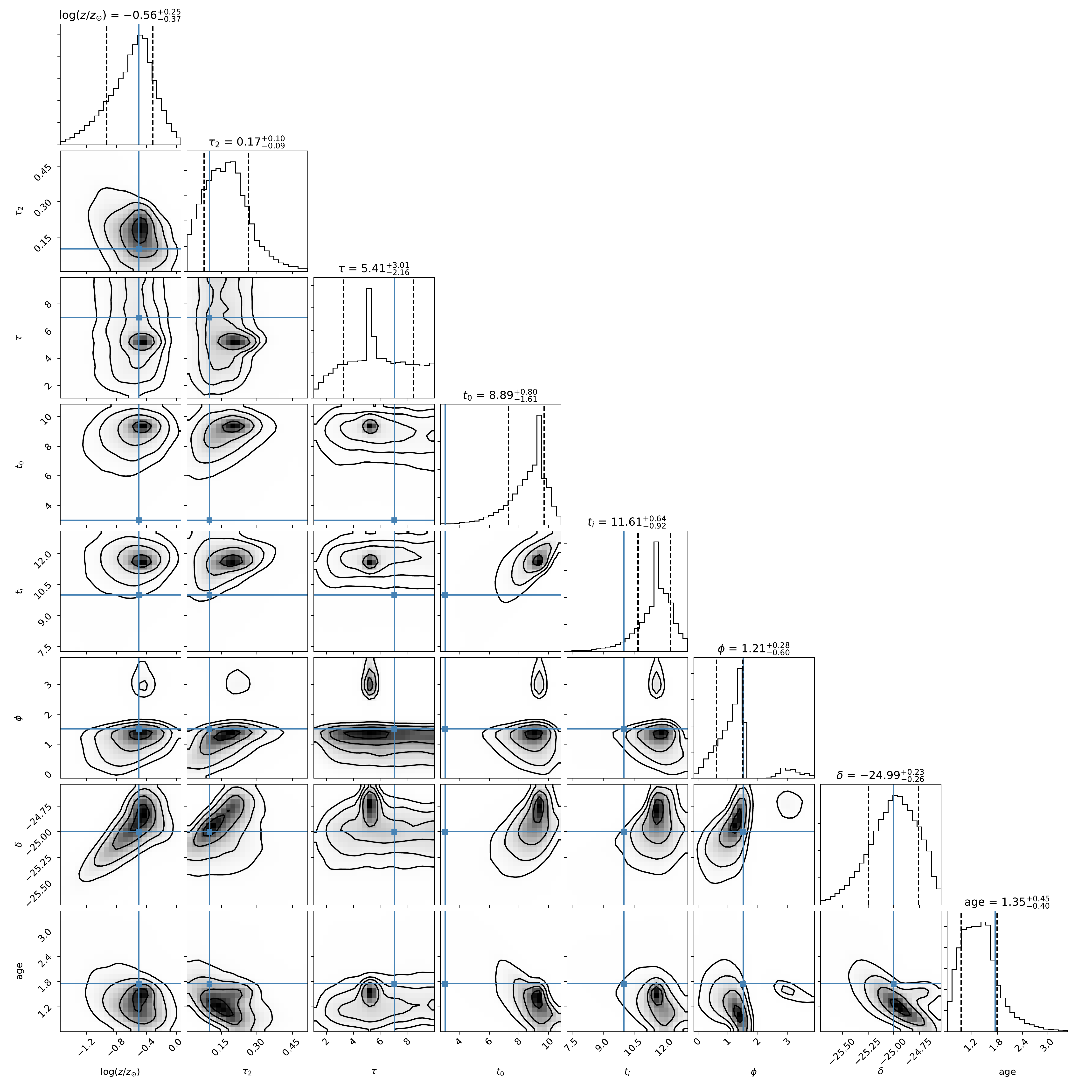}
        \figsetgrpnote{Corner plot of posterior distributions from the self-consistency validation, model 3.}
    \figsetgrpend
    
    \figsetgrpstart
        \figsetgrpnum{\thefigure.4}
        \figsetgrptitle{Self-consistency Test 4}
        \figsetplot{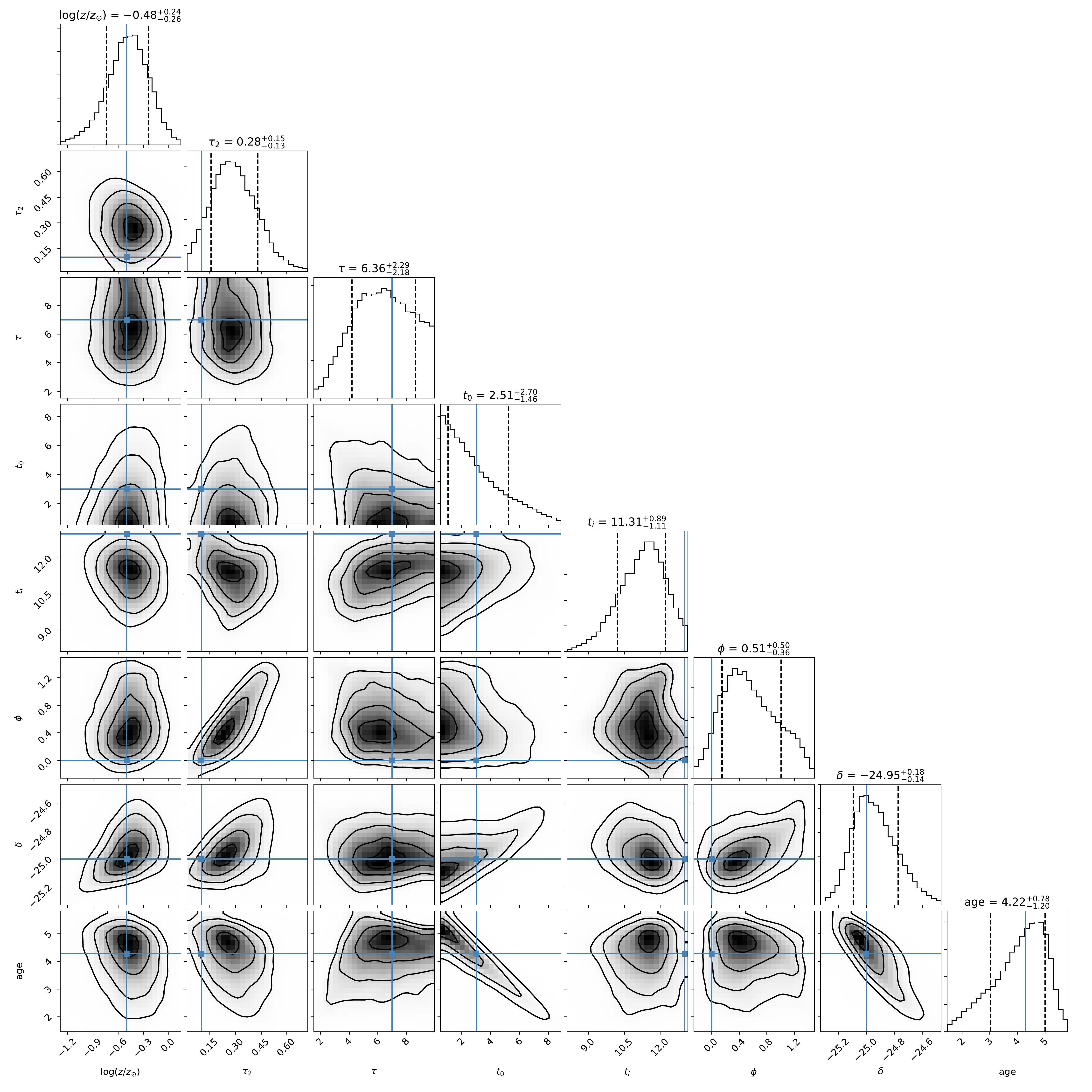}
        \figsetgrpnote{Corner plot of posterior distributions from the self-consistency validation, model 4.}
    \figsetgrpend
    
    \figsetgrpstart
        \figsetgrpnum{\thefigure.5}
        \figsetgrptitle{Self-consistency Test 5}
        \figsetplot{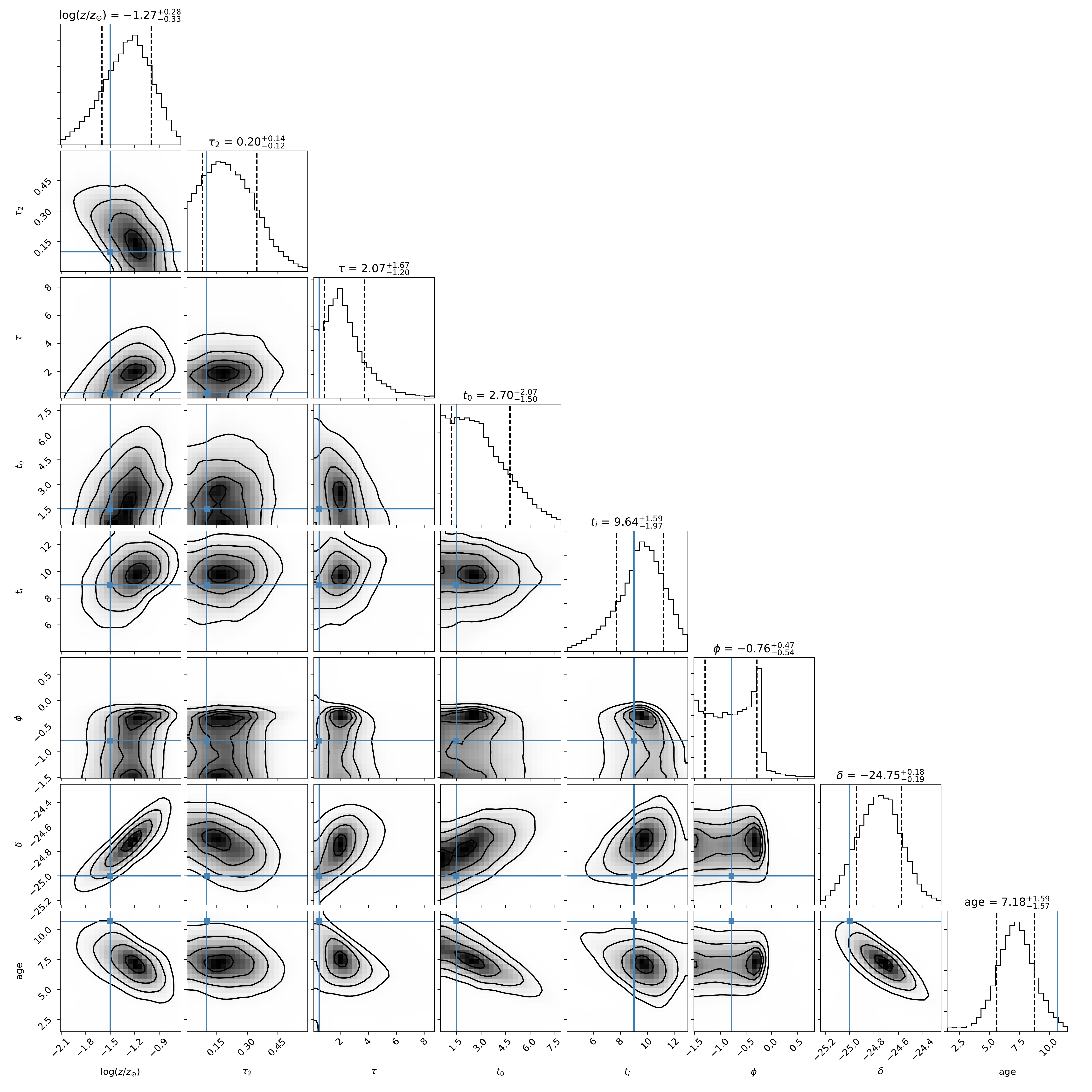}
        \figsetgrpnote{Corner plot of posterior distributions from the self-consistency validation, model 5.}
    \figsetgrpend
    
    \figsetgrpstart
        \figsetgrpnum{\thefigure.6}
        \figsetgrptitle{Self-consistency Test 6}
        \figsetplot{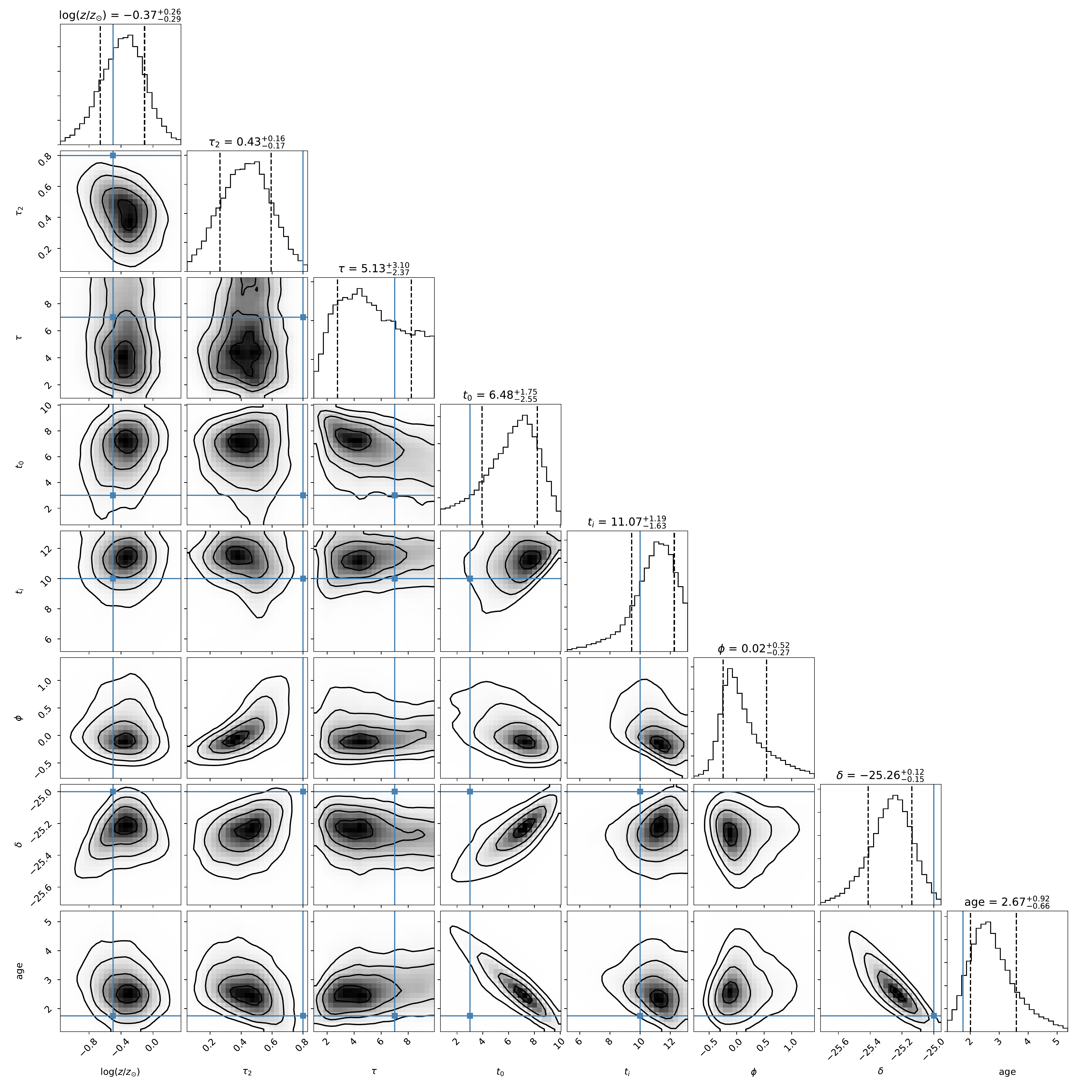}
        \figsetgrpnote{Corner plot of posterior distributions from the self-consistency validation, model 6.}
    \figsetgrpend
    
    \figsetgrpstart
        \figsetgrpnum{\thefigure.7}
        \figsetgrptitle{Self-consistency Test 7}
        \figsetplot{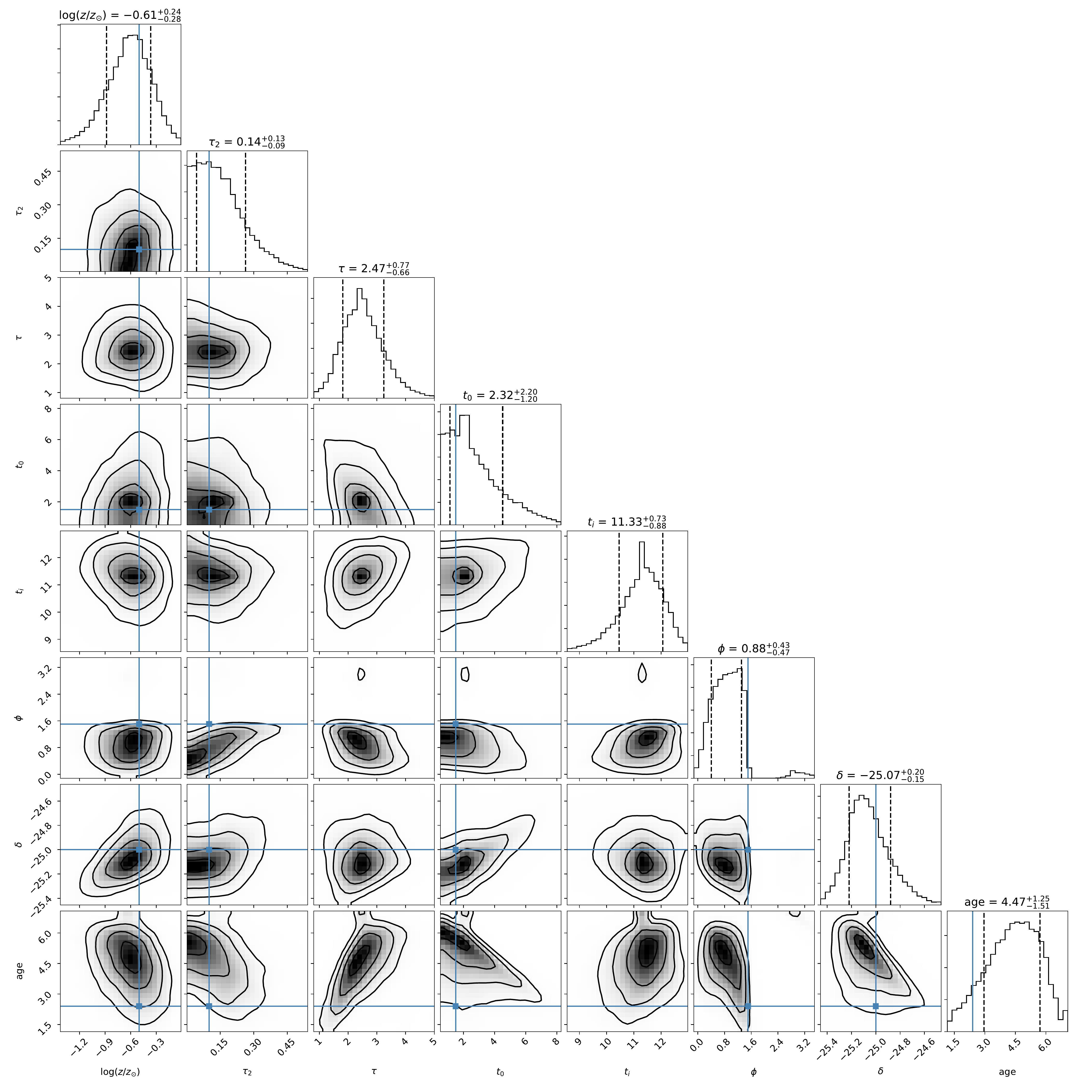}
        \figsetgrpnote{Corner plot of posterior distributions from the self-consistency validation, model 7.}
    \figsetgrpend
    
    \figsetgrpstart
        \figsetgrpnum{\thefigure.8}
        \figsetgrptitle{Self-consistency Test 8}
        \figsetplot{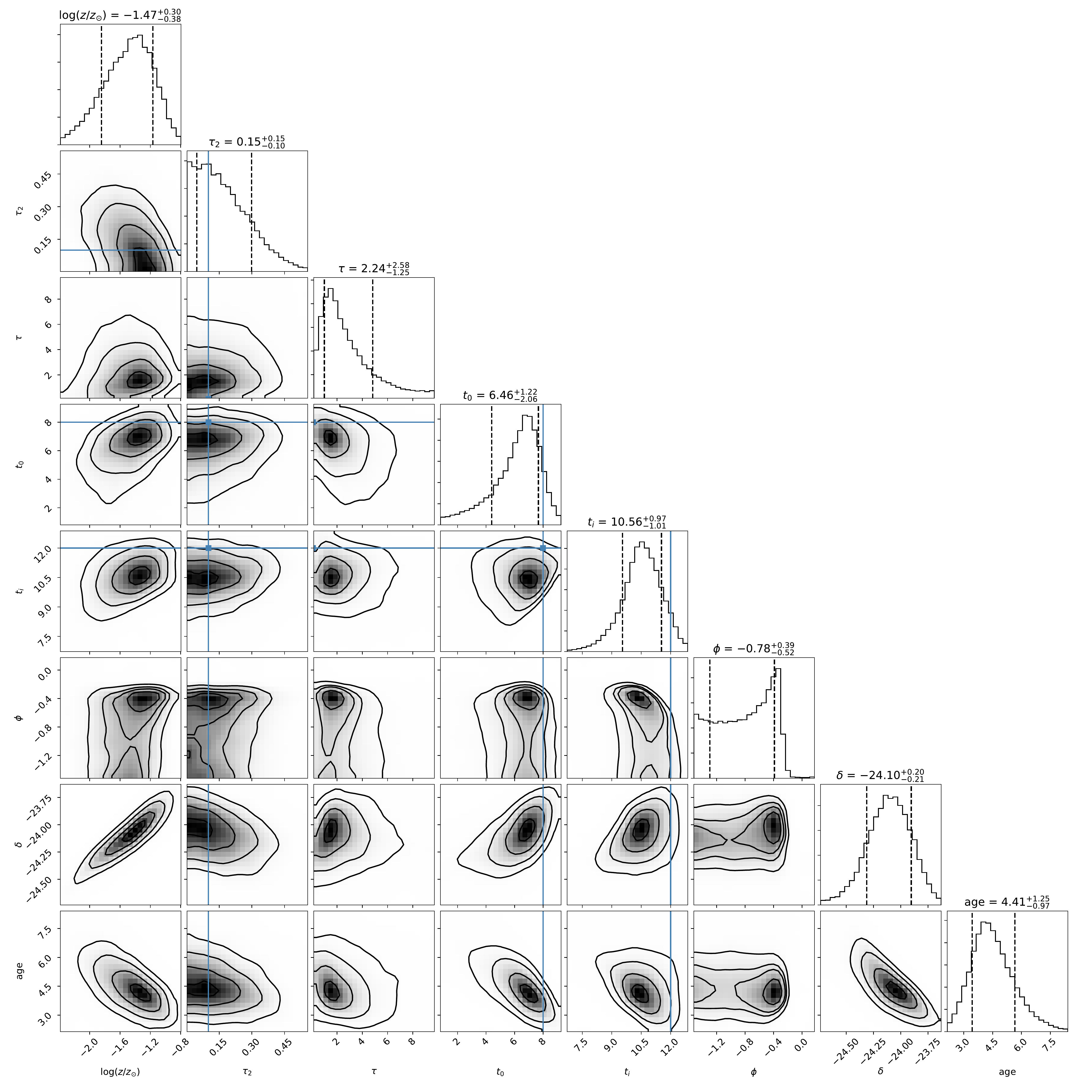}
        \figsetgrpnote{Corner plot of posterior distributions from the self-consistency validation, model 8.}
    \figsetgrpend

\figsetend

\subsection{Comparison to Previous Work}

\begin{figure}
    \centering
    \includegraphics[width=3in]{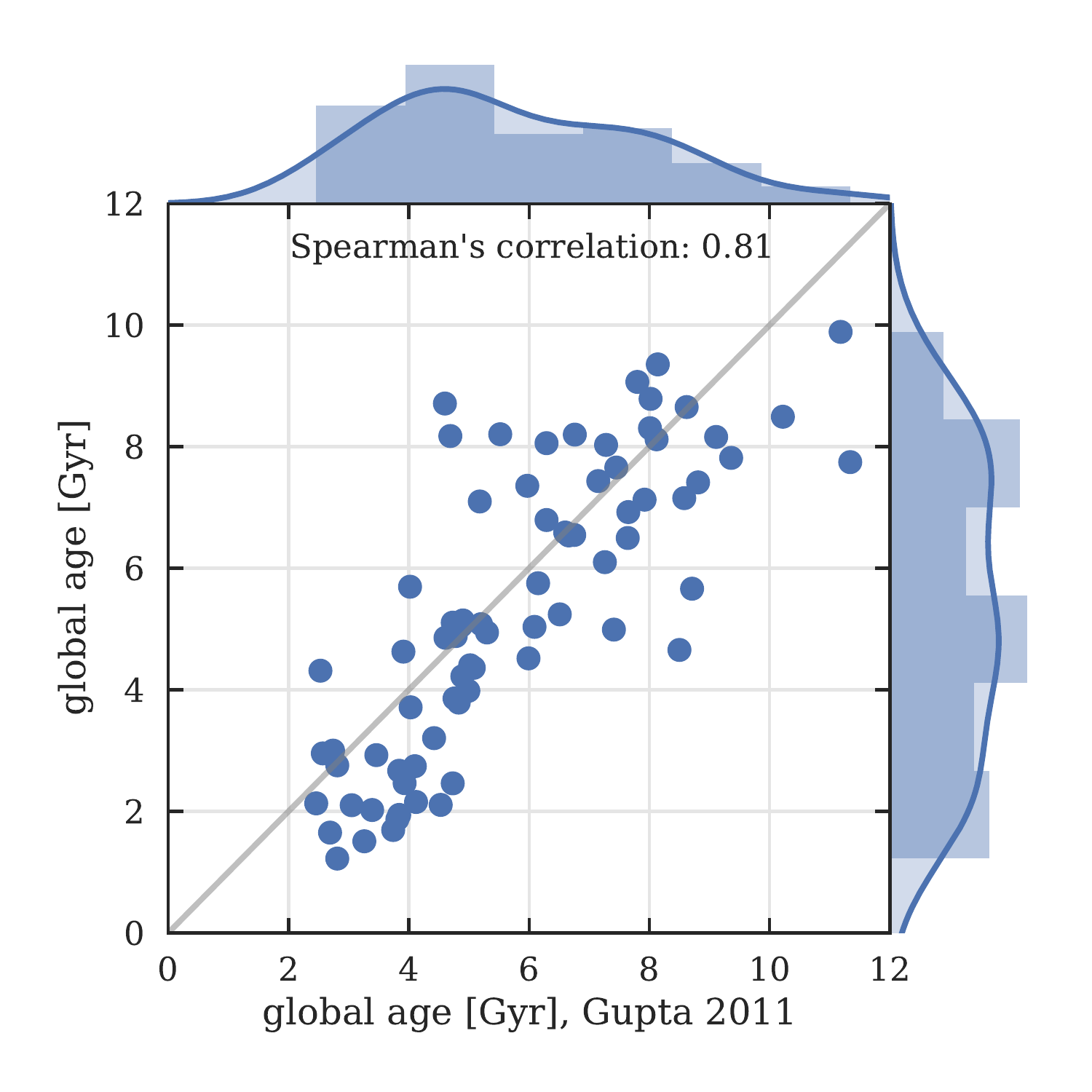}
    \caption{A comparison between the age estimated by the method described in this paper and the results presented in \citetalias{Gupta2011}. They agree with a $\sim 2 \un{Gyr}$ scatter, except for the six hosts that our method estimated to be $\lesssim 2 \un{Gyr}$ but \citetalias{Gupta2011} estimated to be up to $\sim 4 \un{Gyr}$. For the youngest populations ($\lesssim 4 \un{Gyr}$), our method systematically estimates a lower age. This is expected because our method's chosen star formation history is better at modeling young stellar populations. The 3$\sigma$ significance critical value for $N=76$ is a Spearman's correlation of $\pm0.35$, so it is very unlikely for this distribution, with a correlation coefficient of 0.81, to arise if these two methods were not correlated.
    }
    \label{fig:RosevGupta}
\end{figure}

The final validation was to recalculate the global host ages originally presented in \citetalias{Gupta2011}. A direct comparison between ages from \citetalias{Gupta2011} and the results from our analysis can be seen in \cref{fig:RosevGupta}. Most points fit along the one-to-one line with a scatter of around $2\un{Gyr}$ implying that our analysis is consistent with the previous work. However, there are six hosts that our method estimates to be $\lesssim 2 \un{Gyr}$, whereas none of the ages in \citetalias{Gupta2011} were that young. At $\lesssim 4\un{Gyr}$, our method systematically estimates a younger age. This is because our star formation history model allows for more recent star formation than the simple $\tau$ models permits. 
In their discussion on this topic, \cite{Simha2014} claimed that the star formation model used in \citetalias{Gupta2011} can overestimate young populations by $\sim 2 \un{Gyr}$. This overestimation can be seen in \cref{fig:RosevGupta}.


\section{Results} \label{sec:results}



Our analysis generates a probability distribution for the age of a stellar population. Probability distributions can be summarized by a median and 68\% confidence intervals, especially if the distribution is Gaussian. For non-normal distributions, particularly ones with long tails or multiple modes, the distribution can still be summarized by a median age, but its interpretation is not as clear. 
%
To accurately \replaced{display our estimated star formation histories}{represent the estimated age posterior probabilities shown in} \cref{fig:HRvAgeCGCzless02,fig:HRvAgeCLCzless02,fig:HRvAgeCLCsplit},
we plotted the results of 100 random samples from both the age and Hubble residual distributions for each \sn{}.
\added{This technique results in a probability density plot of finding a \sn{} at a given age and \hr{}.}


We test for correlations between parameters with the Spearman's rank-order correlation coefficient. This is the non-parametric version of the more common Pearson's correlation coefficient. There are several differences between Pearson's and Spearman's correlations. The most important difference for our study is that the Spearman's correlation has a high absolute value for any monotonic relationship rather than just linear relationships. This means a linear, exponential, or a single step function would all rank highly with the Spearman's correlation, but not necessarily with the Pearson's correlation. Since several previous host galaxy studies have seen steps in \hr{s} (or more generally sigmoid functions) it is reasonable to use a statistical measurement that is sensitive to these non-linear correlations. See \citet[section 4.2.3]{Wall2012} for more information on the Spearman's correlation.

A large absolute rank-order correlation refers to a tighter association of points, indicated by a small scatter around the correlation. Significance of a Spearman's correlation can be described by a standard $p$-value, or the probability under the null hypothesis of obtaining a result equal to or more extreme than what was observed. 
Given a sample size and Spearman's correlation a $p$-value can be calcuated.
If we let 3$\sigma$ be our significance limit, then for our main sample of 103 objects, the Spearman's critical correlation value would be $\pm 0.30$. That is, there is a 0.2\% chance of seeing a Spearman's correlation value of $>0.3$ or $<-0.3$ from our data set assuming no underlying correlation. 

\subsection{Global Environments}

\begin{figure}
    \centering
    \includegraphics[width=3in]{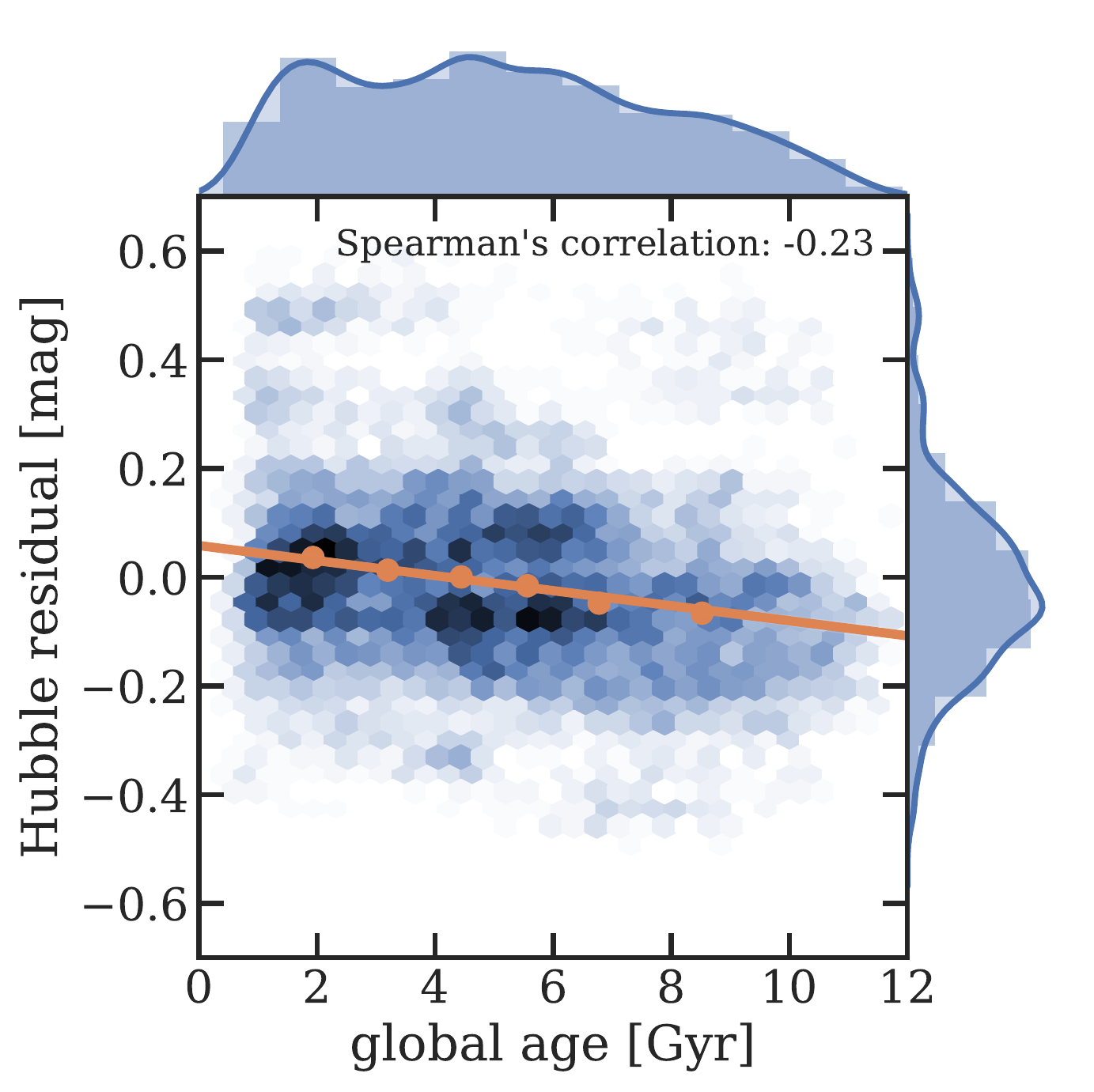}
    \caption{A 2D density plot (darker colors indicate a higher density) \replaced{comparing Hubble residuals of \sn{} versus the}{depicting the probability of finding a \sn{} at a given \hr{} and} average age of its host galaxy (global age\replaced{) for the sample derived from \citetalias{Campbell2013} (see \cref{tab:global})}{, \cref{tab:global}}. 
    The presented data is a representative sampling of the underline probability distributions for these two parameters for each \sn{} in our data set. A linear fit of the data is shown as a orange line. The orange dots represent the mean of six evenly-filled bins of the underlying data.
    The observed correlation, with a Spearman's correlation of \globalCorr, is only a \globalCorrSig{} significance. The data \replaced{shows a tranision}{shows a possible transition} or ``step'' around \ageStepLocation{}.}
    \label{fig:HRvAgeCGCzless02}
\end{figure}

First, we compare the Hubble residuals with ages derived from the global photometry of the hosts using \citetalias{Campbell2013} (\cref{tab:global}) sample. The comparison is presented in \cref{fig:HRvAgeCGCzless02}. This data set has a low \globalCorrSig{} correlation, as defined by the Spearman's value of \globalCorr{}. 
In addition, the distribution in Hubble residual-age space appears to show a distinctive ``step'' between 7 and $\ageStepLocation{}$.

\subsection{Local Environments}



\begin{figure}
    \centering
    \includegraphics[width=3in]{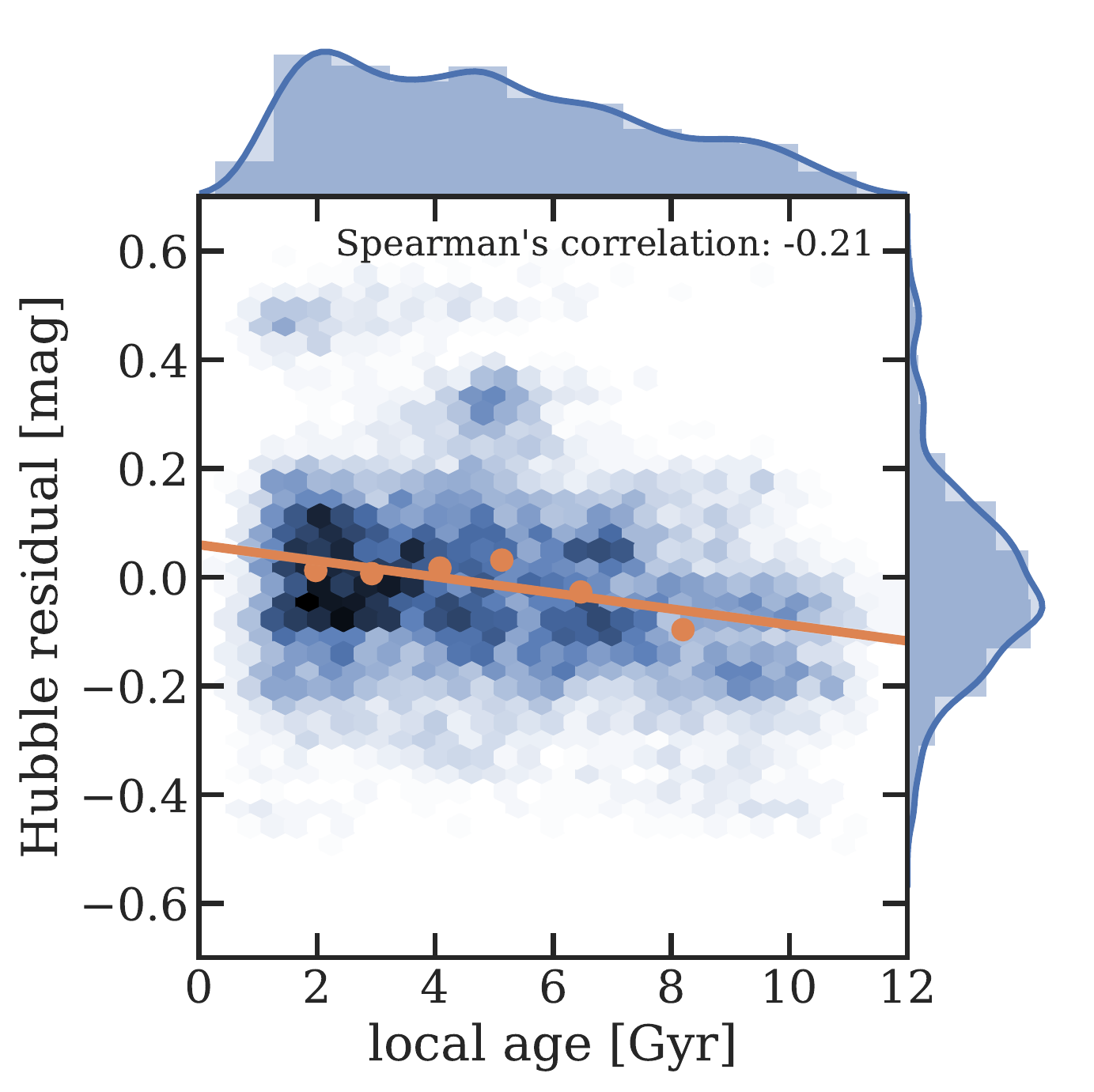}
    \caption{Comparing Hubble residuals of \sn{} versus the average age of the local stellar environment for the sample derived from \citetalias{Campbell2013} (see \cref{tab:local}).
    The data is presented the same way as \cref{fig:HRvAgeCGCzless02}. The Spearman's coefficient (\localCorr) is only slightly different than the one seen in \cref{fig:HRvAgeCGCzless02} but is insignificant only have a \localCorrSig{} significance.
    The data also has a stronger ``step'' at $\sim \ageStepLocation{}$ to the global analysis.  There seems to be significantly fewer \sn{} with a Hubble residual at $\gtrsim 0.0 \un{mag}$ with an age $\gtrsim \ageStepLocation{}$.
    }
    \label{fig:HRvAgeCLCzless02}
\end{figure}

Finally, we compare the Hubble residuals versus average local environment age for the data set derived from \citetalias{Campbell2013} (\cref{tab:local}). These results are presented in \cref{fig:HRvAgeCLCzless02}. Using an age derived from the local environment only slightly changes the Spearman's correlation between these two parameters, But this correlation, \localCorr, only has a \localCorrSig{} significance. 
The overall age distribution and apparent ``step'' at $\sim \ageStepLocation{}$ are not significantly changed by switching to a local environment analysis. At first glance, the local age does not appear to contain any additional information not already present in the global age.
\citet{Jones2018} \deleted{also} found \replaced{no}{a marginally} significant difference between their global and local analyses.

\added{Because a fraction of our sample is photometrically classified, there may be some CC contamination that would be found preferentially in the upper left of \cref{fig:HRvAgeCGCzless02,fig:HRvAgeCLCzless02}. This contamination might contribute to the observed correlattions, but at a level that is small compared with the \globalCorrSig{} and \localCorrSig{} trends.}


\section{Analysis and Discussion}

\subsection{Comparison Between Local and Global Ages}

\begin{figure}
    \centering
    \includegraphics[width=3in]{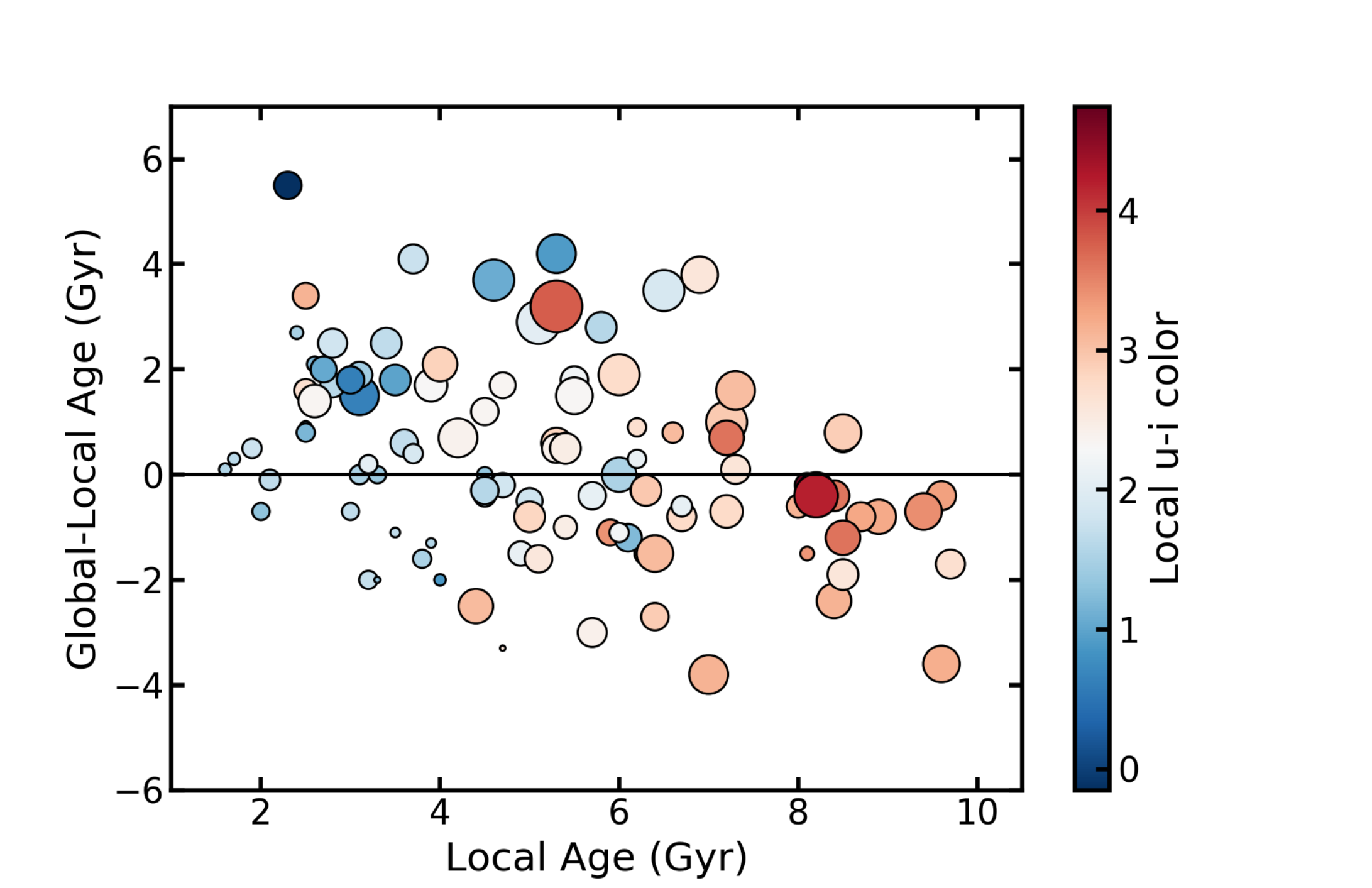}
    \caption{The difference in ages derived from global and local photometry as a function of local age. The size of the marker indicates the stellar mass of the host where the smallest circles show log(M/M$_\odot$)=7.5 and the largest circles indicates log(M/M$_\odot$)=11.5. The marker color represents the $u-i$ color index at the location of the supernova.
    }
    \label{fig:global-local}
\end{figure}

The global age estimated for spiral galaxies is an average of several stellar populations. The prompt component of \sn{} are expected to be found in the youngest regions of a galaxy. So we may see that the population age at the supernova location would tend to be younger than the global age of the galaxy. In \cref{fig:global-local} we show the difference between the estimated global and local ages. For massive galaxies with local ages less than 4~Gyr there is a tendency for the supernova to be located in a younger than average spot in the host. The effect is less apparent for low mass galaxies, but that is likely because the size of the region measured for the local environment is a large fraction of the size of a small galaxy.

Between local ages of 4 to 8~Gyr the difference between global and local age estimates show a large scatter with no apparent trend. Beyond 8~Gyr the scatter between the local and global ages is reduced, probably because the stellar population in ellipticals is fairly uniform. In these older hosts the global age estimates tend to be 1 to 2~Gyr younger than the local ages. The reason for this difference is not clear, but it may be due to activity at the center of ellipticals contaminating the stellar colors.

Although we see no statistical difference between the use of local and global ages when comparing with Hubble residuals, our results suggest that using local photometry to characterize the supernova environment has some benefit over the global average. For example, the SED measured local to the supernova can indicate a population that is a factor of two younger than the global average in large star-forming galaxies. When feasible, the measurement of the local environment, particularly of younger populations, provides a more accurate representation of the progenitor age than simple averaging the light from the host.

\subsection{Investigating the Age Step}

\begin{figure}
    \centering
    \includegraphics[width=3in]{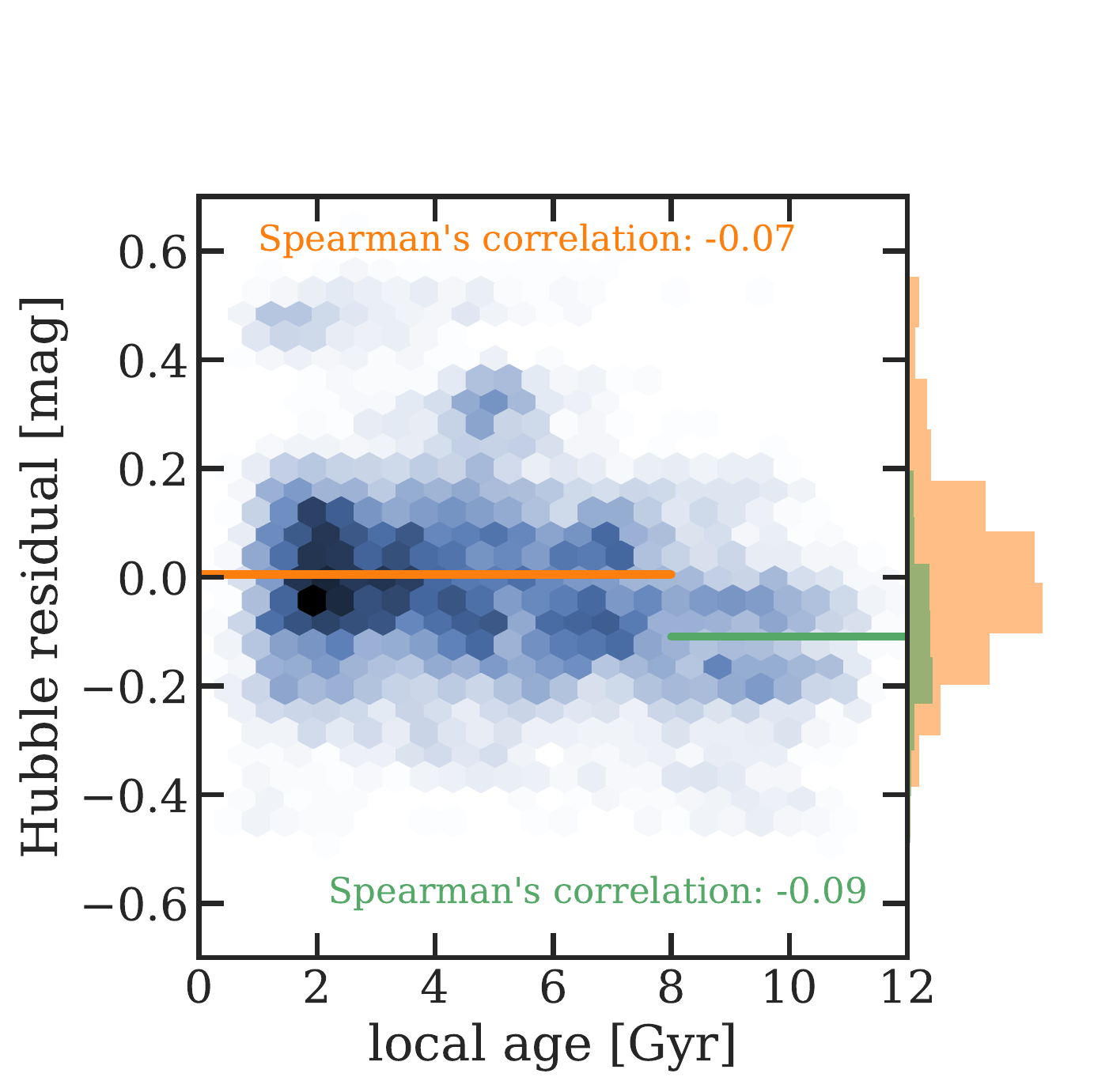}
    \caption{The same data as \cref{fig:HRvAgeCLCzless02} but with the marginalized distributions split by age. This clearly shows that the younger population ($\leq \ageStepLocation$, orange) has a higher mean Hubble residual (\meanYoung) whereas the older population ($>\ageStepLocation$, green) has a lower Hubble residual mean (\meanOld). These mean values are shown as colored lines. Due to these different means, the resulting age step is \ageStep{} (\ageStepSig).
    The Spearman's correlations for each piece is significantly lower than it was for the whole data set. Therefore, individually there is no meaningful trend within each subpopulation.
    }
    \label{fig:HRvAgeCLCsplit}
\end{figure}

The \hr{s} in the \citetalias{Campbell2013} sample \replaced{appears to have}{has a mono-tomically decreasing trend (at \globalCorrSig{}), that appears to be} a break or step at an age of $\sim \ageStepLocation$. \Cref{fig:HRvAgeCLCsplit} plots the same data as \cref{fig:HRvAgeCLCzless02} but this time splits the data into two age bins: $\leq \ageStepLocation$ and $> \ageStepLocation$. Both age ranges show very small Spearman's correlations (\youngCorr{} and \oldCorr{} respectively) that are consistent with a flat distribution. The correlation decreasing in significance when the data set is split in two, implies that the monotonic function seen in \cref{fig:HRvAgeCLCzless02} is really a step-like function with a transition at $\sim \ageStepLocation$. This function may have a transition width making it more like a continuous sigmoid function with a transition faster than our age resolution. In both the local and global age measurements, we find a significant age step in \hr{s}.

The amplitude of the jump between these two populations split at $\sim$~\ageStepLocation{} appears to be more than 0.1~mag. The younger population has a mean Hubble residual just over zero (\meanYoung), whereas the older population has a mean Hubble residual of \meanOld. This makes the resulting \replaced{age step}{difference in the means} equal to \ageStep{} (\ageStepSig). This is almost two times larger than the commonly used mass step of $0.06\un{mag}$. 
\replaced{The size of this step}{A step of this size} may affect precision cosmological measurements since the fraction of each subpopulation is expected to change with redshift. For example, one would not expect to find stellar populations as old a \ageStepLocation{} at redshifts greater than $\sim 1$ given the current standard cosmology.
Such an age step may also impact local measurements of the Hubble constant as the peak luminosity of \sn{} have so far been calibrated with Cepheid variables found exclusively in star forming galaxies. With the age step being detected at a $3\sigma$ significance, \cref{fig:HRvAgeCLCsplit} suggests a need for an additional \sn{} luminosity correction based on their host's local or global stellar age.



\subsection{Age as the Cause of the Mass Step}

\citet{Childress2014} argues for a link between host stellar mass and the delay time between stellar formation and \sn{} explosion, with results summarized in their Figure 4 where the \sn{} progenitor age distribution is divided into host galaxy stellar mass bins.
They find, with reasonable assumptions of star formation histories and \sn{} delay times, that there is a natural division in host mass and age between prompt and ``tardy'' \sn{}. Prompt \sn{} occur in lower mass galaxies with ages $\lesssim 6 \un{Gyr}$ while tardy events continue in high mass galaxies with ages $\gtrsim 6 \un{Gyr}$. Projecting their model onto the host mass axis results in an overlapping distribution of prompt/tardy explosions with a transition near 10$^{10.5}\un{M}_{\odot}$. This transition in progenitor age may correspond to the mass step observed at 10$^{10}\un{M}_{\odot}$. Projection of the \citet{Childress2014} model onto the age axis provides a clean separation between the prompt and tardy \sn{} with a dearth of events between 4 and \replaced{7}{8}~Gyr. Our age estimates are not consistent with a deficit of supernovae exploding in that range, but we do see a shift in the \sn{} light curve properties around a stellar age of about \replaced{$7\un{Gyr}$}{\ageStepLocation}.

We have estimated the host galaxy masses in our sample to test if our measurements agree with the prediction of Figure 4 in \cite{Childress2014}.
We use the {\tt kcorrect} code (v4\_3)\footnote{Available through \url{http://kcorrect.org} or \url{http://github.com/blanton144/kcorrect}.} described in \citet{Blanton2007}. This code utilizes spectral fitting templates based on the stellar population synthesis models of \citet{Bruzual2003}, which are calculated using the \citet{Chabrier2003} initial mass function. We input the SDSS model magnitudes and the pipeline redshifts for each galaxy (presented in \cref{tab:global}) to calculate the $k$-corrections and stellar mass-to-light ratios. The stellar masses are output in units of M$_{\odot} h^{-1}$, and we convert them to units of M$_{\odot}$ using the \citetalias{Campbell2013} cosmology described in \cref{sec:cosmo}. The uncertainties on the stellar masses are approximately $\pm$0.3 dex.   The resulting estimated stellar masses are reported in \cref{tab:results}.

\begin{figure}
    \centering
    \includegraphics[width=3in]{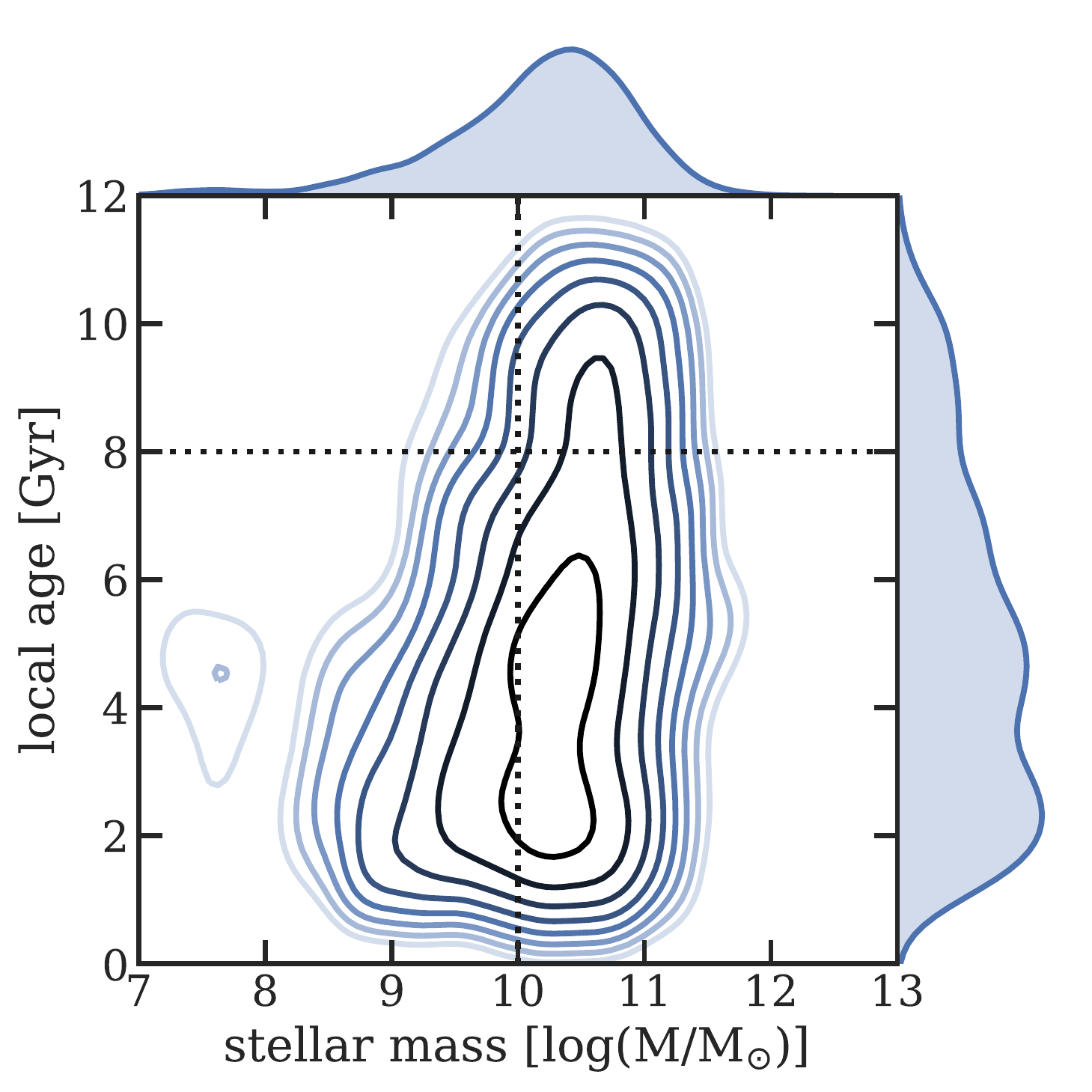}
    \caption{The distribution of \sn{} local environment age versus host stellar mass. The two dotted lines represent the measured mass and age steps, $10^{10} \un{M}_{\odot}$ and $\ageStepLocation{}$ respectively. This is in comparison to the theoretical explanation for the mass step in \cite{Childress2014}. Our ages do not have as large of a range as the theoretical prediction. The expected backwards ``L'' shape and bi-modal features are less pronounced in our data, but not missing entirely.
    }
    \label{fig:AgevStellarMass}
\end{figure}

\Cref{fig:AgevStellarMass} shows the distribution of the local \sn{} age versus stellar mass for our sample. 
For our sample, the distribution of hosts in age-stellar mass space shows similarities and differences with that predicted in \citet{Childress2014}. The observations do show young hosts extending to low stellar masses reproducing the backward ``L'' seen in the \cite{Childress2014} simulation. This indication of ``cosmic downsizing'' is not as pronounced in our data as we are probably still overestimating the ages of the extremely young populations.
Most notably missing, is that the predicted bi-modal age feature is not present. An island of young galaxies is expected between $0.5 - 1.0 \un{Gyr}$ and that is not seen in \cref{fig:AgevStellarMass}.  Our age distribution is relatively flat from 2\un{Gyr} to 5\un{Gyr} and then declines out to 11\un{Gyr}. This does not match the predicted distribution in Figure 4 of \citet{Childress2014}, and there is no clear peak of old hosts. The fact that our sample of supernovae extends to a redshift of 0.2 may contribute to the lack of a clear peak in old age. \citeauthor{Childress2014} does predict that the old-age peak decreases in height and age as the data is taken at higher redshifts simply due to the finite age of the universe. 


\subsection{Principal Component Analysis}

\begin{deluxetable}{c|ccccc||c}
\tablecaption{Results of median age, stellar mass, and PC$_1$\label{tab:results}} 
\tablehead{
    \colhead{SN}
    & 
    \colhead{local age} & \colhead{$\sigma_{a}$} &
    \colhead{global age} & \colhead{$\sigma_{a}$} & \colhead{$\log(\text{M}/\text{M}_{\odot})$} & \colhead{PC$_1$}
    }
    
\startdata
762 & 5.1 & 2.7 & 8.0 & 3.9 & 11.1 & 0.07 \\
1032 & 5.8 & 2.6 & 8.6 & 4.3 & 10.5 & -1.78 \\
1371 & 8.9 & 4.3 & 8.1 & 1.3 & 10.7 & -0.82 \\
1794 & 4.0 & 1.4 & 2.0 & 0.9 & 8.8 & 2.20 \\
2372 & 5.9 & 1.5 & 4.8 & 2.9 & 10.2 & -0.07
\enddata
\tablecomments{Stellar masses have a $0.3\un{dex}$ uncertainty and PC$_1$ was calculated with local environment ages. Full table is available online.} 
\end{deluxetable}

Light curve shape and color of \sn{} have been shown to be strongly correlated with peak luminosity. Other modest trends with population age, host mass, and gas metallicity have been reported and here we have identified a rather significant jump in \hr{s} with local and global stellar age. Because these environmental observables are highly correlated with each other, it is interesting to ask if there is a linear combination of observables that have a clear and significant correlation with \sn{} Hubble residual. As an initial test, we performed a principal component analysis (PCA) on our data set.

PCA is a linear algebra tool that transforms the basis of a matrix of data to orthogonal axes where the new axes are maximally aligned and sorted with the intrinsic scatter of the data.\footnote{For more information, see  \citet[section 7.3]{Ivezic2014}, \citet[section 4.5]{Wall2012}, or visit \url{https://towardsdatascience.com/pca-using-python-scikit-learn-e653f8989e60}} Therefore the first principal component contains the most amount of information and the last principal component contains the least. Thus, it is possible to reduce the dimensions of a problem by retaining only the principal components that comprise most of the variance (or relative information) in the original data set. In addition, the first principal component will identify a linear combination of the original variables that accounts for most of the variance in the data. Interpretation of PCA is difficult because the results are sensitive to noise, specific parameterizations, normalization, and other implementation details. However, PCA can still be useful to understand how multiple parameters work together within a data set. In the search for a trend in Hubble residuals versus \sn{} parameters, we applied PCA on the parameters of SALT2 stretch ($x_1$) and color ($c$), as well as host mass and age. We calculated PCA coefficients first using local ages and then with global ages. \hr{s} were not included as a PCA parameter. Only after the PCA did we search for correlations between the resulting principal components and the \hr{s}.

\begin{deluxetable}{ccc}
\tablewidth{0pt}
\tablecaption{Normalization parameters applied before PCA\label{tab:norm}} 
\tablehead{
    \colhead{}
    & 
    \colhead{$\mu$} & \colhead{$\sigma$}
    }
    
\startdata
$x_1$ & -0.177 & 1.015\\
$c$ & 0.0100 & 0.0829\\
$\log({\rm M}/{\rm M_\odot})$ & 10.15 dex & 0.69 dex\\
age & 5.22 Gyr & 2.11 Gyr\\
\enddata
\end{deluxetable}

Before running PCA, we normalized all parameters by removing the mean and scaling to unit variance resulting in normalized input parameters ($x_1'$, $c'$, $m'$, and $a'$). An example, for $x_1$, is defined as:
\begin{equation}\label{eqn:norm}
    x_1'={(x_1- \mu_{x_1})/\sigma_{x_1}}\ \ .
\end{equation}
The means and standard deviations used in the normalization process can be seen in \cref{tab:norm}.

\begin{deluxetable*}{c|cccc|c||cccc|c}
\tablewidth{0pt}
\tablecaption{PCA coefficients using local and global ages\label{tab:pca}} 
\tablehead{
    \colhead{}
    & 
    \colhead{$x_1$} & \colhead{$c$} & \colhead{$\log(\text{M}/\text{M}_{\odot})$} & \colhead{age\tablenotemark{a}}
    &
    \colhead{\% variance}
    &
    \colhead{$x_1$} & \colhead{$c$} & \colhead{$\log(\text{M}/\text{M}_{\odot})$} & \colhead{age\tablenotemark{b}}
    &
    \colhead{\% variance}
    }
    
\startdata
%
%
$PC_1$ & 0.56 & -0.10 &  -0.54 &  -0.63 & 44 &  0.45 & -0.13 & -0.60 & -0.64 & 47\\
$PC_2$ & -0.16 & 0.96 &  -0.21 & -0.12 & 25 & -0.21 &  0.94 & -0.26 & -0.10 &  25\\
$PC_3$ & -0.65 & -0.26 & -0.71 &  0.07 & 18 & -0.85 & -0.32 & -0.39 & -0.19 &  19\\
$PC_4$ &  0.49 & 0.09 &  -0.40 &  0.77 & 11 &  -0.16 &  0.07 &  0.65 & -0.74 &  8\\
%
%
\enddata
\tablecomments{All observables are normalized via \cref{eqn:norm}.}
\tablenotetext{a}{the median age of the local environment posterior}
\tablenotetext{b}{the median global age posterior}
\end{deluxetable*}

From these four observables, PCA yields four principal components (PC$_i$). \Cref{tab:pca} shows the linear combination of the observable variables that make up each principal component as well as the explained variance.
Substituting global age for local age resulted in some minor differences between the PCA coefficients and variance, but the overall conclusions are very similar between the two analyses.
PC$_2$ accounts for a quarter of the variance and is dominated by the SALT2 color parameter. The other two PCA components contribute only a small portion of the variance.

As a first \replaced{analsysis}{analysis}, it is important to see if there are any correlations between \hr{} and these \sn{}-host galaxy prinicpal components.
Looking at the PCA done with local age, PC$_3$ and PC$_4$ versus \hr{} have extremely low Spearman's coefficients of \pcThreeCorr{} and \pcFourCorr{} respectively. PC$_2$, dominated by the SALT2 color term, when compared with \hr{} is also a scatter plot with a low Spearman's correlation of \pcTwoCorr. Surprisingly there is an increase in the scatter at the high PC$_2$ (high color) domain. Using the PCA done with the global age, PC$_2$, PC$_3$ and PC$_4$ have similar Spearman's correlation coefficients.
Figures showing the relationship between \hr{} and each principal component (4 per analysis method, 8 total figures) are available as a Figure Set in the online journal.

\begin{figure*}
    \centering
    \includegraphics[width=5.8in]{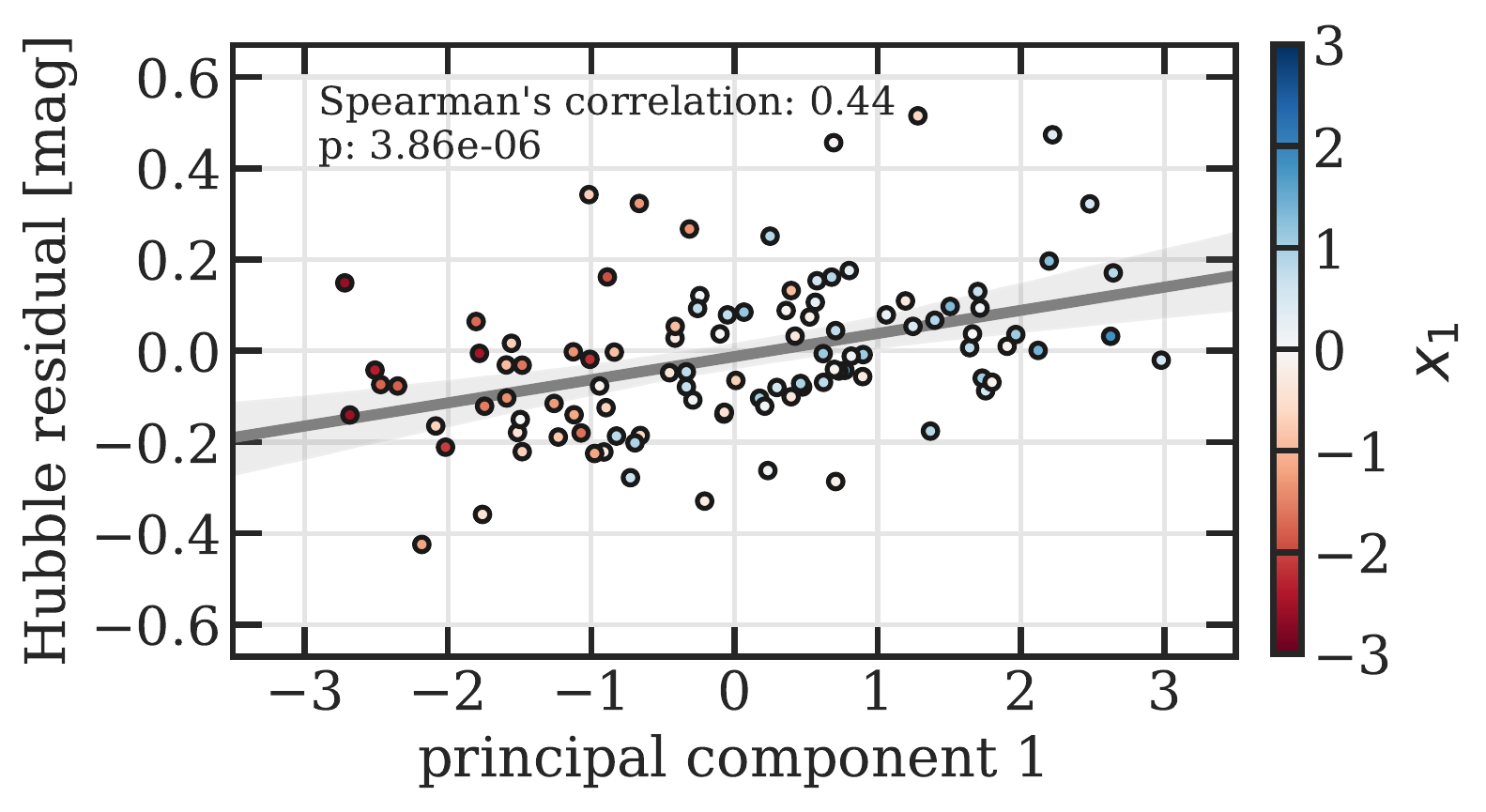}
    \caption{The trend between Hubble residual and principal component 1 (using local age) is clearly visible. With a Spearman's correlation coefficient of \pcOneCorr{} this is a strong correlation. The p-value of \pcOneP{} corresponds to a \pcOneSig{} significance. It is extremely unlikely to measure a correlation like this from uncorrelated variables. The color of each data point represents its SALT2 stretch value, red colors being a negative value and blue colors being positive values. PC$_1$ is not simply an $x_1$ effect because the red points are distributed across a significant range of PC$_1$ and two blue points ($x_1 \approx 1.5$) have an unexpected PC$_1$ of $\sim -1$. 
    The best fit linear regression has a slope of \pcOneSlopeFull{} and an intercept of \pcOneInterceptFull{}.
    The corresponding figures for each principal component (8 total images) are available as a Figure Set in the online journal.
    }
    \label{fig:pca}
\end{figure*}
\figsetstart
\figsetnum{\thefigure}
\figsettitle{Hubble residuals versus principal component}

    \figsetgrpstart
        \figsetgrpnum{\thefigure.1}
        \figsetgrptitle{Principal Component 1, local age}
        \figsetplot{HRvPC1.pdf}
        \figsetgrpnote{There is a very strong and significant correlation between Hubble residual and principal component 1 using local age.}
    \figsetgrpend


    \figsetgrpstart
        \figsetgrpnum{\thefigure.2}
        \figsetgrptitle{Principal Component 2, local age}
        \figsetplot{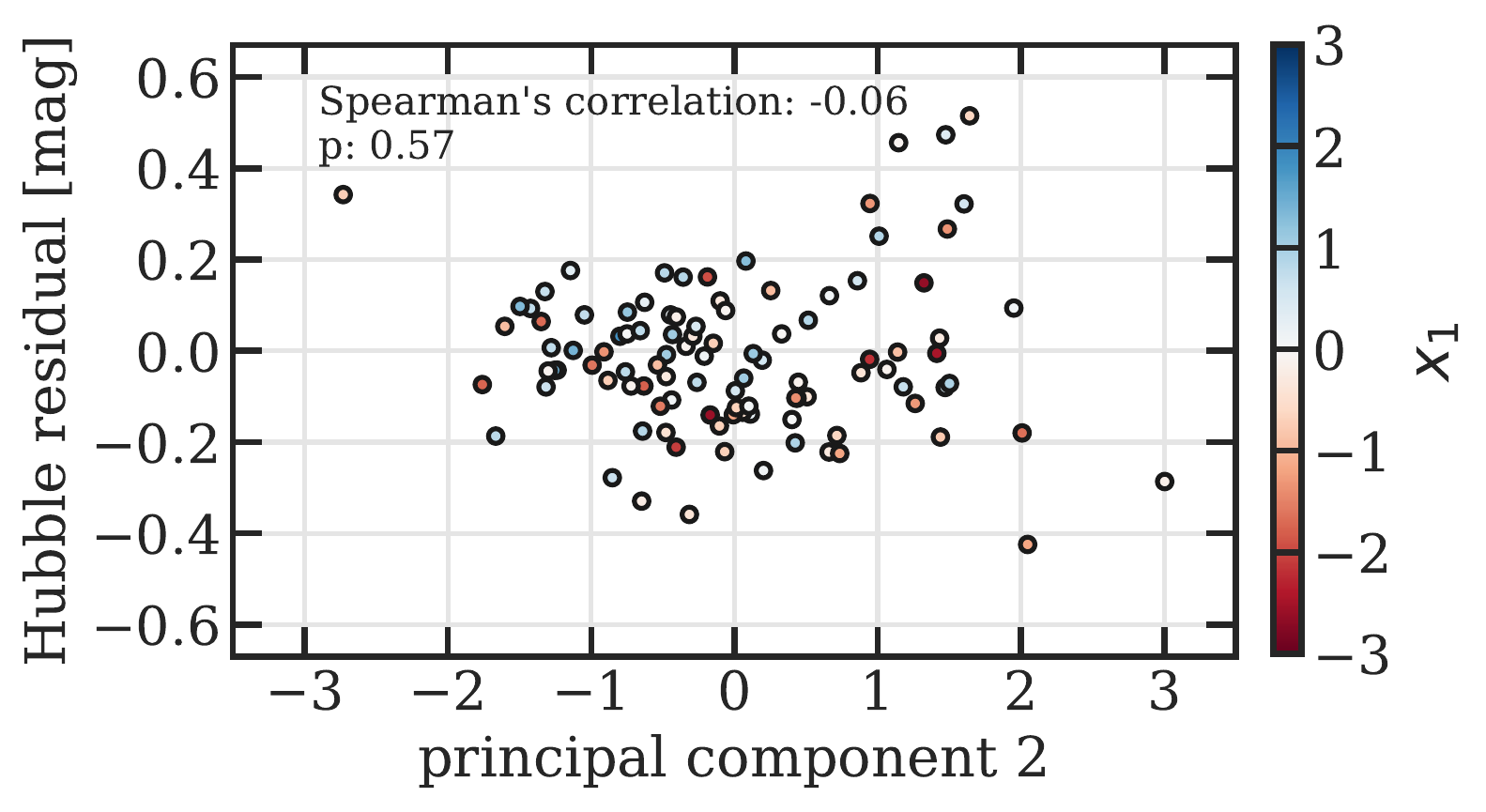}
        \figsetgrpnote{There is no correlation between Hubble residual and principal component 2 using local age.}
    \figsetgrpend
    
    \figsetgrpstart
        \figsetgrpnum{\thefigure.3}
        \figsetgrptitle{Principal Component 3, local age}
        \figsetplot{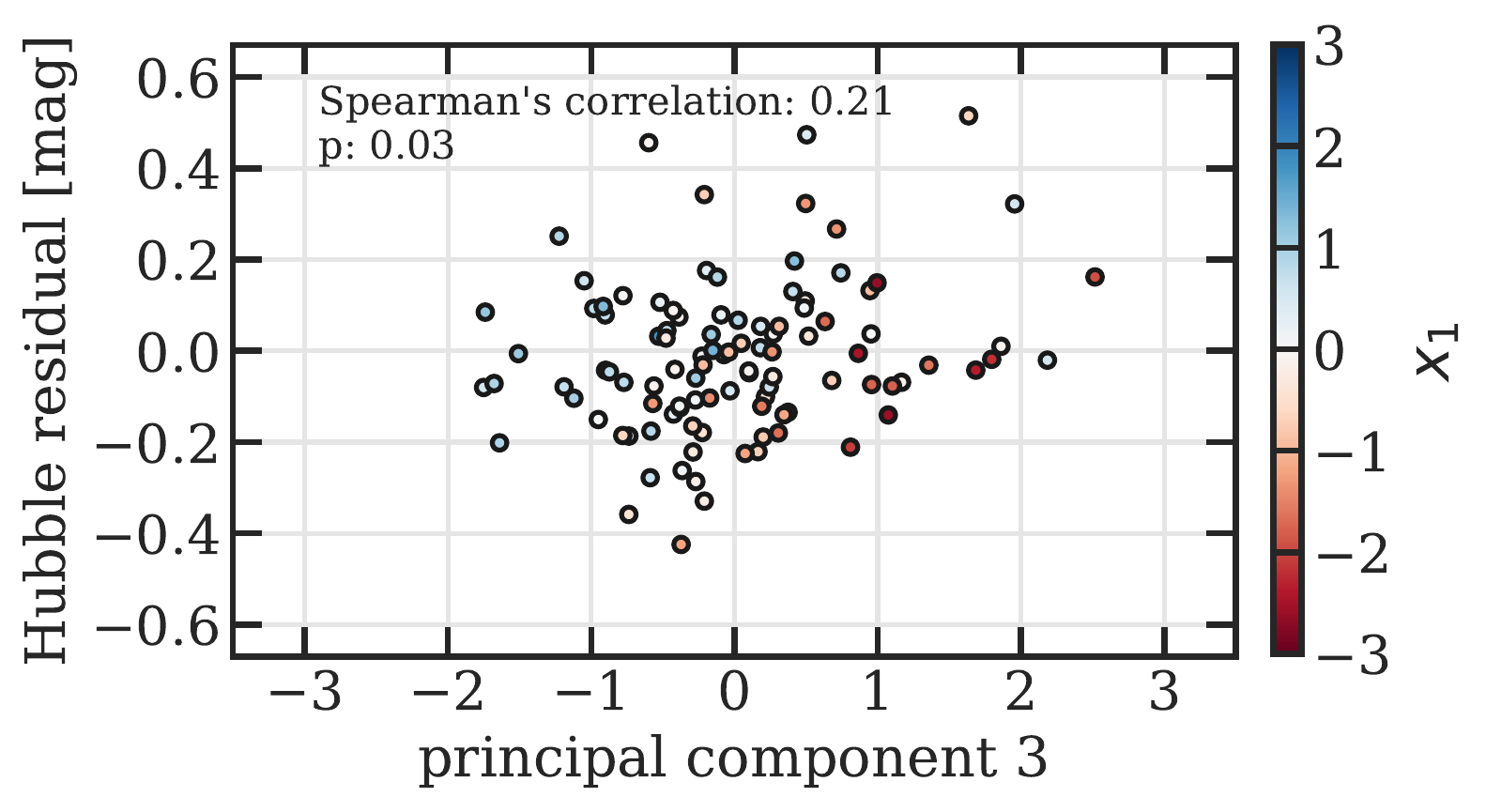}
        \figsetgrpnote{There is no correlation between Hubble residual and principal component 3 using local age.}
    \figsetgrpend
    
    \figsetgrpstart
        \figsetgrpnum{\thefigure.4}
        \figsetgrptitle{Principal Component 4, local age}
        \figsetplot{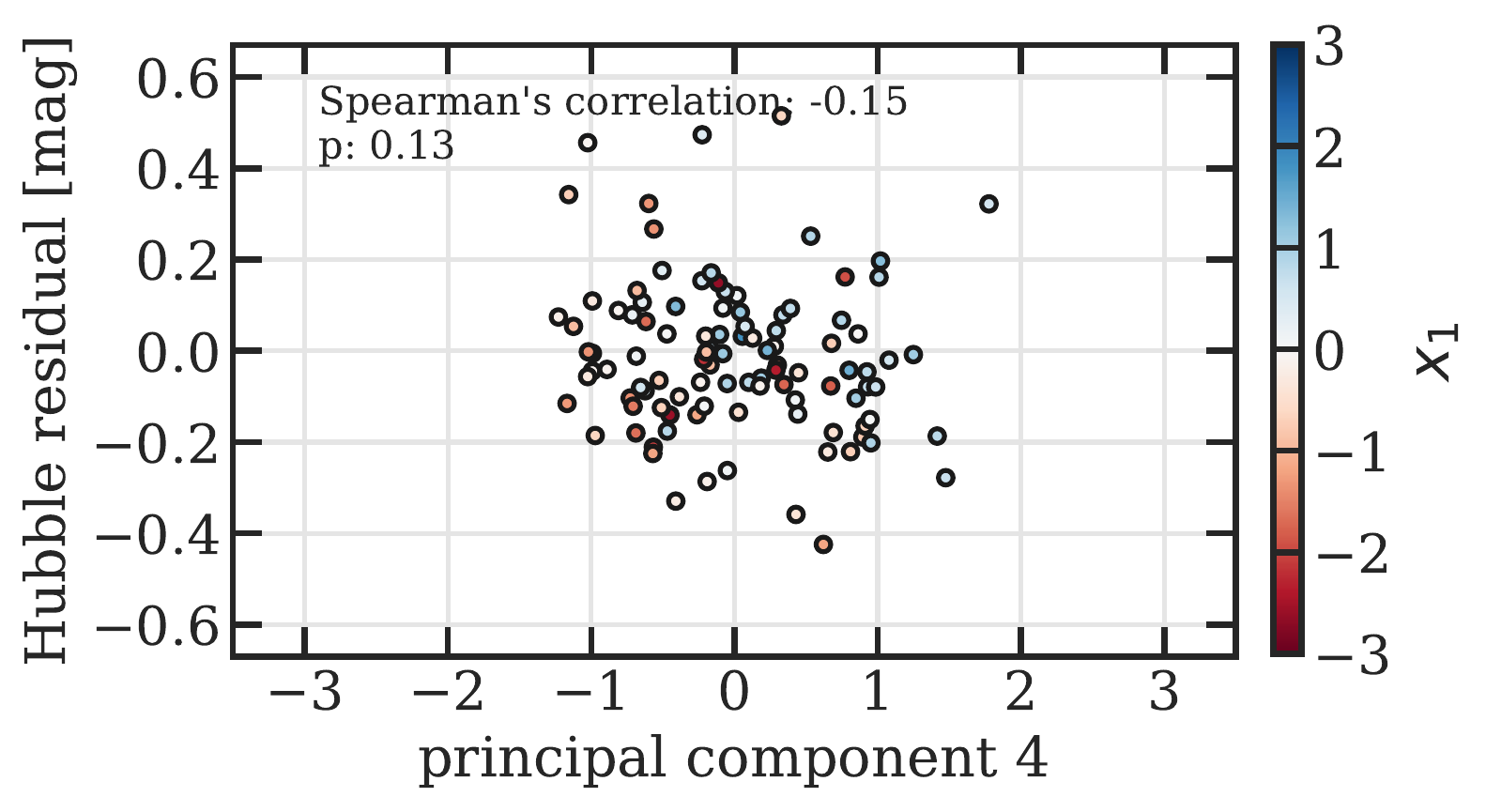}
        \figsetgrpnote{There is no correlation between Hubble residual and principal component 4 using local age.}
    \figsetgrpend
    
    \figsetgrpstart
        \figsetgrpnum{\thefigure.5}
        \figsetgrptitle{Principal Component 1, global age}
        \figsetplot{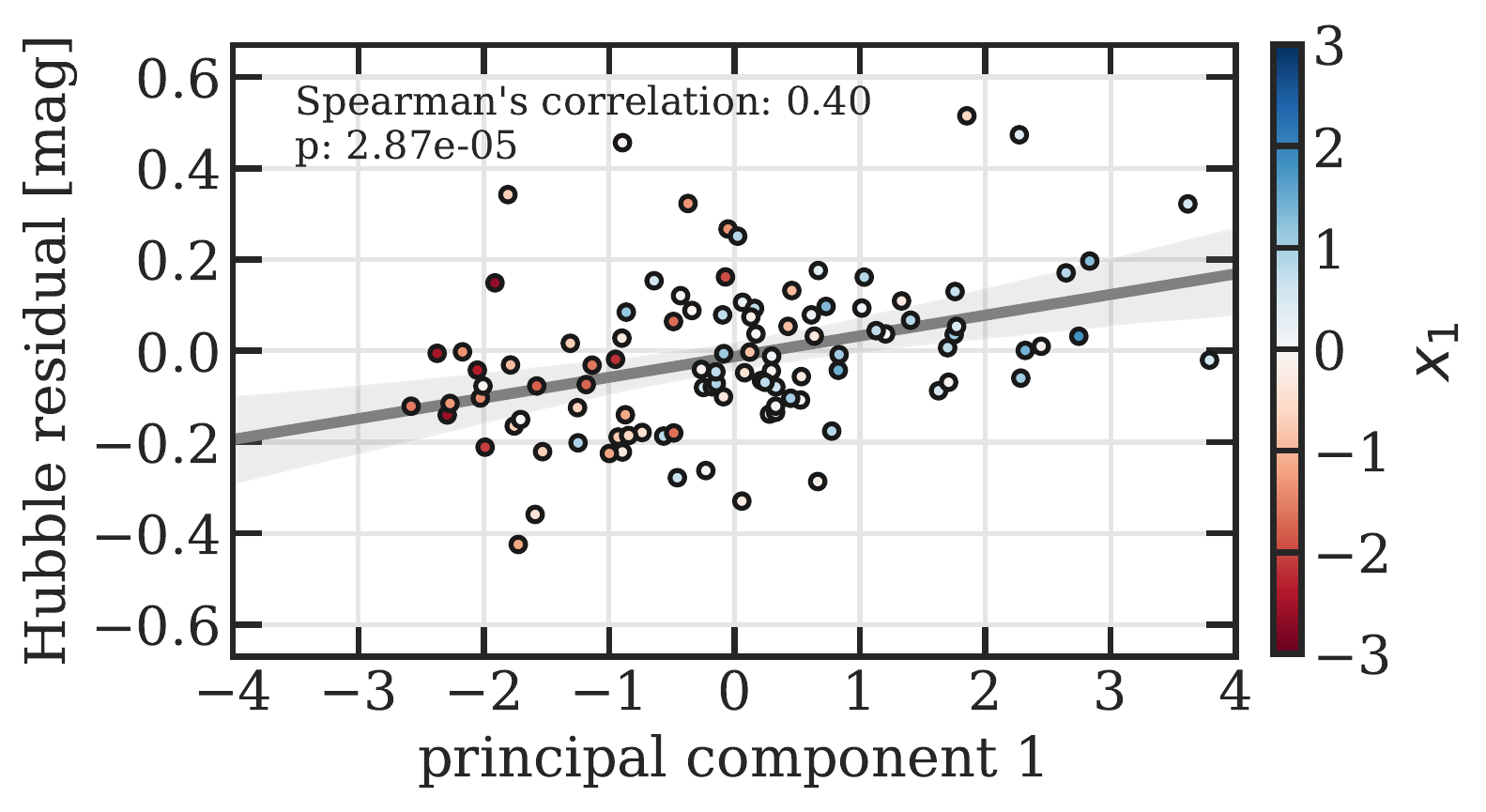}
        \figsetgrpnote{There is a very strong and significant correlation between Hubble residual and principal component 1 using global age.}
    \figsetgrpend

    \figsetgrpstart
        \figsetgrpnum{\thefigure.6}
        \figsetgrptitle{Principal Component 2, global age}
        \figsetplot{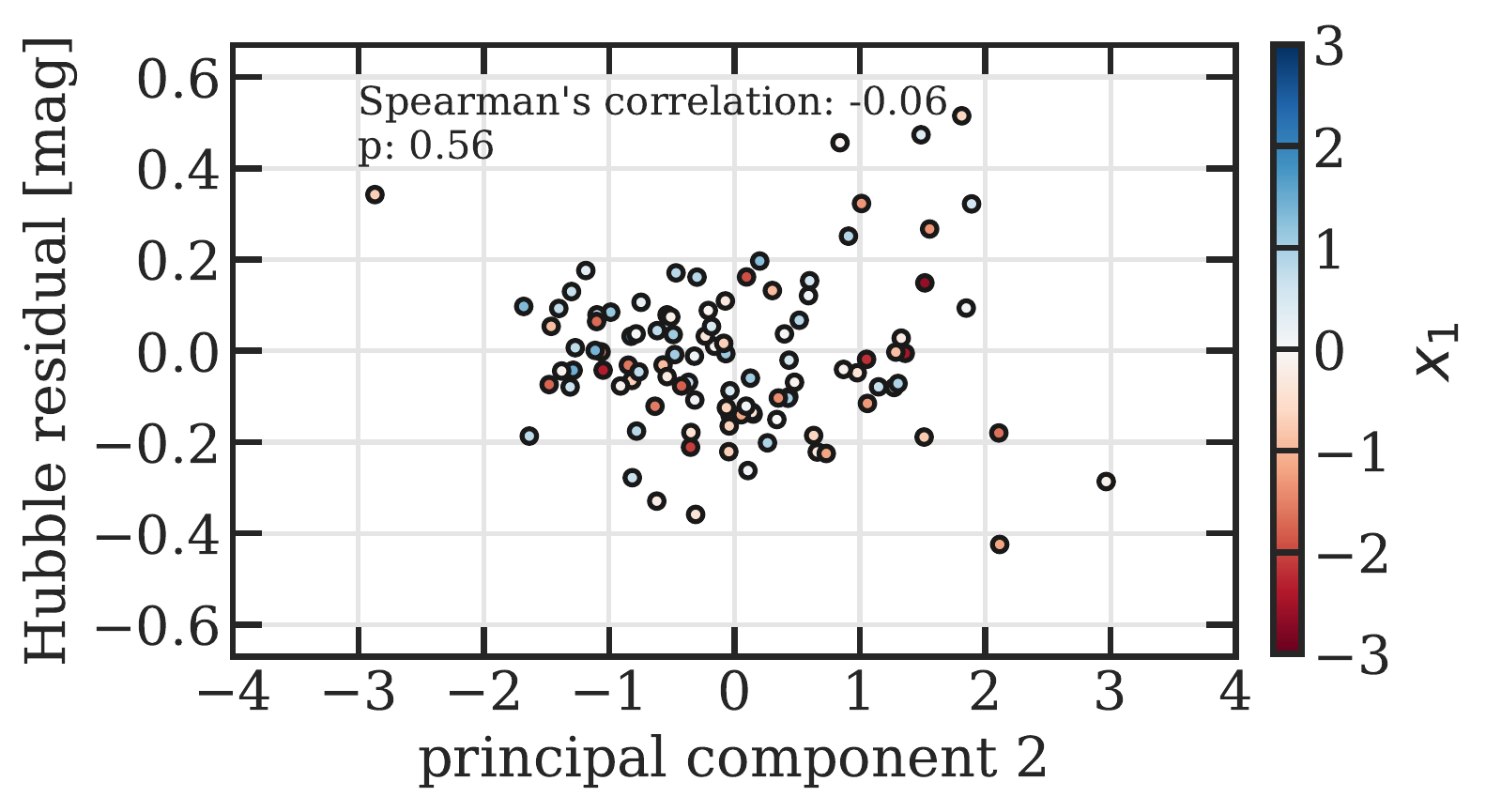}
        \figsetgrpnote{There is no correlation between Hubble residual and principal component 2 using global age.}
    \figsetgrpend

    \figsetgrpstart
        \figsetgrpnum{\thefigure.7}
        \figsetgrptitle{Principal Component 3, global age}
        \figsetplot{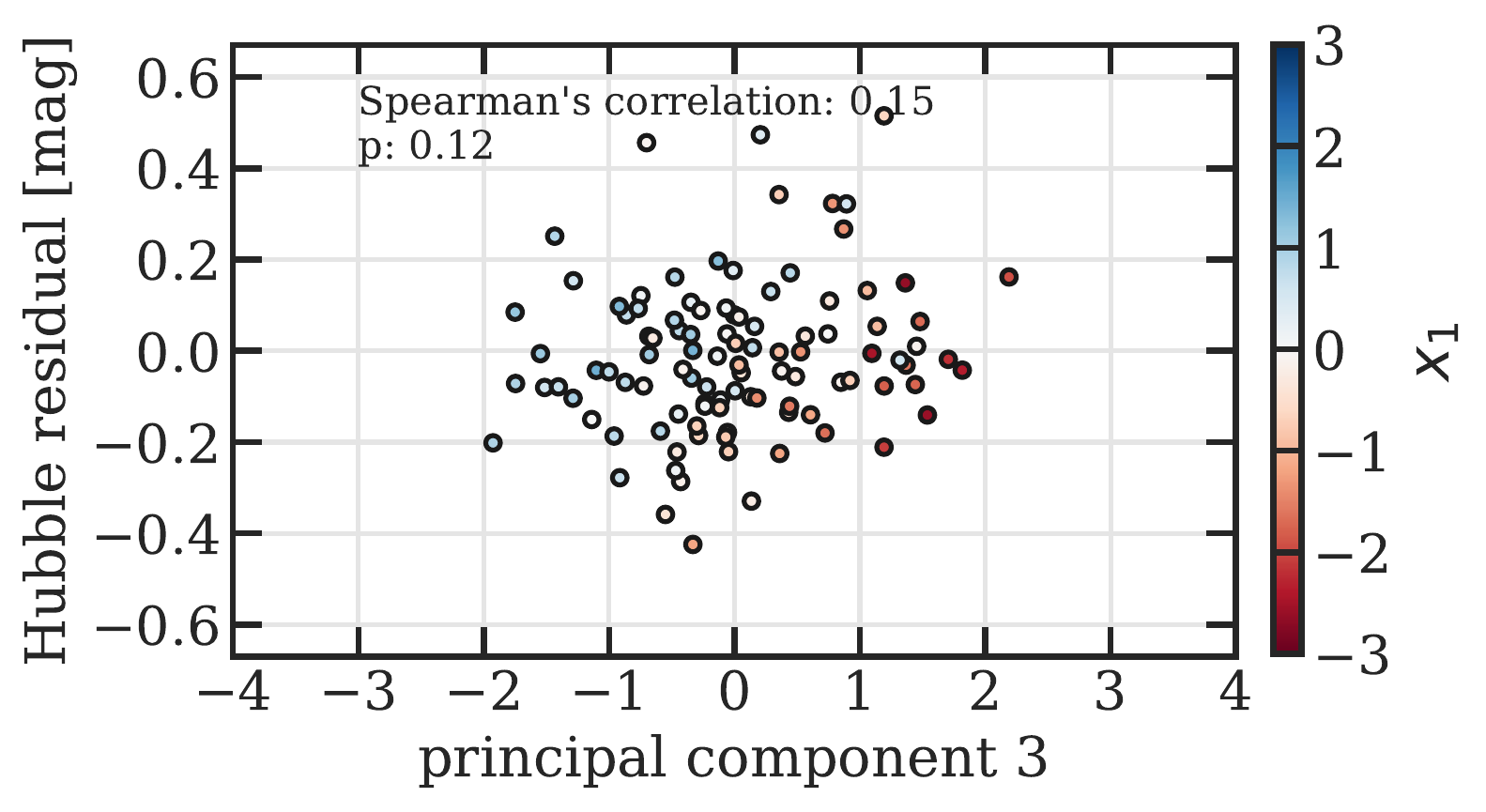}
        \figsetgrpnote{There is no correlation between Hubble residual and principal component 3 using global age.}
    \figsetgrpend

    \figsetgrpstart
        \figsetgrpnum{\thefigure.8}
        \figsetgrptitle{Principal Component 4, global age}
        \figsetplot{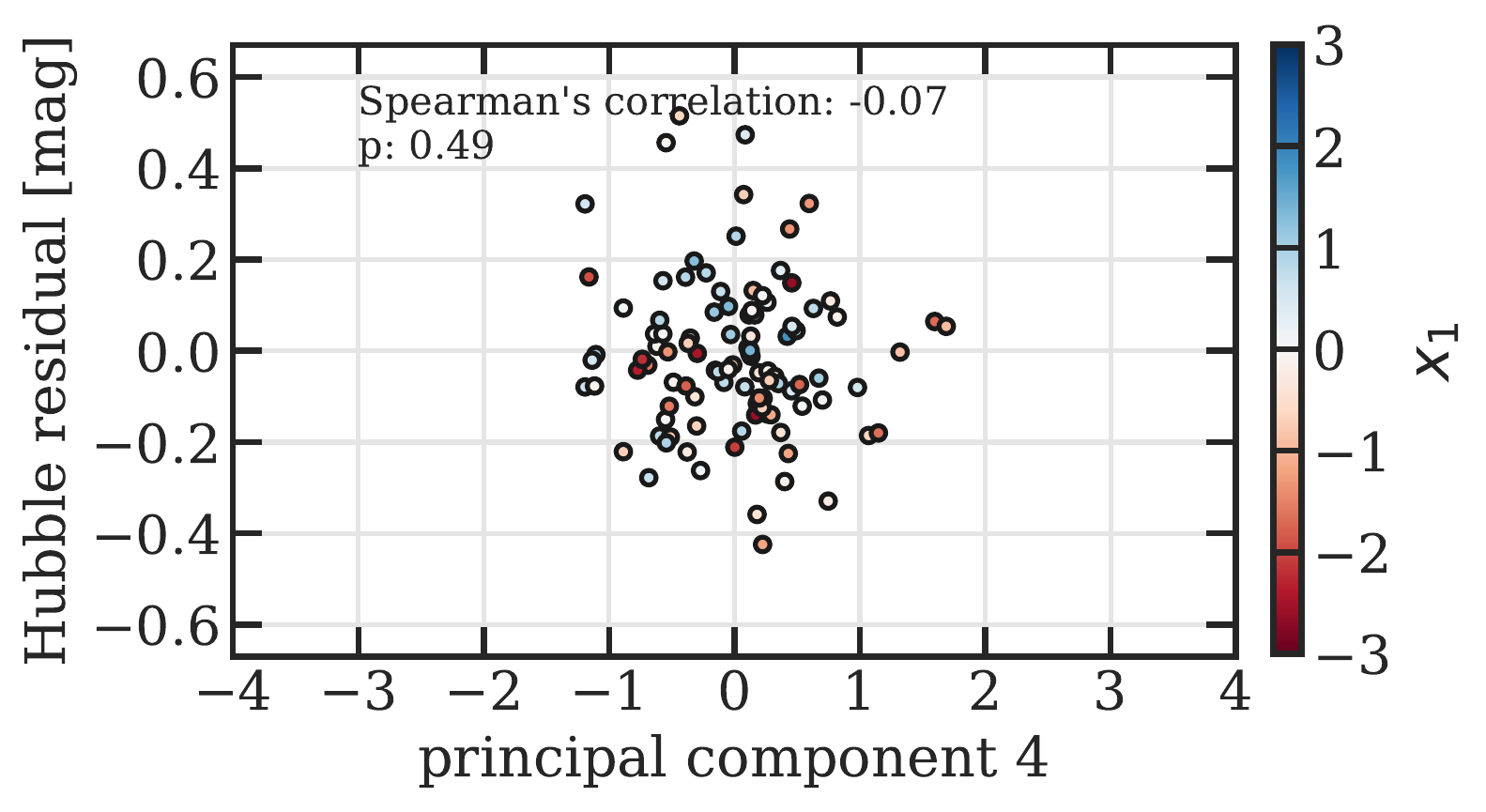}
        \figsetgrpnote{There is no correlation between Hubble residual and principal component 4 using global age.}
    \figsetgrpend
\figsetend

\begin{figure*}
    \centering
    \gridline{\fig{HRvPC1_global.pdf}{0.45\textwidth}{(a)}
          \fig{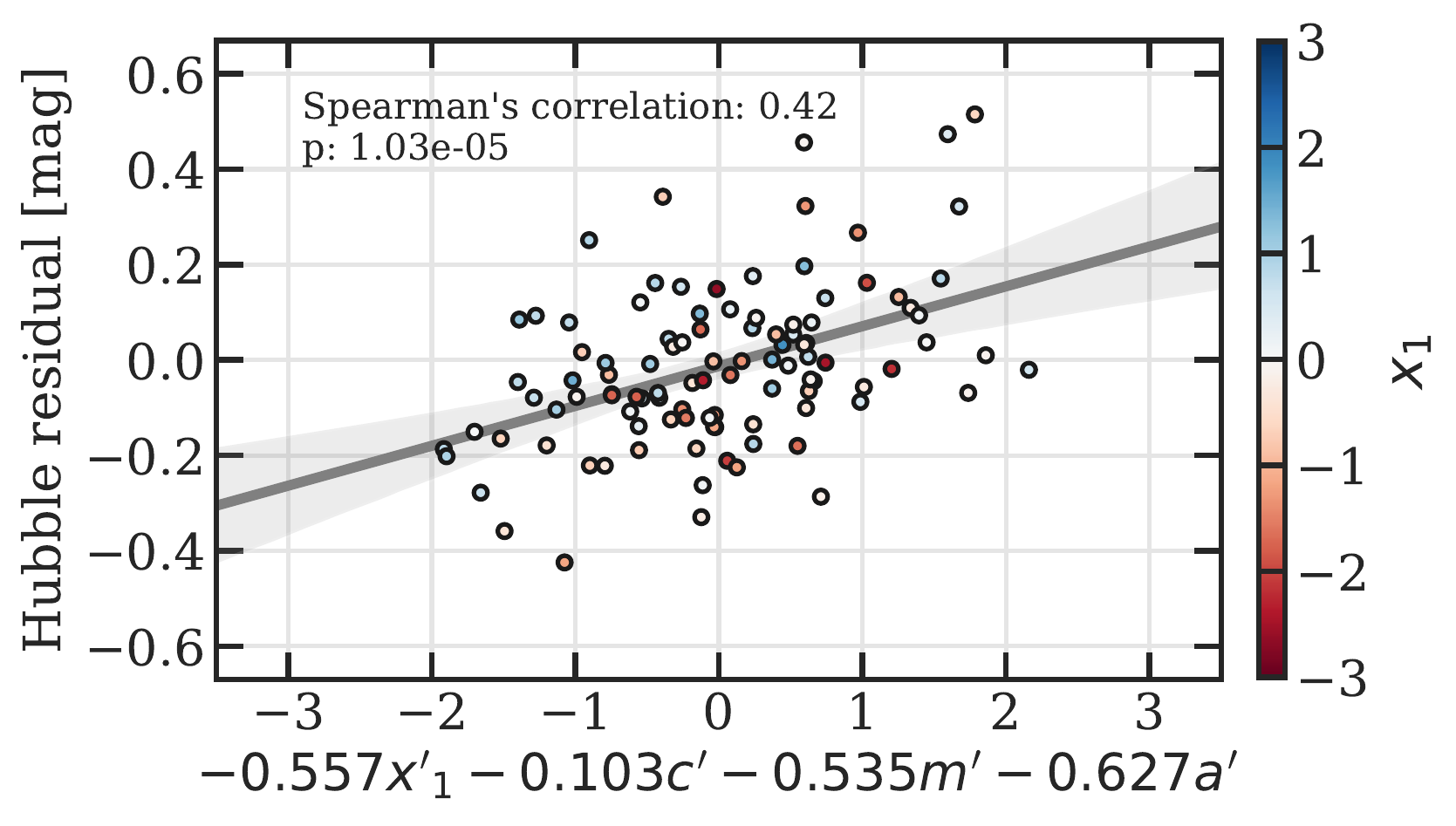}{0.45\textwidth}{(b)}}
    \caption{\textbf{(a)} Same as \cref{fig:pca}, but this figure uses the PCA with global age. In this case, the correlation has a \pcOneGlobalSig{} significance with a linear regression with a slope of \pcOneGlobalSlopeFull{} and an intercept of \pcOneGlobalInterceptFull. 
    \textbf{(b)} Same as \cref{fig:pca}, but calculated after reversing the sign on the $x_1$ parameter. In this parameterization the trend with \hr{} has a larger slope and a smaller $\chi^2$ value than PC$_1$.
    }
    \label{fig:pca-global}
\end{figure*}%

\Cref{fig:pca} shows the relationship between \hr{s} and PC$_1$ using the estimated local age. The correlation when PCA is applied using the global host age can be seen in \cref{fig:pca-global} (a). PC$_1$ with either local or global ages shows a very strong correlations with \hr{s}. 
The first principle component, using local age estimates, is defined as:
\begin{equation}
{\rm PC}_1 = 0.56 x_1' - 0.10 c' - 0.54 m' - 0.63 a'~~,
\end{equation}
or it can be approximated by ignoring the color term since it barely contributes to PC$_1$.

\begin{deluxetable}{c|cccc}
\tablewidth{0pt}
\tablecaption{\added{Statistical summary of bootstrap resampling (N=\bootstrapN) of PC$_{1}$}\label{tab:bootstrapresults}} 
\tablehead{
    \colhead{}
    & 
    \colhead{$x_1$} & \colhead{$c$} & \colhead{mass} & \colhead{age}
    }
    
\startdata
$\mu$ & 0.54 & -0.09 & -0.52 & -0.61\\
$\sigma$ & 0.08 & 0.17 & 0.11 & 0.06\\
min & -0.29 & -0.95 & -0.90 & -0.89\\
max & 0.85 & 0.93 & 0.31 & 0.06\\
\enddata
\end{deluxetable}

\begin{figure}
    \centering
    \includegraphics[width=3.3in]{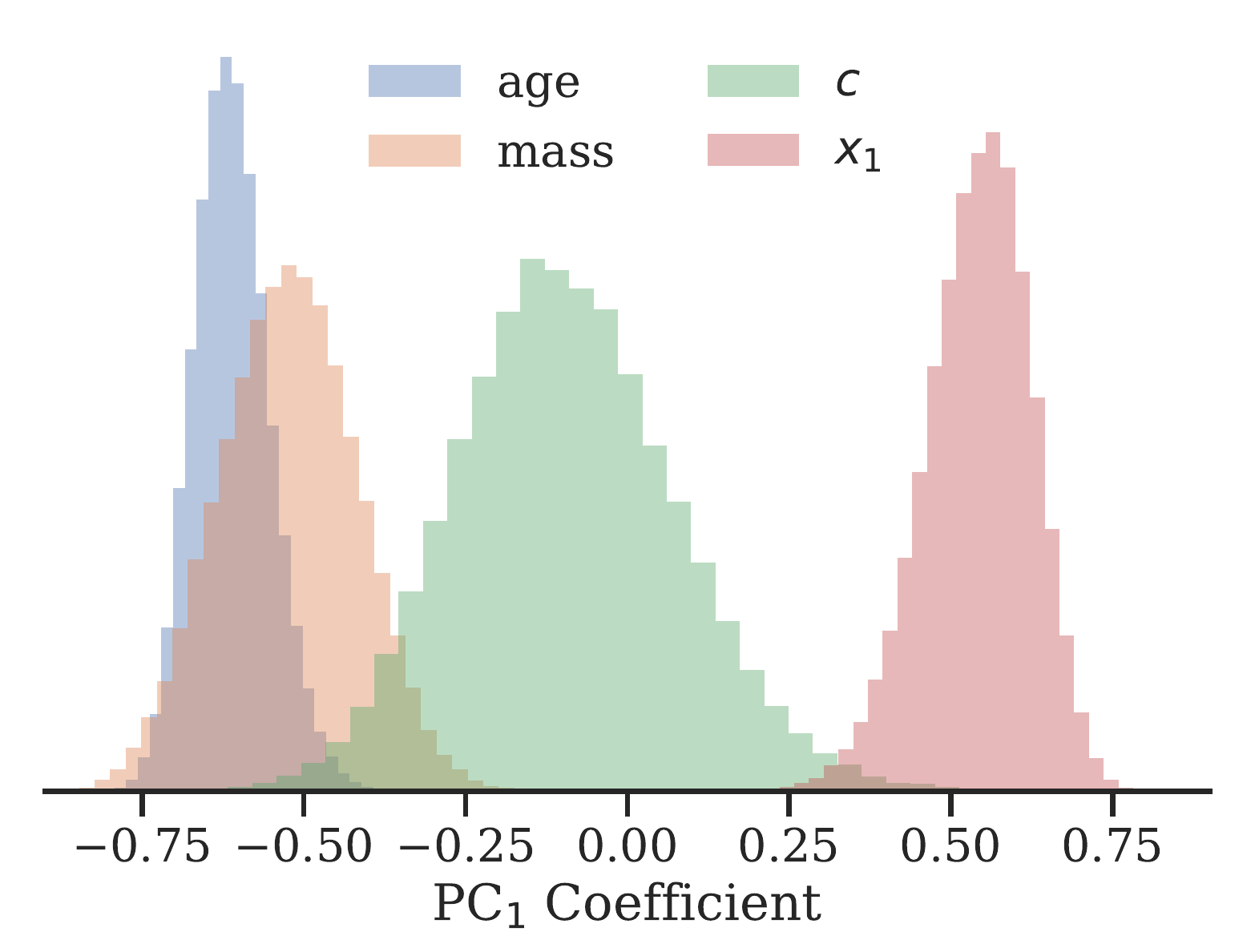}
    \caption{\added{Results of a bootstrap resampling to test the robustness of the PCA, specifically PC$_1$. This was performed using the local age data set. It is evident that the age (blue), mass (orange), and $x_1$ (red) coefficients are inconsistent with zero for \bootstrapN data sets.
    On the other hand, the uncertainty the coefficient for color (c, green) shows that it is consistent with zero.
    A statistical summary of these distributions can be found in \cref{tab:bootstrapresults}.}
    }
    \label{fig:pca-bootstrap}
\end{figure}

\added{Uncertainties for these coefficients can be obtained via bootstrap resampling \citep[section 6.6]{Wall2012}. Here we run a PCA on \bootstrapN data sets that were created randomly (with replacement) from our original data set using local ages. Since principal components are invariant to being multiplied by $-1$, a constraint was made that the bootstrap eigenvector needed to have a positive dot product with the original eigenvector. If this was not true, the bootstrap eigenvector's direction was reversed. The resulting distribution of coefficients for PC$_1$ can be seen in \cref{fig:pca-bootstrap} and a statistical summary can be seen in \cref{tab:bootstrapresults}. As expected, the color coefficient is consistent with zero. Interestingly, the coefficient for local age is more constrained than for stellar mass.}

In PC$_1$, the SALT2 color coefficient is very small, implying supernova color is not a strong contributor to this \hr{} correlation. 
The mass, age, and stretch parameters in our PCA analysis are similar in amplitude and likely are of similar importance in any further standardization of \sn{} distances.
The SALT2 stretch parameter, $x_1$, has a surprising large contribution to PC$_1$ given that the \sn{} have already been corrected for light curve shape. 
\Cref{fig:pca} suggests a correlation between stretch and \hr{}, but this correlation was removed at the start of the analysis. Instead, PC$_1$ shows that the value of $x_1$ is related to mass and age. Attempting correct for the stretch-luminosity relationship requires including host properties as all three have an influence on \hr{s}.
Our results suggest that the $\alpha$ parameter derived from the SALT2 fit is not ideal, because the affects of host mass and population age are not distributed uniformly with stretch.

The \hr{}-PC$_1$ correlation in \cref{fig:pca} has a Spearman's correlation of \replaced{0.64}{\pcOneCorr}. This trend is highly significant with a p-value of \pcOneP{} corresponding to a \pcOneSig{} significance. 
\replaced{We saw that t}{T}he trend in \hr{} versus age has a significance of only \globalCorrSig{} (\cref{fig:HRvAgeCGCzless02}), while including mass and stretch greatly increases the significance. This strong correlation suggests that host properties influence the luminosities of \sn{} beyond the currently applied corrections.

\added{This data set did not have any significant correlations between \hr{} and $x_1$ or $c$. But, when $x_1$ is combined with stellar mass and age there is a significant correlation with \hr{}.}

When using global ages, the underling trend has a Spearman's correlation of \pcOneGlobalCorr{} or a \pcOneGlobalSig{} significance. The Spearman's correlation with PC$_1$ from local ages is slightly more significant than for global ages suggesting that measurements near the event provide some improvement in correlating supernova with environment.
 
The best-fit linear regression of this trend is
\begin{equation}\label{eqn:h0-pc1}
    {\rm HR} = \pcOneSlope{} \un{mag} \times {\rm PC}_1 \pcOneIntercept \un{mag},
\end{equation}
with uncertainties in the slope and intercept as \pcOneSlopeUncert{} and \pcOneInterceptUncert{} respectively. 
This trend reduces the $1\sigma$ scatter in \hr{} from \HrSigma{} to \HrSigmaLocalTrend. The best-fit linear regression using global ages values is ${\rm HR} = \pcOneGlobalSlope \times {\rm PC}_1 \pcOneGlobalIntercept$ with uncertainties in the slope and intercept as \pcOneGlobalSlopeUncert{} and \pcOneGlobalInterceptUncert{} respectively. This reduces the $1\sigma$ scatter in the \hr{} to \HrSigmaGlobalTrend{}.
As a note, the PCA does not maximize this correlation or minimize the $\chi^2$ parameter in a linear regression. For example, reversing the sign of the $x_1$ coefficient nearly doubles the slope of the correlation with \hr{} as seen in \cref{fig:pca-global}\replaced{b}{ (b)} and significantly reduces the $\chi^2$ parameter of a linear fit. Understanding the best use of these coefficients will be part of future work already in preparation.

Since the age and mass coefficients make up nearly two-thirds of the correlation amplitude, we see that
a change of $\pm 2\sigma$ in the normalized mass and age values results in a 0.24~mag shift in \sn{} brightness. Cosmic downsizing suggests these two parameters are correlated since young galaxies tend to be small while typical old hosts are massive in the current epoch. Low-mass, young galaxies tend to be metal deficient while old, massive galaxies can be metal rich. Thus, this combination of age and mass may indicate progenitor metallicity influences the peak luminosities of \sn{}. 

\begin{figure}
    \centering
    \includegraphics[width=3in]{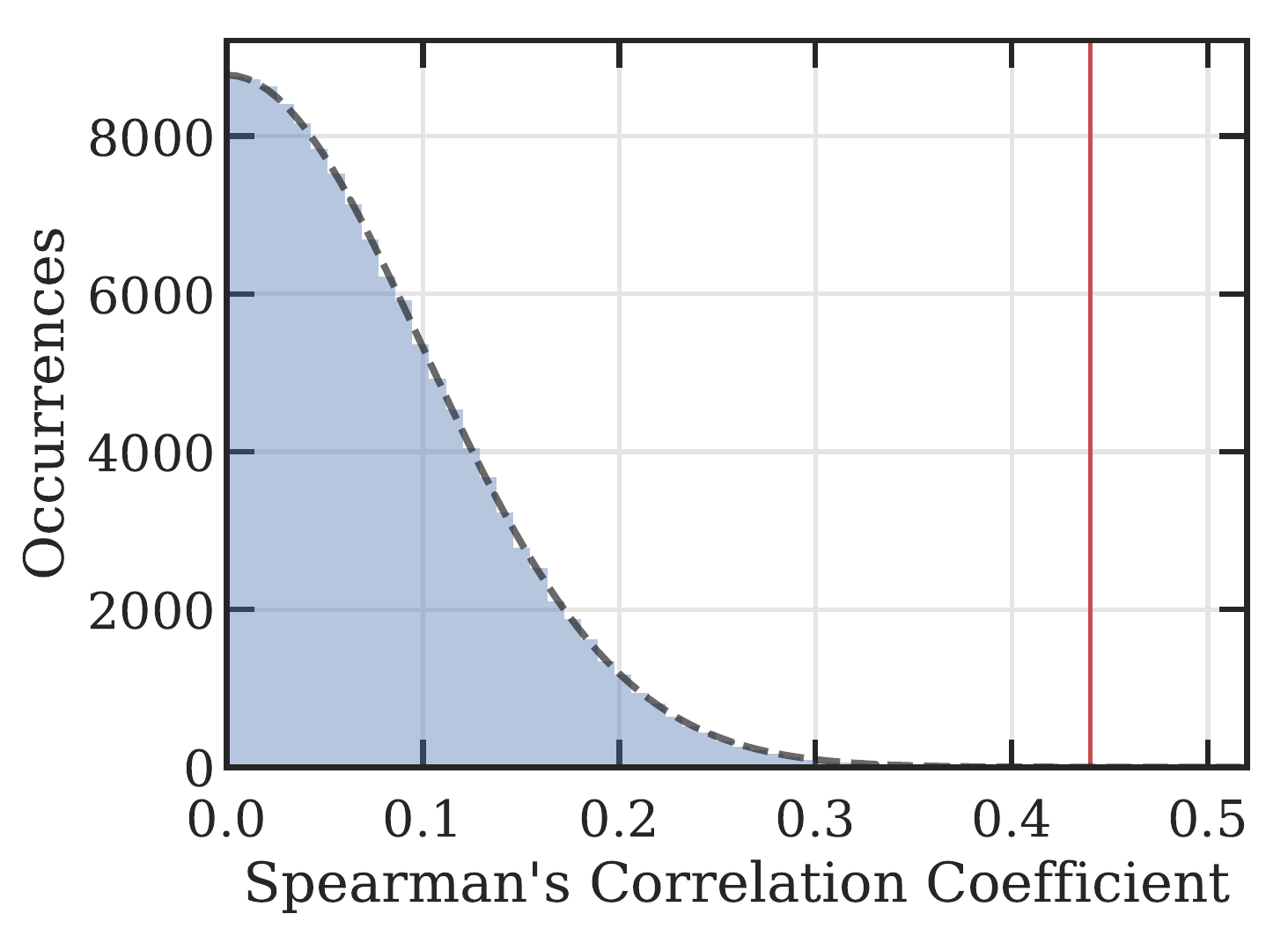}
    \caption{The distribution of 32,000 absolute value Spearman's correlation coefficient for the relationship between \hr{} and principal components of randomized \sn{} $x_1$, $c$, host stellar mass, and local environment age data sets. This distribution was generated from a bootstrap style approach that accounts for any signal PCA might produce when no underline correlation exists. With 32,000 iterations and a maximum Spearman's correlation of \maxCorr{}, the correlation between \hr{} and PC$_1$ (\pcOneCorr{}, red vertical line \cref{fig:pca}) is at least $4\sigma$ significance. This bootstrap style analysis shows that a false positive Spearman's correlation, including any PCA effects, appears to follow a Gaussian distribution with a $\sigma = 0.1$ (black dashed line). This is the expected distribution for Spearman's correlation coefficients for a data set of $N \approx 100$. 
    }
    \label{fig:bootstrap}
\end{figure}

The correlation between \hr{} and PC$_1$ is very significant, and it is unlikely that PCA could generate such a strong correlation from a random distribution. To test the probability that this correlation is caused by chance, we applied a bootstrap style method for hypothesis testing. We took our PCA input matrix ($x_1$, $c$, host mass, local environment age) and shuffled the order along each parameter, creating 103 ``new'' \sn{}. There was no cross shuffling, so a stretch always stays a stretch, it just corresponds to a different \sn{}. 
We applied PCA on each shuffled sample and tested for any correlations with \hr{}. After 32,000 runs, a $4\sigma$ test, the maximum Spearman's correlation coefficient between \hr{} and any principal component was \maxCorr{} while the measured Spearman's test of the non-shuffled values was \pcOneCorr{}. The distribution of the bootstrapped Spearman's correlation coefficients is shown in \cref{fig:bootstrap}.
This bootstrap style analysis shows that the false positive  Spearman's correlation follows a Gaussian distribution with a $\sigma \approx 0.1$. 
This is the expected distribution for Spearman's correlation coefficients for a data set of $N \approx 100$.
It is exceedingly unlikely that the correlation between \hr{} and PC$_1$ found in \cref{fig:pca} would appear at random. As a result, this luminosity-stretch-mass-age relationship is the most significant systematic seen between calibrated \sn{} and host-galaxy environment.

\subsection{Correcting for the Hubble Residual-PC$_1$ Correlation}

Since our PC$_1$ strongly correlates with \hr{s}, it is reasonable to consider a modification to the Tripp formula that is used to correct \sn{} peak luminosities. The new correction coefficients would be the multiplication of the PCA coefficients and the slope of the \hr{}-PC$_1$ trend. Except for the SALT color, the individual components making up PC$_1$ have significant weights, so are included in this modified equation. The stretch parameter in PC$_1$ can be grouped with the original Tripp coefficient, leaving a new term with just host properties. Since PC$_1$ has a positive correlation with \hr{}, the coefficient should come in with a negative sign. From these results, and following the example of others \citep[e.g.][]{Moreno-Raya2016b}, we propose a change in the distance modulus corrections performed by SALT2 by modifying the equation to include PC$_1$. This new equation would be
\begin{equation}
    \mu = m_B - M_B + (\alpha-\alpha') x_1 - \beta c + \gamma~m' + \gamma~a' 
\end{equation}
where $\alpha'\approx\gamma\approx 0.03\un{mag}$. This uses both the approximate form of PC$_1$ and encompasses the very slight renormalization of $x_1$ into $x_1'$. More research is needed to accurately determine $\alpha'$ and $\gamma$.



\begin{figure}
    \centering
    \includegraphics[width=3in]{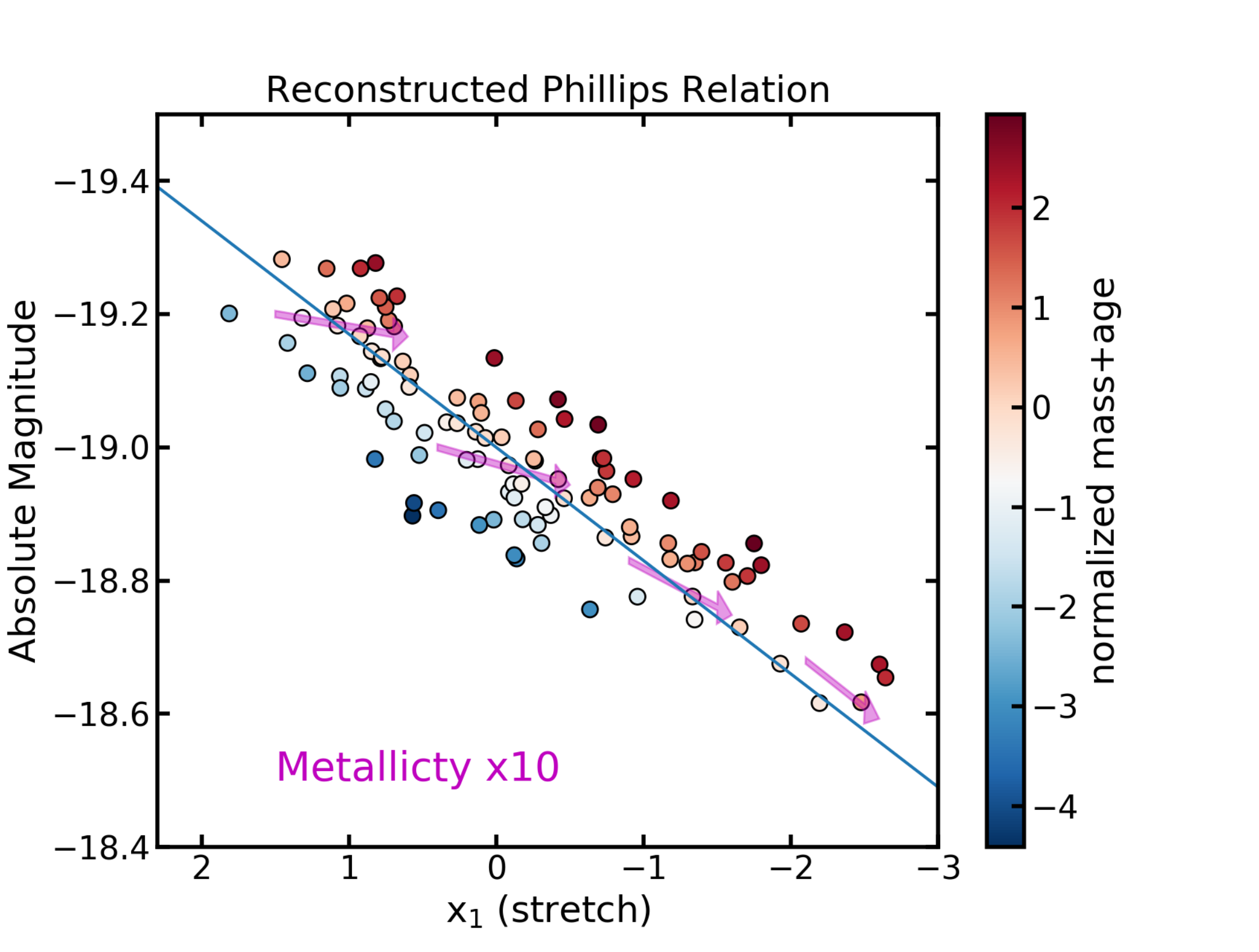}
    \caption{A Phillips relation reconstructed from the PC$_1$ parameters assuming all \sn{} in our sample have a peak ${\rm M_V} = -19.0$. For the reconstruction no color correction was applied and we set the slope to the effective stretch coefficient of $\alpha = 0.16$. The color of the points represents the sum of the mass and age parameters. Blue points are young, low mass galaxies in our sample and red points are high mass, old hosts. The arrows approximate the predictions of Figure~4 in \citet{Kasen2009} and represent how an increase of a factor of 10 in metallicity would impact the Phillips relationship. For luminous events metallicity mainly reduces the stretch at a fixed absolute magnitude. For low-luminosity supernovae the affect of metallicity is to move the points parallel to the stretch correction correlation resulting in little impact on distance estimates. 
    }
    \label{fig:phillips}
\end{figure}

This modified distance correction formula now includes a host galaxy stellar mass correction and a host age term. \citet{Childress2014} attempted to explain the mass step as an age dependency and we see that both appear to have an impact on \sn{} luminosities. Our result is bolstered by \citet{Jones2018} who showed that stellar population color effects are still present after stellar mass corrections. Stellar mass plus population age is similar to the Mannucci relationship \citep{Mannucci2010} that connects galaxy mass and star-formation rate with gas phase oxygen abundance. Our combined age plus mass parameters work in the same sense as the Mannucci relation and maybe a proxy for metallicity. Both \citet{Hayden2013} and \citet{Moreno-Raya2016b} have indicated that metallicty is the underlying galactic variable influencing \hr{s} and our results support this suggestion.

\citet{Kasen2009} shows that varying the metallicity of \sn{} progenitors will shift the slope of the Phillips relationship.  To investigate this we assume our \sn{} have a ${\rm M_{peak}} = -19$. We then add in the Phillips relationship (with $\alpha = 0.16$) and our \hr{}-PC$_1$ relationship. The result is seen in \cref{fig:phillips}. Since, fast decliners are preferentially in larger and older hosts, you see the host galaxy effect on the Phillips relationship is stretch dependant. Put another way, the slope of the Phillips relationship, $\alpha$, is dependant on host galaxy properties of the sample.
These results show that these parameters are interrelated and that there is likely a multi-dimensional relationship between peak absolute magnitude, decline rate, and progenitor metallicity.
Research into this multi-dimensional relationship will be a part of a future work already in preparation.



\section{The Affect on \h{}}

The most precise \sn{} measurement of the local value of \h{} was presented in \cite{Riess2016,Riess2018b}. \citeauthor{Riess2016} rebuilt the distance ladder connecting geometric distances, Cepheid distances and finally \sn{} in the Hubble flow. These works take into account the host galaxy mass step, but this correction has a minor impact \citep[0.7\%,][]{Riess2016} on the resulting \h{}. Our trend in PC$_1$ includes stellar age that has the potential for a significant impact on \h{} since Cepheid hosts tend to have a significantly younger stellar population than the average Hubble flow galaxy.
\added{Here we put a constraint on the influence this \hr{}-PC$_1$ trend, seen in \cref{fig:pca}, may have on the measurement of \h{} presented in \cite{Riess2016}.}


\begin{deluxetable*}{cc|ccccc|ccccc||cc}
\tablecolumns{13}
\tablewidth{0pt}
\tablecaption{Ages, stellar mass, and SALT2 values from the \h{} \sn{} calibration sample\label{tab:h0-cal}} 
\tablehead{
    \colhead{\sn{}} & \colhead{host}
    & 
    \colhead{$x_1$} & \colhead{$\sigma_{x_1}$} & \colhead{$c$} & \colhead{$\sigma_{c}$}
    & 
    \colhead{citation}
    &
    \colhead{local age} & \colhead{$\sigma_{a}$} & \colhead{global age} & \colhead{$\sigma_{a}$} & \colhead{$\log(\text{M}/\text{M}_{\odot})$}
    &
    \colhead{PC$_{1, {\rm local}}$} & \colhead{PC$_{1, {\rm global}}$}
    }

\startdata
1981B & NGC 4536 & -0.32 & 0.14 & 0.030 & 0.010 & J07 & 6.5 & 2.4 & 3.0 & 2.4 & 10.2 & -0.49 & 0.61 \\
1990N & NGC 4639 & 0.63 & 0.04 & 0.014 & 0.004 & J07 & 6.5 & 1.4 & 4.8 & 3.0 & 10.2 & 0.06 & 0.55 \\
1994ae & NGC 3370 & 0.32 & 0.10 & -0.065 & 0.033 & J07 & 5.4 & 1.3 & 5.0 & 1.9 & 10.0 & 0.46 & 0.62 \\
1995al & NGC 3021 & 0.71 & 0.08 & 0.051 & 0.006 & J07 & 6.4 & 1.7 & 6.0 & 1.5 & 10.2 & 0.06 & 0.16 \\
1998aq & NGC 3982 & -0.40 & 0.07 & -0.086 & 0.007 & J07 & 5.2 & 1.4 & 6.1 & 1.4 & 10.0 & 0.10 & 0.01 \\
2002fk & NGC 1309 & 0.22 & 0.04 & -0.101 & 0.003 & S12\tablenotemark{a} & 6.6 & 2.6 & 5.2 & 1.4 & 10.1 & -0.06 & 0.44 \\
2003du & UGC 9391 & 0.30 & 0.04 & -0.100 & 0.004 & J07 & 4.6 & 1.3 & 4.4 & 1.4 & 9.0 & 1.50 & 1.73 \\
2007af & NGC 5584 & -0.45 & 0.02 & 0.053 & 0.004 & H12 & 6.2 & 1.6 & 6.2 & 2.0 & 9.8 & -0.26 & -0.13 \\
2009ig & NGC 1015 & 1.76 & 0.15 & -0.058 & 0.013 & H12 & 8.8 & 4.4 & 7.9 & 2.1 & 10.3 & 0.01 & 0.20 \\
2011by & NGC 3972 & 0.02 & 0.13 & 0.012 & 0.014 & B14\tablenotemark{a} & 4.5 & 3.0 & 4.1 & 2.6 & 9.8 & 0.60 & 0.78 \\
2011fe & M 101 & -0.21 & 0.07 & -0.066 & 0.021 & P13\tablenotemark{b} & 3.1 & 1.0 & 4.9 & 1.2 & 9.9 & 0.88 & 0.46 \\
2012cg & NGC 4424 & 0.45 & 0.04 & 0.080 & 0.020 & V18\tablenotemark{b} & 3.5 & 2.3 & 2.8 & 2.5 & 10.0 & 0.88 & 1.07 \\
2012ht & NGC 3447 & -1.25 & 0.05 & -0.080 & 0.030 & V18\tablenotemark{b} & 4.0 & 1.3 & 4.4 & 1.4 & 9.2 & 0.63 & 0.79 \\
2013dy & NGC 7250 & 0.70 & 0.04 & 0.089 & 0.025 & V18\tablenotemark{b} & 4.6 & 1.2 & 4.2 & 1.4 & 9.2 & 1.33 & 1.51 \\ 
\enddata

\tablecomments{Light curve parameters were estimated from cited light curves using sncosmo (\doi{10.5281/zenodo.592747}). Citation key: J07-\citet{Jha2007}, S12-\citet{Silverman2012}, H12-\citet{Hicken2012}, B14-\citet{Brown2014}, P13-\cite{Pereira2013}, V18-\cite{Vinko2018}}
\tablenotetext{a}{Light curve data supplied by the Open Supernova Catalog \citep{Guillochon2016}.}
\tablenotetext{b}{Citation is for the SALT2 parameters.}

\end{deluxetable*}

To test if the correlation in the PC$_1$ parameter \replaced{will}{could} affect \h{} \added{tension}, we estimated the local age, global age and stellar mass of the \sn{} hosts that have distances calibrated using Cepheid variables. We then compare the average PC$_1$ parameter found for the Cepheid sample with the average for a Hubble flow sample. As a representative Hubble flow data set, we use our analysis of the \citetalias{Campbell2013} galaxies. \added{This assumption makes our estimated shift an upper limit since the work in \citetalias{Campbell2013} does not include more recent corrections of small biases \citep[e.g.][]{Betoule2014,Kessler2017} and the mass step that were performed in \cite{Riess2016}.} The difference in the average PC$_1$ values for the two sets of hosts, multiplied by the slope in \cref{eqn:h0-pc1} provides an estimate of the shift in peak absolute magnitude, $\Delta M$ between the calibration sample and the Hubble flow. The fractional error in distance, and therefore \h{}, due to the differences in age, mass, and stretch between the two samples is then
\begin{equation}
    10^{\Delta M/5} - 1.
\end{equation}

\begin{figure}
    \centering
    \includegraphics[width=3in]{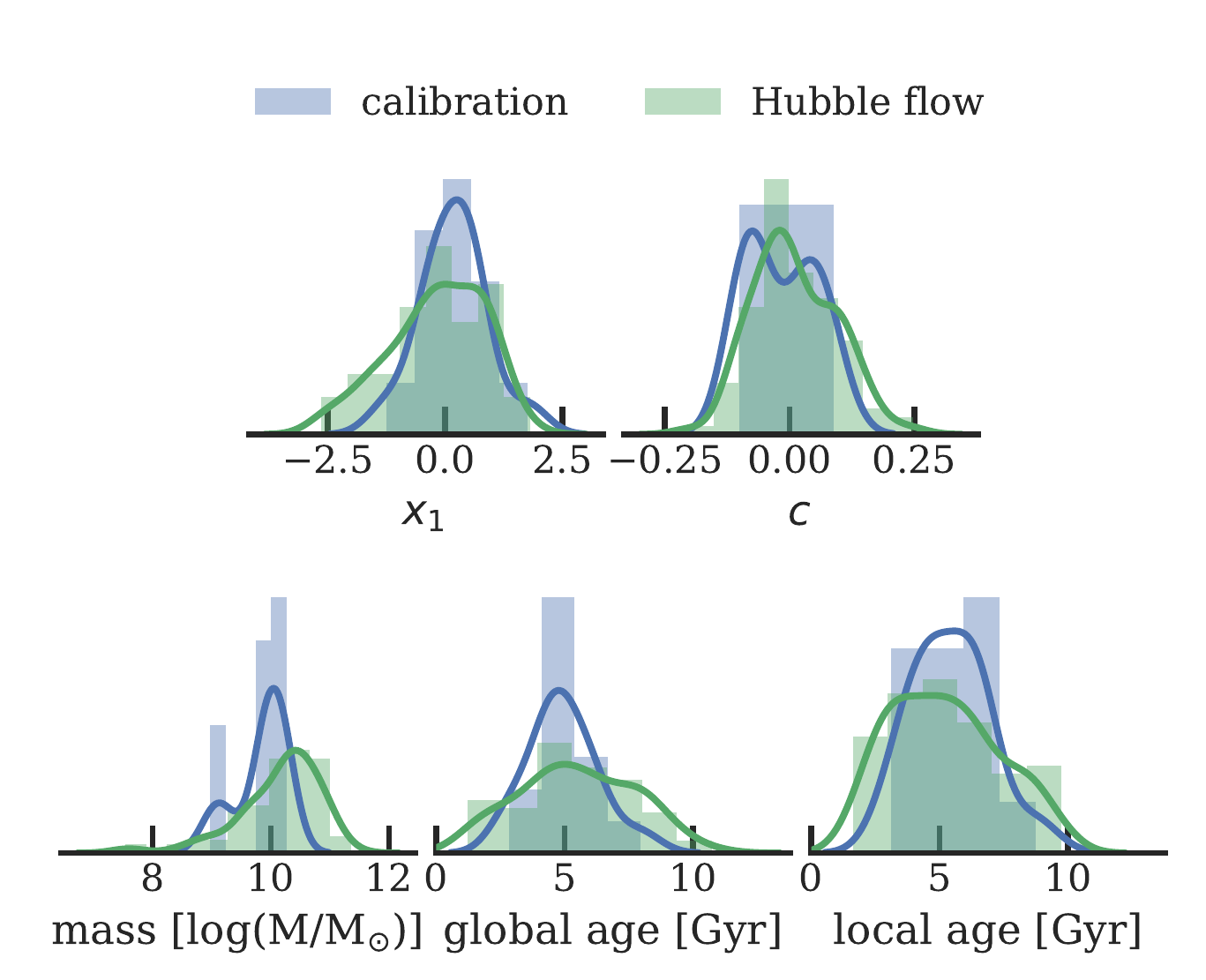}
    \caption{A comparison, between the local calibration \sn{} sample (blue) and the Hubble flow \sn{} data set (green), of the distribution of each parameter (stretch, color, host galaxy stellar mass, and average age). Histograms and kernel density estimations are shown. The  Hubble flow data set has a lower $x_1$ minimum, as expected since it has more passive galaxies. The color distributions appear to be very similar. The calibration galaxies have a lower average mass including an additional peak at $\sim 10^9\un{M}_{\odot}$ that is not seen in the Hubble flow data. This is an expected bias. The calibration age distribution is similar to, but not exactly the same as the Hubble flow sample. The main contrast is the dip in old ($> 7\un{Gyr}$) populations.
    When looking at each variable independently, only mass is drastically different, and the \h{} measurement in \citet{Riess2016} already corrects for this effect.
    }
    \label{fig:H0_comp}
\end{figure}

SDSS imaging was available for 14 of the 19 \sn{} hosts presented in Table 1 of \citet{Riess2016}. The others were outside of the SDSS footprint and without SDSS-$u$ photometry, and therefore we were not able to apply our age estimator in a consistent way. 
For this calibration sample, we followed the same procedures to get the local environment age, global age, and host stellar mass. 
These values, along with the full list of SALT2 parameters can be found in \cref{tab:h0-cal}. A visual comparison of these parameter's distributions can be seen in \cref{fig:H0_comp}.

\begin{figure}
    \centering
    \includegraphics[width=3in]{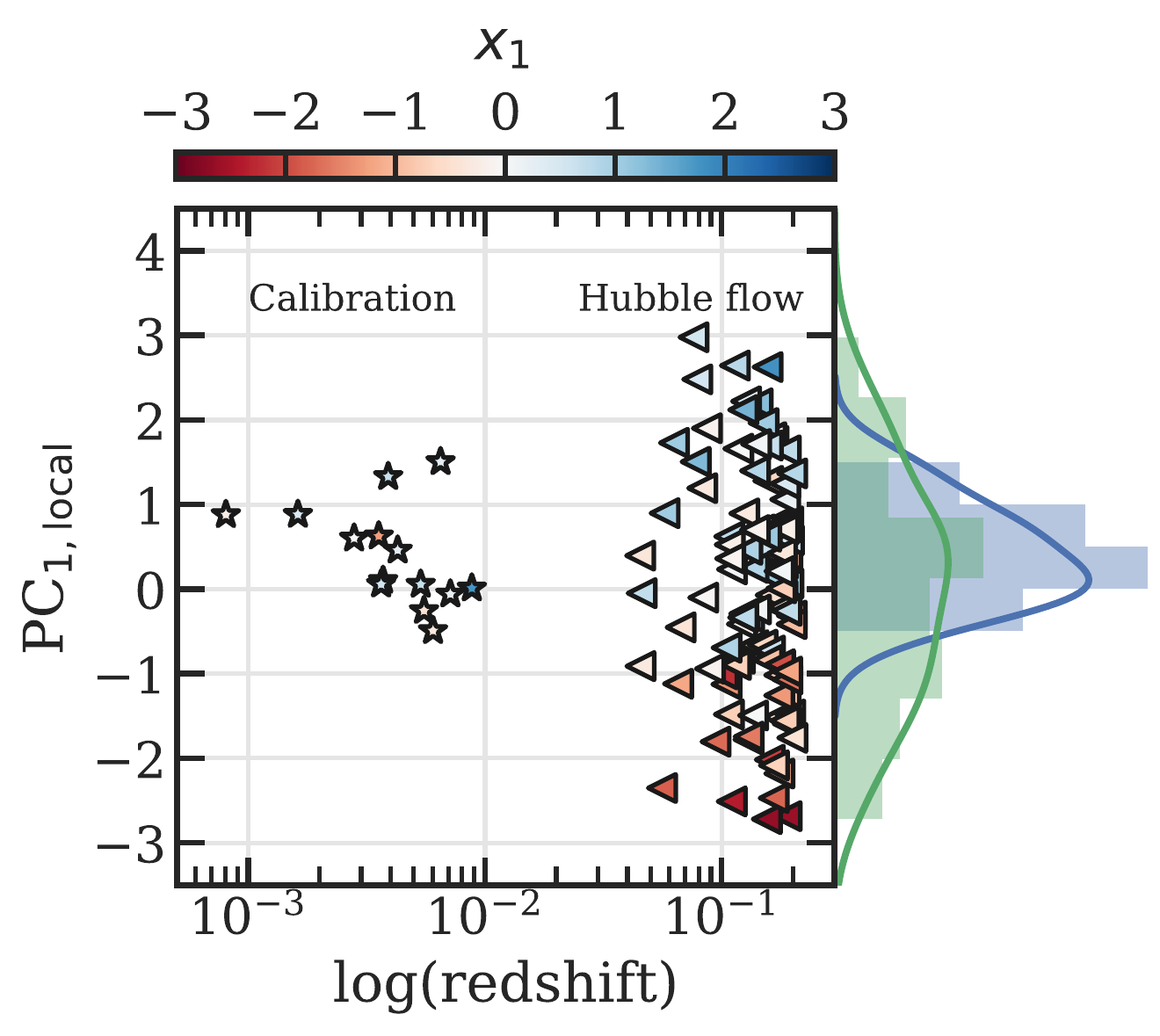}
    \caption{Using local age, a comparison of the two PC$_{1, {\rm local}}$ distributions for the calibration and Hubble flow \sn{} samples. Histograms and kernel density estimations are shown on the vertical axis.  The calibration sample has a higher mean value ($0.407 \pm 0.15$, $2.7 \sigma$) than the Hubble flow sample ($0$ by definition). A KS-test concludes that these samples are different at a $1.9\sigma$ significance. A Mann-Whitney U test says that the calibration sample has a higher mean at a $1.2\sigma$ significance. Using the relationship between \hr{} and PC$_1$, this difference would translate to a \deltaMagLocal{} shift in peak luminosity, or a \hShiftPercentLocal{} effect on \h{}.
    }
    \label{fig:H0_pc1_local}
\end{figure}

\begin{figure}
    \centering
    \includegraphics[width=3in]{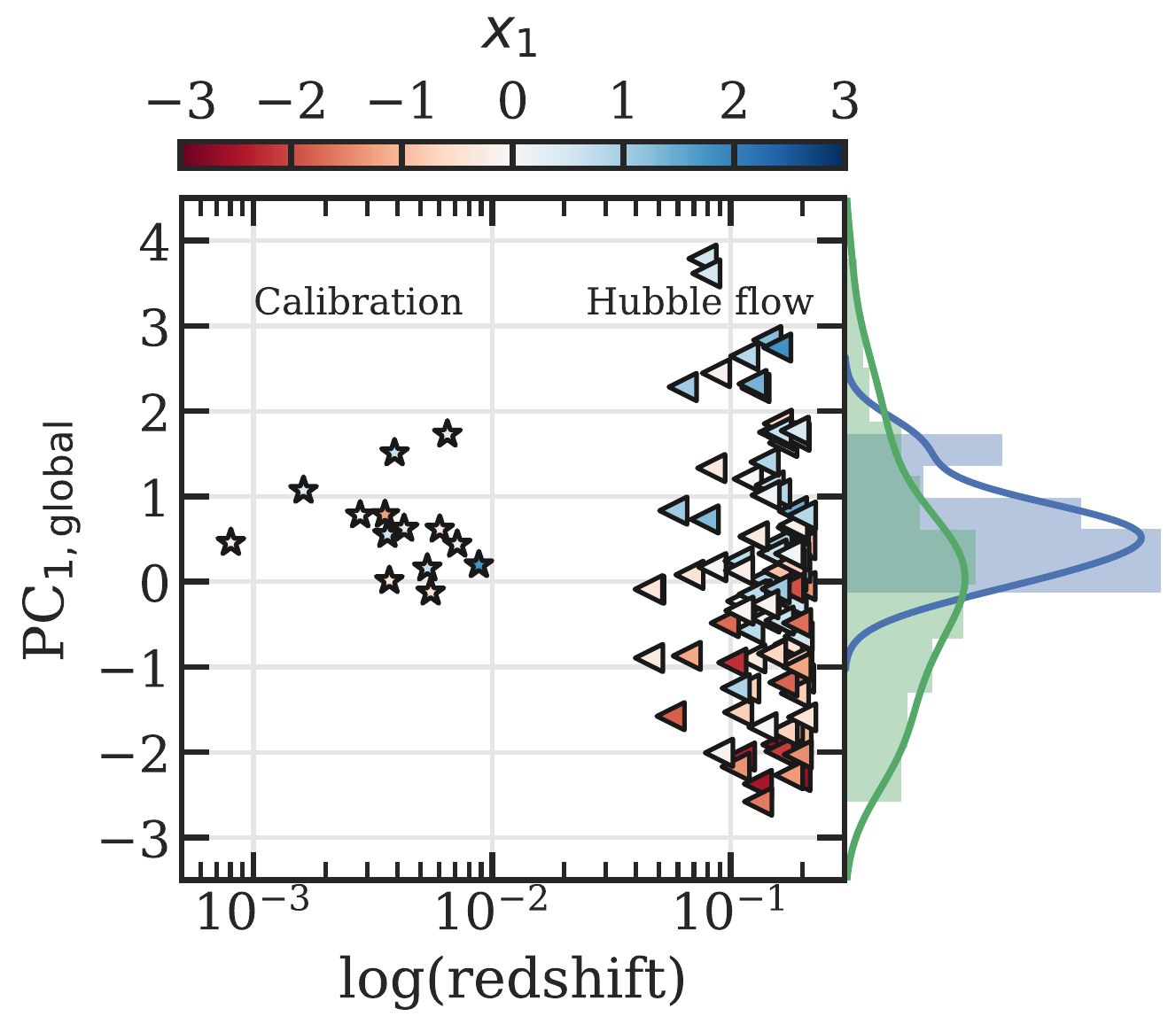}
    \caption{Same as \cref{fig:H0_pc1_local}, but with global ages. Here, the calibration sample mean is $0.63 \pm 0.14$ ($4.6\sigma$). The KS-test see a difference at a $2.7\sigma$ significance, and the Mann-Whitney U test at a $2.3\sigma$ significance. This difference would translate to a \deltaMagGlobal{} shift in peak luminosity, or a \hShiftPercentGlobal{} effect on \h{}.
    }
    \label{fig:H0_pc1_global}
\end{figure}

The individual parameters of PC$_1$ have only mild differences between the \sn{} calibration sample and the Hubble flow hosts. The most extreme fast decliners ($x_1 < -2$) are only in the Hubble flow sample because they are preferentially found in passive galaxies. The mass distributions are significantly different, with a low-mass tail in the calibration sample. The Hubble flow sample has a wider distribution of local ages than the calibration set, same with the global age. What is interesting about PC$_1$ is that all these slight differences work in the same direction. The lower stretch, higher mass and older ages in the Hubble flow combine for a significantly lower average PC$_1$ when compared with the larger stretch, lower mass, younger hosts from the calibration sample.

For the calibration sample, we calculated a PC$_1$ from both the local and global age PCA methodology, PC$_{1, {\rm local}}$ and PC$_{1, {\rm global}}$ respectively. These values can also be seen in \cref{tab:h0-cal}.
A comparison of these two PC$_1$ distributions are shown in \cref{fig:H0_pc1_local,fig:H0_pc1_global}.


Using the local age, the calibration sample in \cref{fig:H0_pc1_local} has a mean PC$_{1, {\rm local}}$ of $0.41 \pm 0.15$ ($2.7 \sigma$), while the Hubble flow PC$_1$ mean is defined as zero. 
To account for the sample sizes, we performed the Kolmogorov-Smirnov test (KS-test) resulting in a 5.5\% chance ($1.9 \sigma$) that these samples are drawn from a common distribution and a Mann-Whitney U test indicates that the calibration sample has a higher mean at a $1.2 \sigma$ level. 
Using the best fit of the correlation between \hr{} and PC$_1$, this corresponds to a difference of \deltaMagLocal{}.

Rerunning this analysis with the global age PCA normalization and methodology, there is a $0.63 \pm 0.14$ ($4.6\sigma$) shift in the PC$_{1, {\rm global}}$ means. The resulting KS-test says there is a 0.8\% chance ($2.7\sigma$) that these two samples are drawn from a common distribution. A Mann-Whitney U test indicates that the calibration sample has a higher mean at a $2.3\sigma$ significance. Applying this shift to the trend seen in \cref{fig:pca-global} (a), there is a shift in peak luminosity
of \deltaMagGlobal{}.

These shifts in peak luminosity produces \added{at most} a \hShiftPercentLocal{} or a \hShiftPercentGlobal{} effect on \h{} respectively. This is \replaced{more than}{about} twice the size \replaced{as}{of} the already accounted for mass step, but is less than the current $1 \sigma$ uncertainty in \h{} (2.3\%).
A large \sn{} systematic effect was found but it had a minimal effect on \h{}. 
For this effect to relieve the full $3.8 \sigma$ tension \citep{Riess2018b}, these two samples would need to have a PC$_1$ shift of \deltaPCNeeded{}, about \fractionLargerPCNeeded{} times larger than currently seen. The \sn{} systematic found in this paper is \replaced{very}{extremely} unlikely to fully resolve the \h{} tension.

\section{Conclusion}

Host galaxy properties have an effect on the absolute magnitude of \sn{}. Host galaxy stellar mass, age, and metallicity have all been shown to be a secondary correction to the Phillips relation. Using a Bayesian method to estimate the age, we were able to look at how Hubble residuals of \sn{} correlate with the mass weighted average age for both the local environment and the galaxy as a whole. This method is better at correctly estimating younger populations than previous methods. A \globalCorrSig{} significant correlation between \hr{} and age was seen. This correlation \replaced{appears as}{may be} an age step of \ageStep{} at $\sim \replaced{7\un{Gyr}}{\ageStepLocation}$. This step is nearly twice the size of the currently used mass step.

Running this analysis on both the local environment and the galaxy as a whole showed that the local age did not show any stronger of a systematic than the global age, but as expected, the local age is younger than the global age of the host for \sn{} in young populations.

We are unable to completely replicate the predictions of \cite{Childress2014} for the distribution of \sn{} hosts in the space of age versus stellar mass. 
The bi-modal \sn{} age distribution is not present in the data set derived from \citetalias{Campbell2013}, but the general trends of old galaxies with high stellar masses and a tail of low-mass young galaxies are seen in the data.

Using PCA on the two SALT2 parameters, host stellar mass, and local environment age, we see a very significant correlation (\pcOneCorr{}) between Hubble residual and the first principal component ${\rm PC}_1 = 0.56 x_1 - 0.10 c - 0.54 m' - 0.63 a'$. This trend was fit with a slope of \pcOneSlopeFull. 
The mixture of parameters making up PC$_1$ suggests that to understand the luminosity variations in \sn{} and to properly correct for them requires simultaneous knowledge of their host and supernova properties.
\added{This data set lacked any significant correlations between \hr{} and $x_1$ or $c$, but the combination of PC$_1$ does have a significant correlation with \hr{}.}
As a result of this significant trend, PC$_1$ should be used as part of an an updated light curve fitter.

The dominant components of PC$_1$ are stretch, mass, and age. Using the Mannucci relationship, PC$_1$ may be implying that $\alpha$ has a metallicity dependence. A theoretical case for this was already made by \citet{Kasen2009}, and the observational trends found in this our work are able to reproduce this predicted effect.

A correlation of this magnitude could have major effects on the precision measurement of \h{}. Looking at the difference in the calibration \added{sample} and \added{a proxy} Hubble flow \replaced{samples}{sample}, we see that these data sets have a meaningful difference in PC$_1$. Using the PCA methodology and normalization from the global age analysis, there is a shift in \added{mean} PC$_{1, {\rm global}}$ of $0.63 \pm 0.14$  (4.6$\sigma$). In addition, a KS-test shows that they are drawn from different underling populations at the $2.7\sigma$ significance level, and a Mann-Whitney U test says the calibration sample has a higher mean at a $2.3\sigma$ significance. Similar differences in the mean were seen using the local age. This difference between these two samples would correspond to a shift in \sn{} peak absolute magnitude of \deltaMagGlobal{}, or \added{at most} a \hShiftPercentGlobal{} shift in \h{}. \added{This analysis only places an upper limit on this effect, because several minor bias corrections were not applied.} \replaced{This only has}{With at most} a $\sim 0.5 \sigma$ effect on \h{},\deleted{ and} this \replaced{very}{is extremely} unlikely to relieve the full $3.8\sigma$ tension between the most recent measurements of the CMB from the Planck collaboration.

A major systematic in \sn{} was discovered, but it had only a small effect on \h{}. This correction should be further investigated and applied to \sn{} used in cosmological studies. Moreover, it appears that even a large \sn{} systematic cannot fully relieve the tension between the local and CMB measurements of \h{}.


\acknowledgements
The authors would like to thank Chris Wotta and Eric Bechter who provided invaluable feedback on improvements to the analysis code's design and Erika Holmbeck for suggestions on the figures. In addition, thank you to David Rubin\replaced{ and Daniel Foreman-Mackey for your}{, Daniel Foreman-Mackey, David Jones, Adam Riess, and the anonymous referee for their} informative comments.
Funding for this research was in part by David Eartly and the Lennox Fellowship. 
This research was supported in part by the Notre Dame Center for Research Computing.

Funding for the SDSS and SDSS-II has been provided by the Alfred P. Sloan Foundation, the Participating Institutions, the National Science Foundation, the U.S. Department of Energy, the National Aeronautics and Space Administration, the Japanese Monbukagakusho, the Max Planck Society, and the Higher Education Funding Council for England. The SDSS Web Site is http://www.sdss.org/.

The SDSS is managed by the Astrophysical Research Consortium for the Participating Institutions. The Participating Institutions are the American Museum of Natural History, Astrophysical Institute Potsdam, University of Basel, University of Cambridge, Case Western Reserve University, University of Chicago, Drexel University, Fermilab, the Institute for Advanced Study, the Japan Participation Group, Johns Hopkins University, the Joint Institute for Nuclear Astrophysics, the Kavli Institute for Particle Astrophysics and Cosmology, the Korean Scientist Group, the Chinese Academy of Sciences (LAMOST), Los Alamos National Laboratory, the Max-Planck-Institute for Astronomy (MPIA), the Max-Planck-Institute for Astrophysics (MPA), New Mexico State University, Ohio State University, University of Pittsburgh, University of Portsmouth, Princeton University, the United States Naval Observatory, and the University of Washington.

\facility{SDSS} 
\software{astropy \citep{Astropy}, astroquery (\doi{10.6084/m9.figshare.805208}), corner.py \citep{Foreman-Mackey2016}, emcee \citep{Foreman-Mackey2012}, FSPS \citep{Conroy2009, Conroy2010}, kcorrect \citep{Blanton2007}, Matplotlib \citep{matplotlib}, Numpy \citep{numpy}, Pandas \citep{pandas}, Python-FSPS \citep{Foreman-Mackey2014}, scikit-learn \citep{scikit-learn}, SciPy \citep{scipy}, Seaborn (\doi{10.5281/zenodo.883859}), sfdmap (\doi{}), SIMBAD \citep{Wenger2000}, sncosmo (\doi{10.5281/zenodo.592747})}

\bibliographystyle{apj}
\explain{There have also been updates and additions to the reference list.}
\bibliography{library}



\listofchanges

\end{document}